\documentclass[11pt]{article}
\usepackage{axodraw}
\usepackage{epsfig}
\usepackage{amsfonts}
\usepackage{amsmath}
\usepackage{bbm,bm}
\usepackage{cite}
\usepackage{jheppub}
\allowdisplaybreaks[1]

\input pix.sty

\newcommand{\nref}[1]{(\ref{#1})}
\newcommand{\kT}{k_\rmii{$T$}}
\newcommand{\qp}{p_{+}} 
\newcommand{\qm}{p_{-}} 
\newcommand{\munLa}{\mu_{\nu_{\rmii{$L$}a}}}
\newcommand{\mueLa}{\mu_{e_{\rmii{$L$}a}}}
\newcommand{\muphin}{\mu_{\phi^{ }_{0}}}
\newcommand{\muphip}{\mu_{\phi^{ }_{+}}}
\newcommand{\muA}{\mu_\rmii{\!$A$}}
\newcommand{\muB}{\mu_\rmii{$B$}}
\newcommand{\muW}{\mu_\rmii{$W^+$}}
\newcommand{\muZn}{\mu_\rmii{$Z^0$}}
\newcommand{\tmuBL}{\tilde{\mu}_\rmii{$B$+$L$}}

\newcommand{\YBpL}{Y_\rmii{$B$+$L$}}
\newcommand{\YB}{Y_\rmii{$B$}}

\newcommand{\muY}{\mu_\rmii{$Y$}}
\newcommand{\muQ}{\mu_\rmii{$Q$}} 
\newcommand{\muZ}{\mu_\rmii{$Z$}}

\newcommand{\bmuZ}{\bar{\mu}_\rmii{$Z$}}
\newcommand{\bmuQ}{\bar{\mu}_\rmii{$Q$}}  
\newcommand{\bmuB}{\bar{\mu}_\rmii{$B$}}

\newcommand{\lnf}{l^{ }_\rmi{1f}}
\newcommand{\lif}{l^{ }_\rmi{2f}}
\newcommand{\lnb}{l^{ }_\rmi{1b}}
\newcommand{\lib}{l^{ }_\rmi{2b}}

\newcommand{\I}{\rmii{$I$}}
\newcommand{\J}{\rmii{$J$}}

\newcommand{\nS}{n_\rmii{$S$}}
\newcommand{\nG}{n_\rmii{$G$}}
\newcommand{\mW}{m_\rmii{$W$}}
\newcommand{\mZ}{m_\rmii{$Z$}}
\newcommand{\mWt}{m_\rmii{$\widetilde W$}}
\newcommand{\mZt}{m_\rmii{$\widetilde Z$}}
\newcommand{\mQt}{m_\rmii{$\widetilde Q$}}
\newcommand{\mWb}{m_{\bar{\rmii{$W$}}}}
\newcommand{\mZb}{m_{\bar{\rmii{$Z$}}}}
\newcommand{\mQb}{m_{\bar{\rmii{$Q$}}}}
\newcommand{\mH}{m_\rmii{$H$}}

\renewcommand{\vec}[1]{{\bf #1}}

\newcommand{\aL}{a^{ }_\rmii{L}}
\newcommand{\aR}{a^{ }_\rmii{R}}

\newcommand{\Nc}{N_{\rm c}}

\newcommand{\mE}{m_\rmii{E}}

\newcommand{\rmO}{{\mathcal{O}}}
\newcommand{\bmu}{\bar\mu}

\def\lsi{\raise0.3ex\hbox{$<$\kern-0.75em\raise-1.1ex\hbox{$\sim$}}}
\def\gsi{\raise0.3ex\hbox{$>$\kern-0.75em\raise-1.1ex\hbox{$\sim$}}}
\newcommand{\lsim}{\mathop{\lsi}}
\newcommand{\gsim}{\mathop{\gsi}}

\newcommand{\nF}{n_\rmii{F}}
\newcommand{\nB}{n_\rmii{B}}
\newcommand{\rmii}[1]{{\mbox{\tiny\rm{#1}}}}
\newcommand{\rmiii}[1]{{\mbox{\tiny{$\scriptstyle{\rm#1}$}}}}
\newcommand{\re}{\mathop{\mbox{Re}}}
\newcommand{\im}{\mathop{\mbox{Im}}}
\newcommand{\Tint}[1]{{\hbox{$\sum$}\!\!\!\!\!\!\!\int\,}_{\!\!\!\!\raise-0.9ex\hbox{$\scriptstyle{#1}$}}}
\newcommand{\Tinti}[1]{{{\Sigma}\!\!\!\!\raise0.3ex\hbox{$\int$}_\rmii{${#1}$}}}
\newcommand{\Tintip}[1]{{{\Sigma'}\!\!\!\!\!\raise0.3ex\hbox{$\int$}_\rmii{${#1}$}}}

\newcommand{\unit}{{\mathbbm{1}}} 
\newcommand{\bi}{\begin{itemize}}
\newcommand{\ei}{\end{itemize}}
\newcommand{\hide}[1]{ }
\newcommand{\bsl}[1]{\,\slash\!\!\!\!{#1}\,}
\newcommand{\msl}[1]{\,\slash\!\!\!{#1}\,}
\newcommand{\deltabar}{\,\raise-0.02em\hbox{$\bar{}$}\hspace*{-0.8mm}{\delta}}

\makeatletter \@addtoreset{equation}{section} \makeatother
\renewcommand{\theequation}{\arabic{section}.\arabic{equation}}
\makeatletter
\renewcommand\section{\@startsection {section}{1}{\z@}%
                                   {-5.5ex \@plus -1ex \@minus -.2ex}
                                   {2.3ex \@plus.2ex}%
                                   {\normalfont\large\bfseries}}
\renewcommand\subsection{\@startsection{subsection}{2}{\z@}%
                                     {-3.25ex\@plus -1ex \@minus -.2ex}%
                                     {1.5ex \@plus .2ex}%
                                     {\normalfont\normalsize\bfseries}}
\renewcommand\thesection {\@arabic\c@section}
\renewcommand\thesubsection   {\thesection.\@arabic\c@subsection}
\renewcommand{\@seccntformat}[1]{%
\csname the#1\endcsname.\hspace{1.0em}}
\makeatother

\begin{document}

\flushbottom


\begin{flushright}
CERN-TH-2018-232 \\
January 2019
\end{flushright}

\vspace*{0.5cm}

\title{Precision study of GeV-scale resonant leptogenesis}

\author[a]{J.~Ghiglieri,}
\author[b]{M.~Laine}

\affiliation[a]{%
Theoretical Physics Department, CERN, \\ 
CH-1211 Geneva 23, Switzerland}
\affiliation[b]{%
AEC, 
Institute for Theoretical Physics, 
University of Bern, \\ 
Sidlerstrasse 5, CH-3012 Bern, Switzerland}

\emailAdd{jacopo.ghiglieri@cern.ch}
\emailAdd{laine@itp.unibe.ch}

\abstract{%
Low-scale leptogenesis is most efficient in the limit of an extreme mass
degeneracy of right-handed neutrino flavours.  Two variants of this situation
are of particular interest: large neutrino Yukawa couplings, which boost the
prospects of experimental scrutiny, and small ones, which may lead to large
lepton asymmetries surviving down to $T < 5$~GeV.  We study benchmarks of
these cases within a ``complete'' framework which tracks both helicity states
of right-handed neutrinos as well as their kinetic non-equilibrium, and
includes a number of effects not accounted for previously. For two right-handed
flavours with GeV-scale masses, Yukawa couplings up to $|h|\sim 0.7\times
10^{-5}$ are found to be viable for baryogenesis, with $\Delta M/M \sim
10^{-8}$ as the optimal degeneracy.  Late-time lepton asymmetries are most
favourably produced with $\Delta M/M \sim 10^{-11}$.  We show that the system
reaches a stationary state at $T < 15$~GeV, in which lepton asymmetries
can be more than $10^{3}$ times larger than the baryon asymmetry, reach 
flavour equilibrium, and balance against helicity asymmetries.
}


 
\maketitle

%
\section{Introduction}
\la{se:intro}

An extension of the Standard Model through two or three 
generations of right-handed neutrinos, which account for
the observed active neutrino mass differences and mixings, offers for 
a simple explanation of the baryon asymmetry 
in the present universe~\cite{classic}.
The Euclidean Lagrangian is 
\ba
 L^{ }_\rmii{E} & \equiv & 
 L^{ }_\rmi{old-SM} 
 + \bar{\nu}^{ }_\rmii{\!$R$} \msl{\partial} \nu^{ }_\rmii{\!$R$} 
 +  
 \tilde\phi^\dagger \bar{\nu}^{ }_\rmii{\!$R$} h_{ }^{ }\,\ell^{ }_\rmii{\!$L$}
 + 
 \bar{\ell}^{ }_\rmii{\!$L$} h_{ }^\dagger  \nu^{ }_\rmii{\!$R$} \tilde\phi
 +  
 \frac{1}{2}
 \bigl(
 \bar{\nu}_\rmii{\!$R$}^c M^{ }_{ } \nu^{ }_\rmii{\!$R$} 
 + 
 \bar{\nu}_\rmii{\!$R$}^{ } M^{\dagger}_{ } \nu^c_\rmii{\!$R$} 
 \bigr)
 \;,  
\ea
where $h$ is a Yukawa matrix, $M$ a Majorana mass matrix, 
$\ell^{ }_\rmii{\!$L$}$ a left-handed lepton doublet, 
and $\tilde\phi = i \sigma^{ }_2\phi^*$ a conjugated Higgs doublet.  
After a singular value decomposition and field rotation we may 
assume $M = \mathop{\mbox{diag}}(M^{ }_1, M^{ }_2, M^{ }_3)$, 
where $M^{ }_\rmii{$I$} \ge 0$. In the following we focus on 
the minimal case that effectively only 
two generations (with masses $M^{ }_1$, $M^{ }_2$) 
play a role; this is sufficient for explaining 
all known active neutrino properties. 

In its classic implementation~\cite{classic}, leptogenesis assumes
that $M^{ }_{\I} \gg 200$~GeV, so that right-handed neutrinos become
non-relativistic and fall out of equilibrium at a time when baryon
number violating interactions through sphaleron processes
are still in thermal equilibrium~\cite{krs}. 
If the Majorana masses are furthermore assumed to be ``hierarchical'',  
only the lightest among them plays a substantial role in leptogenesis. 
This prototypical example has been studied to great detail by now, including 
the effect of radiative 
corrections~(cf.,\ e.g.,\ refs.~\cite{db1,nora,db2,racker}). The drawback
of this scenario is that it is not falsifiable:  
leptogenesis depends on high-energy
parameters which cannot be uniquely fixed in low-energy experiments
(cf.,\ e.g.,\ ref.~\cite{testable}). 

Falsifiability can be boosted by making the right-handed 
neutrinos light. If we push their mass scale all the way down to 
the vicinity of a cosmologically admissible lower bound
$M^{ }_{\I}\sim 0.1$~GeV~\cite{Neff}, 
right-handed neutrinos could become accessible e.g.\ 
to B-factory type experiments. The price
to pay is that a certain degree of mass degeneracy 
is then needed. We refer to this 
framework~\cite{ars,as,singlet,kinetic,late} as 
``low-scale resonant leptogenesis''. The near-degeneracy can be 
argued to be ``natural'' in the sense that it may originate from 
a slightly broken symmetry (cf.,\ e.g.,\ ref.~\cite{symmetry}). The neutrino 
Yukawa couplings can be tuned relatively large, perhaps
making the framework particularly well suited for experimental 
detection. 
The purpose of the current paper is 
to scrutinize the parameter space of this scenario, 
following many recent 
investigations~\cite{broken,old1,old2,n3,eijima,cptheory,%
asaka,ht,new1,shintaro_new,new2,cpnumerics,new3,new4}. 

Right-handed neutrino oscillations become efficient when 
the oscillation rate of a comoving momentum mode $k$
equals the Hubble rate, i.e.\ around the temperature 
\be
 T^{ }_\rmi{osc} \sim 700 \, \mbox{GeV} 
 \, \left( \frac{M}{\mbox{GeV}} \, \frac{|\Delta M|}{\mbox{eV}} 
 \, \frac{T_\rmi{osc}}{k} \right)^{1/3}
 \;, 
\ee
where 
$M \equiv ({M^{ }_1 + M^{ }_2})/2$ and 
$\Delta M \equiv M^{ }_2 - M^{ }_1$. 
Baryon asymmetry generation through sphaleron processes stops at 
$T^{ }_\rmi{sph} \sim 130$~GeV~\cite{sphaleron}.
If we make $\Delta M$ very small, 
the dynamics relevant for baryogenesis 
takes place at temperatures just above 
$T^{ }_\rmi{sph}$~\cite{shintaro_new}.
Electroweak crossover is at $T\sim 160$~GeV~\cite{crossover,dono}, 
and therefore we may find ourselves on the side 
of the Higgs phase\footnote{%
 We refer to the Higgs phase alternatively as a ``broken'' phase, 
 even if strictly speaking the Standard Model 
 gauge symmetries do not get broken. 
 }
in this situation. 

Our study is based on a quantum field theoretic 
formalism that we have developed in a series of previous
papers~\cite{broken,cptheory,cpnumerics}, drawing upon
earlier investigations~\cite{bb1,bb2,interpolation,dmpheno,bsw}. 
The system is characterized
by a number of 
slow equilibration rates, which are 
mediated by neutrino Yukawa couplings and are of magnitude 
$\sim |h|^2 g^2 T/\pi$, where $g^2 \equiv 4\pi\alpha$ is a generic 
Standard Model coupling, 
as well as by a slow flavour oscillation 
rate, which is of magnitude $\sim |M_2^2 -M_1^2|/k$. 
The slow rates imply that right-handed
neutrinos are neither in chemical, nor in kinetic, 
nor in helicity, nor in flavour equilibrium, and need 
to be tracked through density matrices. The equilibration rates
contain both ``direct'' and ``indirect'' contributions, 
with the former referring to $1\leftrightarrow 2$ and $2\leftrightarrow 2$
decays or scatterings and the latter to rates experienced by 
off-shell left-handed neutrinos, which subsequently ``oscillate''
into right-handed neutrinos thanks to the presence
of the Higgs mechanism at $T \lsim 160$~GeV. 

The plan of this paper is as follows. 
After reviewing the overall theoretical framework in \se\ref{se:overview}
and the parametrization of a charge-asymmetric ensemble in the 
presence of a Higgs mechanism in \se\ref{se:ensemble}, we discuss
the structure of the indirect contribution
in \se\ref{se:indirect}, keeping 
consistently track of both helicity states. All ingredients
appearing in the rate coefficients are computed in 
\se\ref{se:rates}, generalizing previous results in order to account
for both helicity
states and the presence of chemical potentials. The direct 
contributions are discussed in \se\ref{se:direct}, again 
resolving existing results to the chemical potential assignments
relevant for the Higgs phase. The resulting system is solved 
numerically in an approximate form in \se\ref{se:scan}, in order
to identify relevant corners of the parameter space. A more 
precise solution is presented in \se\ref{se:large_h}, 
for a benchmark with large Yukawa couplings, 
and in \se\ref{se:small_h}, 
for a benchmark with small ones. 
We conclude in \se\ref{se:concl}, and 
relegate some technical details to four appendices. 

%
\section{Overview of the framework}
\la{se:overview}

We start by summarizing the form 
of the master equations that were 
derived in ref.~\cite{cptheory} from operator
equations of motion and from arguments based on a separation of time scales. 
The variables considered are the yield parameters for lepton minus 
baryon asymmetries, $Y^{ }_a - \YB^{ }/3$, and the helicity-symmetrized and 
antisymmetrized density matrices for right-handed neutrinos, 
$\rho^{\pm}_{ }(k)$.
The cosmological evolution is conveniently tracked through a variable
$x \equiv \ln(T^{ }_\rmi{max}/T)$, where $T^{ }_\rmi{max}$ is the 
temperature at which we start the evolution, 
and momentum through the co-moving
variable $\kT^{ } \equiv k \, [s(T)/s(T^{ }_\rmi{min})]^{\fr13}$, 
where $s$ is the entropy density and 
$T^{ }_\rmi{min}$ the 
temperature at which we stop the evolution. 
The yield parameters evolve as 
\be
  {Y}_a' - \frac{\YB'}{3}  \; = \;  
  \frac{4}{s} \int_{\vec{k}^{ }_\rmii{$T$}} 
  \!\! \tr 
  \Bigl\{ 
   - \, \nF^{ }(\kT^{ }) [1 - \nF^{ }(\kT^{ })] \,  \widehat A^{+}_{(a)}
   + \bigl[\, \rho^{+}_{ } - \unit \, \nF^{ }(\kT^{ }) \, \bigr]
     \widehat B^{+}_{(a)}
   + \rho^{-}_{ } \widehat B^{-}_{(a)}
 \Bigr\} 
 \;, \la{evol_Ya}
\ee
where the first structure on the right-hand side may be called a washout term
and the latter structures source terms. The trace goes over the flavour
indices and $\nF^{ }$ denotes the Fermi distribution. 
To $\rmO(h_{\I a}^2,\mu^2)$ the coefficients read 
\ba
 \widehat A^{+}_{(a)\I\J}
 & = & 
  \re (h^{ }_{\I a}h^*_{\J a}) \, 
 \bar{\mu}^{ }_a \,
 \widehat Q^{+}_{(a)\{\I\J\}} 
 \;, \la{ap} \\[2mm]
 \widehat B^{+}_{(a)\I\J}
 & = & 
 - i \im (h^{ }_{\I a}h^*_{\J a}) \, 
    \widehat Q^{+}_{(a)\{\I\J\}} 
  + 
  \re (h^{ }_{\I a}h^*_{\J a}) \, 
  \Bigl[ 
  \bar{\mu}^{ }_a \, 
        \widehat R^{+}_{(a)\{ \I \J\} } 
   + {\textstyle\sum_i}\,\bar{\mu}^{ }_i \,
        \widehat S^{+(i)}_{(a)\{ \I\J\} } 
  \Bigr]
 \;, \la{bp} \\[2mm]
 \widehat B^{-}_{(a)\I\J}
  & = &
  \re (h^{ }_{\I a}h^*_{\J a}) \, 
    \widehat Q^{-}_{(a)\{\I\J\}} 
 - i \im (h^{ }_{\I a}h^*_{\J a}) \, 
  \Bigl[ 
    \bar{\mu}^{ }_a \, 
     \widehat R^{-}_{(a)\{ \I \J\} } 
   + {\textstyle\sum_i}\,\bar{\mu}^{ }_i \, 
        \widehat S^{-(i)}_{(a)\{ \I\J\} }
  \Bigr]
 \;, \la{bm}
\ea
where $h^{ }_{\I a}$ are Yukawas coupling 
a sterile neutrino of flavour $I$ to a lepton of generation~$a$; 
$\bar\mu^{ }_i \equiv \mu^{ }_i/T$ 
are rescaled chemical potentials; 
and rate coefficients $Q,R,S$ 
(to be defined in \se\ref{se:indirect}, 
cf.\ \eqs\nref{Q} and \nref{RS})
are normalized as 
$ 
  \widehat Q \; \equiv \; {Q}/({3 c_s^2 H})
$, 
where $H$ is the Hubble rate and $c_s^2$ 
the speed of sound squared. The superscripts $\pm$ indicate a 
symmetrization/antisymmetrization over helicity, and $\{ IJ \}$
indicates a symmetrization over flavour indices. 
Right-handed neutrino density matrices evolve as 
\ba
  ({\rho}^{\pm}_{ })'(\kT^{ })  & = & 
   i \bigl[\widehat H^{ }_{0}, \rho^{\pm}_{ } \bigr]^{ }_{\!\star}
   + 
   i \bigl[\widehat \Delta^{ }_{0}, \rho^{\mp}_{ } \bigr]^{ }_{\!\star} 
   \; + \; 2 \nF^{ }(\kT^{ }) [ 1-  \nF^{ }(\kT^{ })] \, \widehat C^{\pm}_{ }
  \nn[2mm] 
   & - & 
    \bigl\{ 
    \widehat D^{\pm}_{ }  
    \,,\, \rho^{+}_{ } - \unit \nF^{ }(\kT^{ })
    \bigr\}^{ }_{\!\star}
  - \bigl\{ \widehat D^{\mp}_{ } 
    \,,\, \rho^{-}_{ }
    \bigr\}^{ }_{\!\star}
 \;, \la{evol_rho}
\ea
where 
$
 [A,B]^{ }_{\star} \equiv A B - B^\dagger_{ } A^\dagger_{ }
$,  
$
 \{A,B\}^{ }_{\!\star} \equiv A B + B^\dagger_{ } A^\dagger_{ }
$
(with $\rho^{\pm\dagger}_{ } = \rho^{\pm}_{ }$). 
The coefficients read 
\ba
 \widehat H^{ }_{0\I\J} & = & \;
 \frac{ 
  \delta^{ }_{\I\J}  M_\I^2 
   }{6 \kT^{ } c_s^2 H} 
 \nn[2mm] 
 & + & \; 
 \frac{ \; {\textstyle\sum_a} \re (
     h^{ }_{\I a}h^{*}_{\J a} ) [
      T^2\beta^{+}_{(a)}
     +  v^2 \kappa^{+}_{(a)\I\J} ]
                   - i\,{\textstyle\sum_a} \im  (
     h^{ }_{\I a}h^{*}_{\J a} ) [
      T^2\beta^{-}_{(a)}
     +   v^2 \kappa^{-}_{(a)\I\J} ]
   }{6 \kT^{ } c_s^2 H} 
 \;, \hspace*{1mm} \la{H0} \\[2mm]
 \widehat \Delta^{ }_{0\I\J} & = &   
 \frac{
     - i\, {\textstyle\sum_a} \im(
     h^{ }_{\I a}h^{*}_{\J a} ) [
                 T^2 \beta^{+}_{(a)} 
                +  v^2 \delta^{+}_{(a)\I\J} ]                   
     + {\textstyle\sum_a} \re(
     h^{ }_{\I a}h^{*}_{\J a} ) [
                 T^2 \beta^{-}_{(a)}  
                +  v^2 \delta^{-}_{(a)\I\J} ] 
    }{6\kT^{ } c_s^2 H }    
 \;, \hspace*{4mm} \la{D0} \\[2mm]
 \widehat C^{+}_{\I\J} & = & 
 - i {\textstyle\sum_a} 
   \im (h^{ }_{\I a} h^{*}_{\J a})\, \bar{\mu}^{ }_a  
   \, \widehat Q^{+}_{(a)\{\I\J\} }
 \;,  \quad 
 \widehat C^{-}_{\I\J} \; = \; 
 {\textstyle\sum_a} 
 \re (h^{ }_{\I a} h^{*}_{\J a})\, \bar{\mu}^{ }_a  
   \, \widehat Q^{-}_{(a)\{\I\J\} }
 \;,  \la{cp} \\[3mm] 
 \widehat D^{+}_{\I\J} & = & 
 {\textstyle\sum_a}
   \re (h^{ }_{\I a} h^{*}_{\J a})   
   \,  
   \widehat Q^{+}_{(a)\I\J}
  -  i \, {\textstyle\sum_a} \im (h^{ }_{\I a} h^{*}_{\J a})  \, 
  \Bigl[ \bar{\mu}^{ }_a \, \widehat R^{+}_{(a)\I\J}
   + {\textstyle\sum_i}\, \bar{\mu}^{ }_i\, \widehat S^{+(i)}_{(a)\I\J}
  \Bigr] \,  
 \;, \la{dp} \\[2mm]
 \widehat D^{-}_{\I\J} & = & 
  -  i \, 
  {\textstyle\sum_a} 
  \im (h^{ }_{\I a} h^{*}_{\J a}) 
   \, 
   \widehat Q^{-}_{(a)\I\J}
  + 
 {\textstyle\sum_a}
   \re (h^{ }_{\I a} h^{*}_{\J a})   
   \,  
  \Bigl[ \bar{\mu}^{ }_a \, \widehat R^{-}_{(a)\I\J}
   + {\textstyle\sum_i}\, \bar{\mu}^{ }_i\, \widehat S^{-(i)}_{(a)\I\J}
  \Bigr]
 \;, \la{dm}
\ea
where~$v \simeq 246$~GeV is the Higgs expectation value, 
$
 \kappa^{\pm}_{(a)\I\J}
$
and
$
 \delta^{\pm}_{(a)\I\J}
$
are given in \eq\nref{dispersive}, 
and 
$\beta^{\pm}_{(a)}$
is in \eq\nref{dispersive2}. 
There is also an evolution equation for baryon plus lepton asymmetry, 
specified above \eq\nref{evol_YBpL} and parametrized by 
the Chern-Simons diffusion 
rate $\Gamma^{ }_\rmi{diff}$~\cite{sphaleron}.\footnote{%
 Compared with refs.~\cite{cptheory,cpnumerics}, we have displayed 
 a subscript $(a)$ in $Q,R,S$ because these coefficients can depend
 non-linearly on lepton chemical potentials $\mu^{ }_a$ in the broken phase; 
 we have inserted a superscript $(i)$ in $S$ because a larger 
 set of chemical potentials plays a role; and, most importantly, 
 we have included all the mass corrections relevant for the broken 
 phase, parametrized by the coefficients
 $\beta^{\pm}_{ }$, $\kappa^{\pm}_{ }$
 and $\delta^{\pm}_{ }$.
 } 

To close the set of equations, 
the yields appearing on the left-hand side of \eq\nref{evol_Ya}
and the chemical potentials appearing on the right-hand sides 
of \eqs\nref{evol_Ya} and \nref{evol_rho}
need to be related to each other. This ``static'' relation can be 
established as  
$n^{ }_i = \partial p / \partial \mu^{ }_i$, 
where the $\mu^{ }_i$-dependence
of the pressure~$p$ is specified in \se\ref{se:ensemble}
and in more detail in appendix~A. 

As seen from \eqs\nref{ap}--\nref{bm} and \nref{cp}--\nref{dm}, 
the microscopic
information needed for solving the rate equations is contained in the
mass corrections $\beta,\kappa,\delta$ and in the  
rate coefficients $Q,R,S$, which at high temperatures 
are functions of the temperature~$T$, 
the momentum~$k$, 
and the right-handed neutrino masses~$M^{ }_{\I}$. 
At low temperatures $T \lsim 160$~GeV, when we find ourselves in 
the Higgs phase, the coefficients become more complicated, 
depending also on~$v$ and 
on various particle masses. 
In the class of gauges in which 
the Goldstone modes and the gauge fields 
do not couple to each other, 
the coefficients can be expressed as~\cite{broken}
\be
 Q = Q_{ }^\rmi{direct} + Q_{ }^\rmi{indirect}
 \;. \la{dir_indir}
\ee 
Here the direct contributions refer
to $1+n\leftrightarrow 2+n$ 
and to $2\leftrightarrow 2$ processes also
present in the symmetric phase, 
whereas the indirect contributions are
proportional to~$v^2$, and originate from the ``oscillation'' of 
left-handed (active) neutrinos into right-handed (sterile) ones. 
The direct contributions were derived in ref.~\cite{cptheory}, but 
require a modification with respect to their chemical potential
dependence in the Higgs phase
(cf.\ \se\ref{se:direct}). The indirect contributions
require a lengthier analysis, as we need to generalize the results
of ref.~\cite{broken} to include dependences both on helicity 
and on various chemical potentials. 
After specifying the chemical potentials (\se\ref{se:ensemble}), 
we thus first turn to the indirect contributions
(cf.\ \ses\ref{se:indirect} and \ref{se:rates}).

%
\section{Parametrization of the asymmetric ensemble}
\la{se:ensemble}

As shown in \eqs\nref{ap}--\nref{bm} and \nref{cp}--\nref{dm}, 
we aim to compute 
the coefficients entering the rate equations to leading non-trivial order
in chemical potentials. Having non-zero chemical 
potentials at $T \lsim 160$~GeV implies that the Higgs field
and both neutral components of the gauge potentials develop 
expectation values. The Feynman rules pertinent to this situation
are non-standard and somewhat subtle; 
moreover sign conventions can be a source of trouble. 
We summarize in this section
the conventions and Feynman rules that are needed later on. 

With the density matrix 
\be
 \rho^{ }_\rmii{SM} = \frac{1}{Z^{ }_\rmii{SM}}
 \exp\biggl(-\frac{H^{ }_\rmii{SM} - \sum_a \mu^{ }_a L^{ }_a - \muB^{ } B}{T}
 \biggr)
 \;, \la{rho_SM}
\ee
where 
$
 L^{ }_a \equiv \int_\vec{x}
 [ \bar{\ell}^{ }_{\rmii{$L$}_a} \gamma^{ }_0\,
   \ell^{ }_{\rmii{$L$}_a} + 
 \bar{e}^{ }_{\rmii{$R$}_a} \gamma^{ }_0 {e}^{ }_{\rmii{$R$}_a} ] 
$
is the lepton number for generation $a$,
the part of the Euclidean action containing the kinetic 
terms for $\ell^{ }_{\rmii{$L$}_a}$ is 
\be
 S^{ }_\rmii{E} \supset 
 \int_0^{1/T} \! {\rm d}\tau 
 \int_\vec{x} \bar{\ell}^{ }_{\rmii{$L$}_a} \,
 \bigl(
   \gamma^{ }_\mu D^{ }_\mu  
    - \gamma^{ }_0 \, \mu^{ }_a
 \bigr)
 \, \ell^{ }_{\rmii{$L$}_a}
 \;. 
\ee
The covariant derivative acting on $\ell^{ }_{\rmii{$L$}_a}$ reads
\be
 D^{ }_\mu \;\equiv\; 
 \partial^{ }_\mu 
 - \frac{i g^{ }_1 B^{ }_\mu}{2}
 - \frac{ i g^{ }_2\, \sigma_{a}^{ } A^{a}_\mu}{2}
 \;, 
\ee
where $B^{ }_\mu$ is the hypercharge field and 
$\sigma^{ }_a$ are the Pauli matrices. 
Gauge field backgrounds 
(we employ Euclidean conventions for 
$B^{ }_{\mu}$, $A^a_{\mu}$) are denoted by 
\be
 \muY^{ } \; \equiv \; -i g^{ }_1 B^{ }_0
 \;, \quad
 \muA^{ } \; \equiv \; -i g^{ }_2 A^{3}_0
 \;.  \la{muY_muA}
\ee
The resulting chemical potentials for 
$\nu^{ }_{\rmii{$L$}_a}$, $e^{ }_{\rmii{$L$}_a}$
and for other particles are collected in table~\ref{table:mus}.

%
\begin{table}[t]

{ 
\begin{center}
\begin{tabular}{lclcl}
 \hline\hline
 particle species & & left-handed state & & right-handed state \\
 \hline \hline \\[-3mm]
 up-type quarks & &
 $\displaystyle \mu_{u_{\rmii{$L$}}} \equiv \mu^{ }_q + \frac{\muY}{6}
 - \frac{\muA}{2}$
 & &
 $\displaystyle \mu_{u_{\rmii{$R$}}} \equiv \mu^{ }_q + \frac{2\muY}{3} $ 
 \\[3mm]
 down-type quarks & & 
 $\displaystyle \mu_{d_{\rmii{$L$}}} \equiv \mu^{ }_q + \frac{\muY}{6}
 + \frac{\muA}{2}$
 & &
 $\displaystyle \mu_{d_{\rmii{$R$}}} \equiv \mu^{ }_q - \frac{\muY}{3} $ 
 \\[3mm]
 neutrinos & & 
 $\displaystyle \munLa \equiv \mu^{ }_a - \frac{\muY}{2}
 - \frac{\muA}{2}$
 & &
 \\[3mm]
 charged leptons & & 
 $\displaystyle \mueLa \equiv \mu^{ }_a - \frac{\muY}{2}
 + \frac{\muA}{2}$
 & &
 $\displaystyle \mu_{e_{\rmii{$R$}a}} \equiv \mu^{ }_a - \muY^{ } $ 
 \\[3mm]
  \hline \hline \\[-3mm]
 neutral scalars / $Z^{0}_{ }$ & & 
 $\displaystyle \muphin \equiv \frac{\muY}{2}
 + \frac{\muA}{2}$
 & &
 $\displaystyle \muZn \equiv 0$
 \\[3mm]
 charged scalars / $W^+_{ }$ & & 
 $\displaystyle \muphip \equiv \frac{\muY}{2}
 - \frac{\muA}{2}$
 & &
 $\displaystyle \muW^{ } \equiv -\muA^{ }$
 \\[3mm]
  \hline \hline 
\end{tabular} 
\end{center}
}

\vspace*{3mm}

\caption[a]{\small
  Effective chemical potentials carried by 
  Standard Model particles in the chiral limit, 
  obtained from \eq\nr{rho_SM} (we denote $\mu_q \equiv \muB^{ }/3$)
  and from covariant derivatives after the use of \eq\nr{muY_muA}.
  In the symmetric phase $v \ll T$, we impose \eq\nref{symmetric},  
  so only $\muY^{ }$ plays a role. 
  Deep in the broken phase $v \gg T$, 
  when fermion masses and the chiral anomaly 
  lead to rapid transitions between the two chiral states,
  we impose \eq\nref{broken}, guaranteeing that both chiral states 
  have the same chemical potential.
  The same applies to Goldstone modes and
  the corresponding gauge fields. The intermediate regime $v\sim T$
  is more delicate and the assignments above are only 
  suggestive (cf.\ the text). No chemical potential is indicated for 
  right-handed neutrinos, which are not necessarily
  in chemical equilibrium.
 }
\label{table:mus}
\end{table}
%

Now, in the ``symmetric phase'', 
where the Higgs mechanism is not operative, 
the SU$^{ }_\rmii{L}$(2) gauge symmetry is intact, so within 
a perturbative treatment we should have
\be
 \muA^{ } = 0  \qquad (v \ll T) \;. \la{symmetric}
\ee
In contrast, in the ``broken phase'', 
fermion masses induced by Yukawa couplings, as well as 
the chiral anomaly, violate chirality. 
If we assume that these reactions are in chemical equilibrium and
that a quasiparticle description is viable, we should assign the   
same chemical potential to both chiral states.\footnote{%
 Put another way, only by assigning the same chemical potential
 to both chiral states do we obtain simple 
 propagators for massive particles 
 (top, bottom, Higgs, $W^\pm_{ }$, $Z^0_{ }$). 
 If we violate this condition, which 
 happens in the regime $v\sim T$, chemical potentials should probably
 be treated
 as ``insertions'' within perturbation theory, rather than being 
 resummed into propagators. We have not undertaken this rather
 cumbersome treatment. 
 At the same time the violation of \eq\nref{broken} induces
 a certain free energy cost in the landscape 
 parametrized by $v$, $\muY^{ }$ and $\muA^{ }$, 
 and this has been fully accounted for, 
 as explained around \eq\nref{pressure} and in appendix~A.  
 } 
According to table~\ref{table:mus}, this implies that
\be
 \muA^{ } + \muY^{ } = 0  \qquad (v \gg T,\; \mbox{tree-level}) \;. \la{broken}
\ee
In contrast a large chemical potential can be assigned 
to the electromagnetic field ($\equiv \muQ^{ }$), 
which means that we may write
\be
 \muA^{ } \equiv - \muQ^{ } + (1-s^2) \muZ^{ } 
 \;, \quad 
 \muY^{ } \equiv \muQ^{ } +  s^2 \muZ^{ }
 \;, \quad
 |\muZ^{ }| \sim \frac{T^2}{v^2}\, |\muQ^{ }| \ll |\muQ^{ }|
 \;. \la{broken2} 
\ee
Here $s  \equiv \sin(\tilde{\theta}) $ 
denotes a temperature-dependent weak mixing angle  
(cf.\ \eq\nref{thetat}).  

In the following, we keep both $\muA^{ }$ and $\muY^{ }$ non-zero, 
with the motivation of having expressions
that can be extrapolated both to $v \ll T$ and $v \gg T$.
Furthermore this helps to illustrate the challenges 
that arise in the regime $v\sim T$, $|\muZ^{ }|\sim |\muQ^{ }|$.
We are interested in determining rate coefficients and mass corrections
up to linear order in chemical potentials.
With the choice of \eq\nref{broken},  
terms linear in $\muZ^{ }$ arise from 1-loop 
``tadpoles'' mediated by $Z^0_{ }$ exchange; the corresponding 
value of $\muZ^{ }$ is given in \eq\nref{muZ}. 

Next, consider fluctuations around the minimum of the thermal
Higgs effective potential. The covariant derivative
acting on the Higgs field is given by
\be
 D^{ }_\mu\phi \;\equiv\; 
 \biggl( \partial^{ }_\mu 
 + \frac{i g^{ }_1 B^{ }_\mu}{2}
 - \frac{ i g^{ }_2\, \sigma_{a}^{ } A^{a}_\mu}{2}
 \biggr) \phi
 \;. 
\ee
We write the (fluctuating parts of the) Higgs doublet and gauge potentials as 
\be
 \phi 
    \equiv 
 \biggl(
  \begin{array}{c}
    \phi^{ }_+ \\  \phi^{ }_0  
  \end{array}
 \biggr)
    \equiv 
 \frac{1}{\sqrt{2}} \biggl(
  \begin{array}{c}
    \phi^{ }_2 + i \phi^{ }_1 \\ h - i \phi^{ }_3  
  \end{array}
 \biggr)
 \;, \quad
 W^{+}_\mu  \equiv  \frac{A^1_\mu - i A^2_\mu}{\sqrt{2}} 
 \;, \quad
 Z^{ }_\mu  \equiv  \frac{
   g^{ }_1 B^{ }_\mu + g^{ }_2 A^3_\mu 
 }{\sqrt{g_1^2 + g_2^2}}
 \;. \hspace*{5mm} \la{Zp}
\ee
We also denote $W_{\mu}^- \equiv W_{\mu}^{+*}$
and 
$
   Z'_{\mu} \equiv 
   (g^{ }_1 B^{ }_\mu - g^{ }_2 A^3_\mu ) /  
   \sqrt{g_1^2 + g_2^2}
$. 
Feynman gauge fixing is adopted because it 
simplifies the power counting relevant 
for the ultrarelativistic regime~\cite{broken}. 
The gauge constraints are chosen to contain
components of the background fields, 
\be
 S^{ }_\rmii{E} \supset \int_X \frac{1}{2} 
 \Bigl( {\textstyle\sum_{a=1}^{3}} G_a^2 + G^2 \Bigr)
 \;, 
\ee
where 
\ba
 G & \equiv &  
 \partial^{ }_\mu B^{ }_\mu - \frac{g^{ }_1 v \phi^{ }_3}{2}
 \;, \quad
 G_3^{ } \; \equiv \;  
 \partial^{ }_\mu A^{3}_\mu - \frac{g^{ }_2 v \phi^{ }_3}{2}
 \;, 
 \\ 
 G_1^{ } & \equiv &
 \partial^{ }_\mu A^{1}_\mu - i \muA^{ } A^2_0 - \frac{g^{ }_2 v \phi^{ }_1}{2}
 \;, \quad
 G_2^{ } \; \equiv \;
 \partial^{ }_\mu A^{2}_\mu + i \muA^{ } A^1_0 - \frac{g^{ }_2 v \phi^{ }_2}{2}
 \;. 
\ea

With this gauge fixing, the quadratic part of the charged sector is 
\ba
 S^{ }_\rmii{E} & \supset &
 \Tint{P}
 \biggl\{ 
  \,  W_\mu^-(P)\, W^{+}_\mu(P) \,
    \Bigl[ (p^{ }_n - i \muA^{ })^2 + {p}^2 + \mW^2 \Bigr]
 \nn 
 &  + &
 \phi_+^*(P)\, \phi_+^{ }(P) 
    \Bigl[ \Bigl( p^{ }_n + \frac{i\muY^{ }-i\muA^{ }}{2} \Bigr)^2
  + p^2 + \mW^2
    \Bigr]
 \nn 
 & - & 
 {(i\muA^{ } + i\muY^{ })}\,   \mW^{ } 
  \bigl[
    \phi_+^*(P) \, W^{+}_0(P)  + 
    \phi_+^{ }(P) \, W^{-}_0(P)  
 \, \bigr]
 \; \biggr\} 
 \;,
\ea
where $p^{ }_n$ denotes a bosonic Matsubara frequency 
and $ P \equiv (p^{ }_n,\vec{p})$. It is observed that
with \eq\nref{broken} (or, more generally, 
to linear order in $\muA^{ }+\muY^{ }$), the gauge propagator
obtains a simple form 
(here
$
 \Tinti{\!P}\, \deltabar(P) \equiv 1
$): 
\be
 \bigl\langle W^{ }_\mu(P)\, W^{*}_\nu (Q) \bigr\rangle 
  =  
 \frac{\delta^{ }_{\mu\nu}\, \deltabar(P - Q)}
 {(p^{ }_n - i \muA^{ })^2 + {p}^2 + \mW^2}
 \; + \; 
 \rmO( \muA^{ } + \muY^{ } )^2 \times \rmO(\mW^2)
 \;. \la{propW}
\ee
Similarly, $\phi^{ }_{+}$ can be assigned
the chemical potential 
$\muphip  = (\muY^{ } - \muA^{ })/2$ as given in table~\ref{table:mus}. 

An analogous consideration can be carried out in the neutral sector. 
The mass splitting between the Higgs field $h$ and the neutral Goldstone
$\phi^{ }_3$ complicates matters, so that 
the quadratic part now reads
\ba
 S^{ }_\rmii{E} & \supset &
 \Tint{P}
 \biggl\{ 
  \,\fr12  Z_\mu^{ }(-P)\, Z^{ }_\mu(P) \,
    \Bigl[ p^{2}_n + {p}^2 + \mZ^2 \Bigr]
 \nn 
 &  + &
 \phi_0^*(P)\, \phi_0^{ }(P) 
    \biggl[ \Bigl( p^{ }_n + \frac{i\muA^{ }+i\muY^{ }}{2} \Bigr)^2 + p^2 + 
    \frac{\mH^2 + \mZ^2 }{2}
    \biggr]
 \nn 
 & + &
 \frac{\mH^2 - \mZ^2}{4}
 \bigl[\,\phi^{ }_0(-P)\,\phi^{ }_0(P) + 
  \phi^{*}_0(-P)\,\phi^{*}_0(P)\, \bigr] 
 \nn 
 & + & 
 {(i\muA^{ } + i\muY^{ })}\, 
 \frac{ 
 \mZ^{ }
  }{\sqrt{2}} \,
  \bigl[
     \phi_0^{ }(-P) + 
     \phi_0^{*}(P)  
  \,\bigr]
 \, Z^{ }_0(P)
 \; \biggr\} 
 \;. \la{SEZ}
\ea
The coupling of the temporal gauge
field component to the scalars disappears for 
$\muA^{ }+\muY^{ } = 0$, and the $Z$ propagator reads 
\be
 \bigl\langle Z^{ }_\mu(P)\, Z^{ }_\nu (Q) \bigr\rangle 
  =  
 \frac{\delta^{ }_{\mu\nu}\, \deltabar(P + Q)}
 {p^{2}_n + {p}^2 + \mZ^2}
 \; + \; 
 \rmO( \muA^{ } + \muY^{ } )^2 \times \rmO(\mZ^2)
 \;. \la{propZ}
\ee
For the neutral scalar field $\phi^{ }_0$, a simple propagator
parametrized by $\muphin^{ } = (\muA^{ }+ \muY^{ })/2$ 
can only be obtained if $\mH^{ }= \mZ^{ }$.

In order to fix the values of $\muA^{ }$ and $\muY^{ }$, we need to extremize
the corresponding effective potential~\cite{khlebnikov}. 
The effective potential equals
minus the pressure. Since the chemical potentials are small compared
with the temperature, only the leading non-trivial order is needed, 
and we can indeed treat chemical potentials as insertions. 
Restricting to leading order in Standard Model couplings, the
result can be represented as 
a smooth interpolating function which has correct leading-order 
limits at $\pi T \ll \mW^{ }$ and $\pi T \gg \mW^{ }$~\cite{broken}:
\ba
 p(T,\mu) - p(T,0) & \approx & 
 \sum_a
 \chi^{ }_\rmii{F}(m_{\nu_a}) 
 \, \biggl[ 
  \frac{\mu_a^2}{2}
 - \frac{\muA^{ }\mu^{ }_a}{2}
 - \frac{\muY^{ }\mu^{ }_a}{2}
 + \frac{\muA^2}{8}
 + \frac{\muA^{ }\muY^{ }}{4}
 + \frac{\muY^2}{8}
 \biggr]
 \nn 
 & + & 
 \sum_a
 \chi^{ }_\rmii{F}(m_{e_a}) 
 \, \biggl[ 
   \mu_a^2
 + \frac{\muA^{ }\mu^{ }_a}{2}
 - \frac{3\muY^{ }\mu^{ }_a}{2}
 + \frac{\muA^2}{8}
 - \frac{\muA^{ }\muY^{ }}{4}
 + \frac{5 \muY^2}{8}
 \biggr]
 \nn 
 & + & 
 \sum_{i = u,c,t}
 \chi^{ }_\rmii{F}(m_i) 
 \, \biggl[ 
   3 \mu_q^2
  - \frac{3 \muA^{ }\mu_q}{2} 
  + \frac{5 \muY^{ }\mu_q}{2}
  + \frac{3 \muA^2}{8}
  - \frac{\muA^{ }\muY^{ }}{4}
  + \frac{17\muY^2}{24}
 \biggr]
 \nn 
 & + & 
 \sum_{i = d,s,b}
 \chi^{ }_\rmii{F}(m_i) 
 \, \biggl[ 
   3 \mu_q^2
  + \frac{3 \muA^{ }\mu_q}{2} 
  - \frac{\muY^{ } \mu_q}{2}
  + \frac{3 \muA^2}{8}
  + \frac{\muA^{ }\muY^{ }}{4}
  + \frac{5\muY^2}{24}
 \biggr]
 \nn 
 & + & 
  \Bigl[ \chi^{ }_\rmii{B}(\mH^{ }) + \chi^{ }_\rmii{B}(\mZ^{ }) \Bigr]
  \, \frac{(\muA + \muY)^2}{16}
  \; + \; 
  \chi^{ }_\rmii{B}(\mW^{ })
 \, \biggl[
   \muA^2
 + \frac{(\muA^{ } - \muY^{ })^2}{8} 
 \biggr]
 \nn 
 & + & 
  \frac{ v^2 (\muA^{ } + \muY^{ })^2}{8}
 \;. \la{pressure}
\ea
Here the susceptibilities are defined as
\ba
 \chi^{ }_\rmii{F}(m)
   \; \equiv \; 
 \int_\vec{p} \bigl[ - 2 \nF'(E) \bigr] 
 \;\; \stackrel{m\to 0}{\to} \;\; \frac{T^2}{6}
 \;, \quad
 \chi^{ }_\rmii{B}(m)
   \; \equiv \;
 \int_\vec{p} \bigl[ - 2 \nB'(E) \bigr] 
 \;\; \stackrel{m\to 0}{\to} \;\; \frac{T^2}{3}
 \;, \la{chi} 
\ea
where $\nF^{ }$ and $\nB^{ }$ are the Fermi and Bose distributions. 
The neutrino masses 
$m^{ }_{\nu_a}$ serve as a symbolic indicator of the origin
of the contribution. 
The relations between
chemical potentials and lepton and baryon asymmetries  
following from \eq\nref{pressure}
are given in appendix~A. 
Corrections, which are of $\rmO(g)$, have so far only been 
determined for the symmetric phase~\cite{kubo,sangel}.

%
\section{General structure of the indirect contribution}
\la{se:indirect}

As can be inferred from 
\eqs\nref{ap}--\nref{bm} and \nref{cp}--\nref{dm}, 
the rate coefficients $Q$ are related to C-even 
and $R,S$ to C-odd processes. In the symmetric phase, $R,S$ 
could be determined from a Taylor expansion
in chemical potentials. In contrast,
the dependence on chemical potentials 
is non-linear in the broken phase, 
so we need to generalize the definitions. 

At $\rmO(h_{\I a}^2)$, 
all rate coefficients can be related to the Euclidean 2-point correlator
of the operators to which the right-handed neutrinos couple: 
\be
 \Pi^{ }_\rmii{E}(\tilde K)
 \; \equiv \; 
 \int_X e^{i \tilde K \cdot X}
 \bigl\langle 
 (\tilde{\phi}^\dagger  \ell^{ }_{\rmii{$L$}_a})(X)  
 \, 
 (\bar{\ell}^{ }_{\rmii{$L$}_a} \tilde{\phi})(0) 
 \bigr\rangle
 \;, \quad 
 \tilde K = (k^{ }_n - i \mu^{ }_a,\vec{k})
 \;, \la{PiE}
\ee
where $k^{ }_n$ is a fermionic Matsubara frequency, 
$X = (\tau,\vec{x})$, and 
$
 \tilde K \cdot X = 
 (k^{ }_n - i \mu^{ }_a)\tau + \vec{k}\cdot\vec{x}
$. 
In the language of the canonical formalism, the expectation
value is taken with respect to the density matrix in \eq\nref{rho_SM}.
In perturbation theory, 
$\mu^{ }_a,\muB^{ }\neq 0$ induce expectation values for gauge field
zero modes, which effectively act as additional chemical potentials
(cf.\ \se\ref{se:ensemble}). 

The central object is the spectral function corresponding to 
\eq\nref{PiE}. It is the imaginary part of the 
retarded correlator $\Pi^{ }_\rmii{R}(\mathcal{K})$,\footnote{%
 The real part of $\Pi^{ }_\rmii{R}(\mathcal{K})$ is also important, 
 cf.\ the discussion around \eqs\nref{dispersive}
 and \nref{dispersive2}.
 } 
which in turn 
is an analytic continuation of $\Pi^{ }_\rmii{E}({\tilde K})$: 
\be
 \rho^{ }_a (\mathcal{K}) \; \equiv \; 
 \im \Pi^{ }_\rmii{R}(\mathcal{K}) \; \equiv \;  
 \left. 
  \im \Pi^{ }_\rmii{E}({\tilde K})
 \right|^{ }_{\tilde{k}^{ }_n \to -i [k^{ }_0 + i 0^+]}
 \;. \la{rho_a}
\ee
The rate coefficients are obtained by taking 
matrix elements of $\rho^{ }_a (\mathcal{K})$, 
\be
  \Omega^{ }_{(a\tau)\I\J}
 \; \equiv \; 
  \frac{
  \bar{u}^{ }_{\vec{k}\tau\J}
  \, \aL \, \rho^{ }_a(\mathcal{K}^{ }_\J) \, \aR \, 
  u^{ }_{\vec{k}\tau\I} 
  }{\sqrt{\omega^k_\I \omega^k_\J}} 
 \;, \la{Phi}
\ee
where 
 $\mathcal{K}^{ }_\J \equiv (\omega^k_\J,\vec{k})$, 
$
 \omega^k_\J
 \; \equiv \; \sqrt{k^2 + M_\J^2}
$, 
$\aL, \aR$ are chiral projectors, 
and $u^{ }_{\vec{k}\tau\I}$ is an on-shell spinor 
for sterile flavour $I$ in the helicity state $\tau = \pm$. 
As we work at $\rmO(h_{\I a}^2)$ in neutrino Yukawa couplings 
and the mixing of active and sterile neutrinos
was already accounted for within the reduction of the
non-equilibrium problem into the correlators in 
\eq\nref{Phi}, the mixing can be omitted 
in the definition of the on-shell spinors $u^{ }_{\vec{k}\tau\I}$. 

In the equations of ref.~\cite{cptheory}, another version of 
\eq\nref{Phi} also appears, in which the chiral projectors and 
helicity states are interchanged ($\aL\leftrightarrow\aR,\tau\to -\tau$)
and the four-momentum is simultaneously put to $-\mathcal{K}^{ }_\J$.
In the chiral limit, flipping the helicity is compensated for
by exchanging the chiral projectors, however changing 
the sign of $\mathcal{K}^{ }_{\!\J}$ does have an effect. 
Specifically, without chemical potentials
the real (imaginary) part of the neutrino 
self-energy is odd (even)  
in $\mathcal{K}^{ }_{\!\J}\to -\mathcal{K}^{ }_{\!\J}$, whereas 
a single insertion of a chemical potential reverses these
properties. This implies that the 
substitution $\mathcal{K}^{ }_{\!\J}\to -\mathcal{K}^{ }_{\!\J}$
corresponds to $\mu\to -\mu$, and we can write
\ba
   Q^{ }_{(a\tau)\I\J}
 & \equiv & 
 \fr12 \Bigl[ 
   \left. \Omega^{ }_{(a\tau)\I\J} \right|^{ }_{\mu}
 + \left. \Omega^{ }_{(a\tau)\I\J} \right|^{ }_{-\mu} \, 
 \Bigr]
 \;, \la{Q} \\
 \bar{\mu}^{ }_a R^{ }_{(a\tau)\I\J}
 + \sum_i\bar{\mu}^{ }_i\, S^{(i)}_{(a\tau)\I\J}
 & \equiv & 
 \fr12 \Bigl[ 
   \left. \Omega^{ }_{(a\tau)\I\J} \right|^{ }_{\mu}
 - \left. \Omega^{ }_{(a\tau)\I\J} \right|^{ }_{-\mu} \, 
 \Bigr] 
 \;. \la{RS}
\ea
The dependence of $Q,R,S$ on the flavour index $a$ 
vanishes in the symmetric phase, where a Taylor
expansion in chemical potentials is viable. 

Let us now focus on the indirect contribution in the 
language of ref.~\cite{broken}, obtained by replacing 
$\tilde{\phi}$ by its vacuum expectation value $(v/\sqrt{2},0)^T_{ }$, 
where $v \simeq 246$~GeV:
\be
 \Pi^\rmi{indirect}_\rmii{E}({\tilde K}) 
 \; \equiv \; 
 \frac{v^2 \bigl\langle {\nu}^{ }_{\rmii{$L$}_a}(-{\tilde K})
 \bar{\nu}^{ }_{\rmii{$L$}_a}(0) \bigr\rangle }{2}
 \;. \la{DeltaE}
\ee 
Then
\be
 \rho^\rmi{indirect}_a(\mathcal{K}) 
 = \frac{v^2 \im \Delta_{ }^{-1}(-\mathcal{K}-i u 0^+)}{2}
 \;, \la{adv}
\ee
where $\Delta$ is an analytic continuation of the 
inverse neutrino propagator and $u \equiv (1,\vec{0})$
is the plasma four-velocity.\footnote{%
 The real part of $\Delta^{-1}_{ }$ also plays a role, 
 leading to a ``dispersive'' correction as elaborated upon
 in appendix~A of ref.~\cite{cptheory}. Following an analysis
 similar to that leading to \eqs\nref{final_minus} and 
 \nref{final_plus}, we find that 
 this amounts to the 
 terms $\propto v^2$ in \eqs\nref{H0} and \nref{D0}, with 
 \be
 \kappa^{\pm}_{(a)\I\J}
  \approx
 \biggl[ 
 \frac{1}{2} +  
 \frac{M^{ }_{\I}M^{ }_{\J} (M_{\J}^2 + 2 k b) 
 }
 {2[(M_{\J}^2 + 2 k b)^2 + k^2 \Gamma^2_{\!u} ]}
 \biggr]^{ }_{\mu^{\pm}_{ }}
 \;, \quad
 \delta^{\pm}_{(a)\I\J}
  \approx 
 \biggl[
 \frac{1}{2} - 
 \frac{ M^{ }_{\I}M^{ }_{\J} (M_{\J}^2 + 2 k b) 
 }
 {2[(M_{\J}^2 + 2 k b)^2 + k^2 \Gamma^2_{\!u} ]}
 \biggr]^{ }_{\mu^{\pm}_{ }}
 \;, \la{dispersive}
 \ee
 where the coefficients are from \eq\nref{Sigma_form}, and 
 $
 \mu^{\pm}_{ }
 $ 
 indicates a symmetrization/antisymmetrization
 with respect to chemical potentials. 
 In the degenerate ($M^{ }_{\I} = M^{ }_{\J}$)
 vacuum ($b = \Gamma^{ }_{\!u} =0$) limit, 
 $\kappa^+_{ } = 1$, 
 $\kappa^{-}_{ } = \delta^{+}_{ } = \delta^{-}_{ } = 0$, 
 whereas in the temperature regime $T \gsim 30$~GeV
 most relevant for us, the helicity-flipping factors 
 ``$\frac{1}{2}$'' dominate.
 Equivalent mass corrections, 
 apart from the chemical potential dependence, 
 were obtained in ref.~\cite{eijima}. 
 }
Suppressing chiral projectors, let us write $\Delta$ as 
\be
 \Delta^{ }_{ }(-\mathcal{K} - i u 0^+)
 \; \equiv \; 
 - \bsl{\mathcal{K}} - \bsl{\Sigma}(\mathcal{K})
 \;, \quad
 \bsl{\Sigma}(\mathcal{K}) \; = \; 
 \re\bsl{\Sigma}(\mathcal{K}) + i \im\bsl{\Sigma}(\mathcal{K})
 \;. \la{S_R}
\ee
Assuming that the self-energy is proportional to either 
$\bsl{\mathcal{K}}$ or $\msl{u}$~\cite{weldon},  we find 
\be
 \im \Delta_{ }^{-1}(-\mathcal{K}-i u 0^+)
 = 
 \frac{2 (\bsl{\mathcal{K}} + \re\bsl{\Sigma})\, 
  ( \mathcal{K} + \re\Sigma ) \cdot\im\Sigma  
 - \im\bsl{\Sigma}\, 
 \bigl[(\mathcal{K} + \re \Sigma)^2 - (\im\Sigma)^2\bigr] }
 {\bigl[( \mathcal{K} + \re \Sigma)^2 - (\im\Sigma)^2 \bigr]^2
 + 4 \bigl[( \mathcal{K} + \re\Sigma ) \cdot\im\Sigma \bigr]^2}
 \;. \la{imS_Rinverse}
\ee

The self-energy $\Sigma$ can be parametrized as 
\be
 \bsl{\Sigma}(\mathcal{K}) \; = \; 
 \bsl{\mathcal{K}}\, \Bigl( a + \frac{i \Gamma^{ }_{\!\mathcal{K}}}{2} \Bigr)
  + 
 \msl{u}\, \Bigl( b + \frac{i \Gamma^{ }_{\!u}}{2} \Bigr)
 \;, \la{Sigma_form}  
\ee
where the coefficients are defined as real. 
All the coefficients in \eq\nref{Sigma_form}
are proportional to~$g^2$. 
We may expect that the coefficient $a$ can be 
omitted, given that it is subleading compared with the tree-level
term $\bsl{\mathcal{K}}$ in \eq\nref{S_R}, but for completeness we
keep it for the moment and verify that it indeed does not
contribute in \eqs\nref{dispersive}, 
\nref{final_minus}, or \nref{final_plus}. 

Inserting \eq\nref{Sigma_form} into \eq\nref{imS_Rinverse}, 
using the momentum  $\mathcal{K}^{ }_{\J}$ 
as needed in \eq\nref{Phi}, 
and counting $M_{\J}^2 \sim g^2 T^2$, we find that
\be
 \rho^\rmi{indirect}_a(\mathcal{K}^{ }_{\J}) 
 = 
\frac{v^2}
 {2[(M_{\J}^2 + 2 k b)^2 + k^2 \Gamma^2_{\!u} ]}
 \,  \Bigl\{ 
 \beta^{ }_{\mathcal{K}}\, \bsl{\mathcal{K}}^{ }_{\!\!\J}
  + \beta^{ }_{u} \msl{u}
 \Bigr\} 
 \;. \la{rho_params}
\ee
Here, in an expansion in $M^{ }_{\J}/k$, the coefficients read
\ba
 \beta^{ }_{\mathcal{K}} & \approx & 
  \biggl( 
    \frac{M_{\J}^2}{2}
  \biggr) \,
   \Gamma^{ }_{\!\mathcal{K}}
  +
  \biggl[ 
    k\,(1+2a) + b + \frac{M_{\J}^2}{2k}    
  \biggr]\,
   \Gamma^{ }_{\!u} 
 \;, \la{betaK} \\[2mm] 
 \beta^{ }_{u} & \approx & 
   b\, \Bigl( 
    b\, k + M_{\J}^2
  \Bigr)\,
   \Gamma^{ }_{\!\mathcal{K}}
  +
  \fr12 \biggl[ 
   b^2 - M_{\J}^2\,(1+2a) + \frac{\Gamma^{ }_{\! u}
   (
    \Gamma^{ }_{\!u} 
   + 2 k \Gamma^{ }_{\!\mathcal{K}}
   )}{4}
  \biggr]\,
   \Gamma^{ }_{\!u} 
 \;. \hspace*{7mm} \la{betau}
\ea
Terms up to $\rmO(g^4 T^2)$ have been retained in $\beta^{ }_{\mathcal{K}}$
and up to $\rmO(g^6 T^3)$ in $\beta^{ }_{u}$; this is because  
$\beta^{ }_{\mathcal{K}}$ is weighted by a coefficient
of $\rmO(g^2T)$ in \eq\nref{proj_plus}. 

The matrix elements 
of \eq\nref{rho_params}, needed in \eq\nref{Phi}, read
\ba
 \bar{u}^{ }_{\vec{k}(-)\J}
  \, \aL \, 
  \bigl(   \beta^{ }_{\mathcal{K}}\, \bsl{\mathcal{K}}^{ }_{\!\!\J}
  + \beta^{ }_{u} \msl{u} 
  \bigr) 
  \, \aR \, 
  u^{ }_{\vec{k}(-)\I}
 & \;\approx\; & 
 M^{ }_{\I} M^{ }_{\J} 
 \biggl( 
   \beta^{ }_{\mathcal{K}} 
  + \frac{ \beta^{ }_{u} }{2 k}
 \biggr) 
 \;,  \la{proj_minus} \\[2mm]  
 \bar{u}^{ }_{\vec{k}(+)\J}
  \, \aL \, 
  \bigl(   \beta^{ }_{\mathcal{K}}\, \bsl{\mathcal{K}}^{ }_{\!\!\J}
 + \beta^{ }_{u} \msl{u} 
  \bigr) 
  \, \aR \, 
  u^{ }_{\vec{k}(+)\I}
 & \;\approx\; & 
   \beta^{ }_{\mathcal{K}} M_{\J}^2 \,
   \biggl( 1 + \frac{M^{2}_{\I} - M^2_{\J}}{8k^2} \biggr)
  + 
   \beta^{ }_{u} \,\biggl( 2 k + \frac{M^{2}_{\I}
  +  M^{2}_{\J} }{4k} \biggr)
 \;. \hspace*{3mm}  \nn 
 \la{proj_plus} 
\ea
Inserting \eqs\nref{betaK} and \nref{betau}
and working to leading order in $g^2$, we find 
the helicity-conserving and helicity-flipping coefficients
\ba
 \Omega^\rmi{indirect}_{(a-)\I\J} 
 & \approx & \frac{v^2 \, M^{ }_{\I}M^{ }_{\J}\, }
 {2[(M_{\J}^2 + 2 k b)^2 + k^2 \Gamma^2_{\!u} ]}
 \,  \bigl\{
   \Gamma^{ }_{\!u} 
  \bigr\} 
 \;, \la{final_minus} \\[2mm] 
 \Omega^\rmi{indirect}_{(a+)\I\J} 
 & \approx & \frac{v^2}
 {8 k^2}
 \, \bigl\{
   \Gamma^{ }_{\!u} 
   + 2 k \Gamma^{ }_{\!\mathcal{K}}
  \bigr\} 
 \;. \hspace*{7mm} \la{final_plus}
\ea
Parametrically, the helicity-flipping rate
$\Omega^\rmi{indirect}_{(a+)}$ is suppressed by $\rmO(g^2)$
with respect to the helicity-conserving rate 
$\Omega^\rmi{indirect}_{(a-)}$, and does
not contain the possibility of resonant enhancement
(the latter observation conforms with refs.~\cite{lello,eijima}). 
On the other hand, we find that in general 
$
 2 k \Gamma^{ }_{\!\mathcal{K}} > \Gamma^{ }_{\!u}
$
at high temperatures, cf.\ \se\ref{se:rates}.
This anticipates the situation in the symmetric phase, where  
$\Omega^{ }_{(a-)\I\J}$ is suppressed by 
$\sim M^{ }_{\I}M^{ }_{\J}/(gT)^2$ with respect to 
$\Omega^{ }_{(a+)\I\J}$~\cite{cptheory}. 

%
\section{Determination of rate coefficients for the indirect contribution}
\la{se:rates}

%

We now turn to the determination of the coefficients
$b$, $\Gamma^{ }_{\!u}$ and $\Gamma^{ }_{\!\mathcal{K}}$
that are defined through \eqs\nref{S_R} and \nref{Sigma_form} and that 
parametrize the indirect contribution to masses and rate coefficients
through \eqs\nref{dispersive}, \nref{final_minus} and \nref{final_plus}, 
respectively. 
Three different regimes are considered (remaining always in the broken phase): 
``high temperatures'', 500~GeV~$\, \gsim \, \pi T \gg \mW^{ }$; 
``intermediate temperatures'', $\pi T \sim \mW^{ }$; 
and 
``low temperatures'', 15~GeV~$\lsim\, \pi T \ll \mW^{ }$.

The starting point, \eq\nref{S_R}, 
involves a specific analytic continuation, 
and some care is needed for implementing it properly in the broken phase. 
We first note that in \eq\nref{rho_a} the combination
$k^{ }_n - i \mu^{ }_a$ is analytically continued to $-i [k^{ }_0 + i 0^+]$, 
whereas in \eqs\nref{DeltaE} and \nref{adv} the combination
$- k^{ }_n + i \mu^{ }_a$ is analytically continued to $i [k^{ }_0 + i 0^+]$.
The latter can be re-interpreted as 
$k^{ }_n + i \mu^{ }_a$ analytically continued to the ``advanced'' 
frequency $-i [k^{ }_0 - i 0^+]$, and subsequently taken 
with an inverted sign of~$k^{ }_0$. 
In other words, 
denoting by 
$\Phi(\mathcal{K},...) |^{ }_{
 k^{ }_n + i \mu^{ }_a
 \to -i [k^{ }_0 - i 0^+]
}$ the {\em advanced} self-energy
{\em before} the last sign inversion,
and factoring
out gauge couplings corresponding to $Z^0$ or $W^\pm$ exchange, 
we can write 
\be
 \Delta(-\mathcal{K}-i u 0^+)  =  
  - \bsl{\mathcal{K}}
 + (g_1^2 + g_2^2)\, \Phi(-\mathcal{K},...)
 + 2 g_2^2\, \Phi(-\mathcal{K},...) 
 \;. \la{splitup}
\ee
Here $...$ stands for masses and chemical potentials pertinent to 
the channel in question. 

Now, there is a complication with this setup, 
arising because in the broken phase
most particles feel a gauge field background, parametrized through
$\muY^{ }$ and $\muA^{ }$ via \eq\nref{muY_muA}, 
cf.\ table~\ref{table:mus}. 
Whenever possible it is very convenient to ``resum'' 
this gauge field background 
into the corresponding propagators. 
But then we must make sure that the relationship corresponding
to chemical equilibrium, 
\be
 \mu^{ }_3  =  \mu^{ }_1 + \mu^{ }_2
 \;, \la{mu_rule}
\ee 
is respected in any $1 \leftrightarrow 2$ reaction. 
Thus the Matsubara frequencies of the corresponding
particles should read 
$\tilde{k}^{ }_n = k^{ }_n + i \mu^{ }_3$, 
$\tilde{q}^{ }_n = q^{ }_n + i \mu^{ }_1$, 
and $\tilde{p}^{ }_n = p^{ }_n + i \mu^{ }_2$,  
with 
$
 \tilde{k}^{ }_n = \tilde{q}^{ }_n + \tilde{p}^{ }_n
$, 
and the analytic continuation needed for computing
$\Phi(\mathcal{K},...)$ with resummed propagators reads 
\be
 k^{ }_n + i \munLa \to -i [ k^{ }_0 - i 0^+ ]
 \;, \la{analytic_new}
\ee
replacing the unresummed analytic continuation 
$
 k^{ }_n + i \mu^{ }_a
 \to -i [k^{ }_0 - i 0^+]
$. 

%
\subsection{Real part of the active neutrino self-energy}
\la{ss:b}

Let us first consider the
real part of the advanced self-energy, parametrized by the function $b$
in \eq\nref{Sigma_form}. 
Like in \eq\nref{splitup}, there are two gauge channels, 
and in addition there is a term linear in 
$
 \muZ^{ }
$, 
originating from a 1-loop $Z^0_{ }$-boson tadpole contribution.\footnote{%
 Alternatively, the existence of such a term can be deduced 
 from paying careful attention to the difference between unresummed 
 and resummed analytic continuations, as alluded to around
 \eq\nref{analytic_new}.}
With the sign conventions of \eq\nref{S_R}, this implies that
\be
 b_{ }^\rmi{}  \; = \; 
 (g_1^2 + g_2^2)\, \mathcal{E}_{ }^\rmi{}(\mZ^{ },\munLa,\muZn^{ })
 + 
 2 g_2^2\, \mathcal{E}_{ }^\rmi{}(\mW^{ },\mueLa,\muW^{ })
 - \frac{\muZ^{ }}{2}
 \;, \la{bBorn} 
\ee
where the arguments show the masses 
and chemical potentials appearing in the loop, 
and $\muZ^{ }$ is given in \eq\nref{muZ}. 
According to table~\ref{table:mus} and the definitions
in \eq\nref{broken2}, 
$
 \munLa = \mu^{ }_a - \muZ^{ }/2
$, 
$
 \muZn = 0
$, 
$
 \mueLa = \mu^{ }_a - \muQ^{ } + \muZ^{ }(\frac{1}{2} - s^2)
$, 
$
 \muW = \muQ^{ } - \muZ^{ }(1-s^2)
$. 
We omit the appearance of~$\muZ^{ }$ inside the function $\mathcal{E}$, 
because this contribution is suppressed by $\sim \alpha^{ }/\pi$
compared with the explicit appearance of $\muZ^{ }$
in the last term of \eq\nref{bBorn}. 

In order to determine $\mathcal{E}$, 
the inverse neutrino propagator of \eq\nref{S_R} can be computed
with the gauge propagators of \eqs\nref{propW} and \nref{propZ}. 
For $\Phi$
of \eq\nref{splitup} this implies 
($D \equiv 4 - 2\epsilon$)
\be
 \Phi_{ }^\rmi{Born}(\mathcal{K},m,\mu_1^{ },\mu^{ }_2)  \equiv  
 \left. 
 \frac{D-2}{4}
 \Tint{P} \frac{i (\bsl{\tilde{K}}-\bsl{\tilde{P}}) }
 {[(\tilde{k}^{ }_n - \tilde{p}^{ }_n)^2 + \epsilon_1^2 ]
 (\tilde{p}_n^{\,2} + \epsilon_2^2)}
 \right|^{ }_{\tilde{k}^{ }_n\to -i [k^{ }_0 - i 0^+]}
 \hspace*{-1.5cm} \;. \la{Born}
\ee
Here 
$\epsilon^{ }_1 \equiv |\vec{k-p}|$,  
$\epsilon^{ }_2 \equiv \sqrt{p^2 + m^2}$,
and the chemical potentials are related by \eq\nref{mu_rule}. 
After carrying out the Matsubara sum, taking the real part 
of the advanced propagator, and recalling the conventions
in \eqs\nref{S_R}, \nref{Sigma_form} 
and \nref{bBorn}, 
we obtain
($\mathbbm{P} \equiv$  principal value) 
\ba
 & & \hspace*{-1.5cm} 
 - 8 \pi^2 k \, \mathcal{E}_{ }^\rmi{}(m,\mu^{ }_1,\mu^{ }_2)
 \nn
 & = & 
 \int_0^\infty \! {\rm d}p \, \nF^{ }(p) \,
 \mathbbm{P} \biggl[
   p + \frac{m^2}{8 k} 
   \ln \biggl| 
             \frac{ m^2 - 4 k p }{m^2 + 4 k p } 
       \biggr|  \,   
  + 
  \frac{4 k p\, m^2 \mu^{ }_1}{m^4 - 16 k^2 p^2}
 \biggr]
 \nn 
 & + & 
 \int_{m}^\infty \! {\rm d}\epsilon \, \nB^{ }(\epsilon) \, 
 \mathbbm{P} \biggl\{ 
   p + 
  \frac{m^2}{8 k } 
   \ln \biggl| 
  \frac{ m^2 - 4 k^2 - 4 k p  }
       { m^2 - 4 k^2 + 4 k p  }
       \biggr|\,      
 \nn 
 & + & 
 \frac{k\, m^2 \mu^{ }_2}{p}
 \biggl[
   \frac{(\epsilon + p)^2}
   {4 k^2 (\epsilon + p)^2 - m^4} 
   + 
   \frac{(\epsilon - p)^2}
   {4 k^2 (\epsilon - p)^2 - m^4} 
 \biggr]
 \, 
 \biggr\}^{ }_{p = \sqrt{\epsilon^2 - m^2}}
 + \rmO(\mu_i^2)
 \;. \hspace*{6mm} \la{b_full}
\ea

Eq.~\nref{b_full} can be simplified 
at high and low temperatures. For $\pi T \gg m$, we find 
\be
 \mathcal{E}(m,\mu^{ }_1,\mu^{ }_2)
 \; \stackrel{\pi T \gg m}{\approx} \; 
 \; - \; \frac{T^2}{32k}
 \; + \; \frac{m^2 \mu^{ }_1}{64\pi^2 k^2}
 \, \ln\biggl(\frac{3.5278 k T}{m^2}\biggr)
 \; - \; \frac{m T \mu^{ }_2}{32\pi k^2}
 \;+\; \rmO(\mu_i^2)
 \;.
 \la{b_highT}
\ee
In each structure only the leading term in an expansion
in $m/(\pi T)$ is shown. 
The $\mu$-independent part corresponds to an ``asymptotic'' lepton
thermal mass~\cite{weldon}.
For $\pi T \ll m$, 
\be
 \mathcal{E}^\rmi{}_{ }(m,\mu^{ }_1,\mu^{ }_2)
 \;\stackrel{\pi T \ll m}{\approx}\;
 \frac{7\pi^2 T^4 k}{180 m^4}
 - \frac{\mu^{ }_1 T^2}{24m^2} \;+\; \rmO(\mu_i^2)
 \;. \la{b_lowT}
\ee
The $\mu$-independent part is equivalent
to the classic result from ref.~\cite{raffelt}.
After inserting  
$\muZ^{ }$ from \eq\nref{muZ} 
and recalling that the Fermi constant 
reads $\sqrt{2}G^{ }_\rmii{F} = g_2^2 / (4 \mW^2) = 1/v^2$,
the $\mu$-dependent part of $b$ agrees 
with the function $-c$ as given in 
\eq(3.13) of ref.~\cite{dmpheno}.

%
\subsection{Widths at high temperatures: 
$2\leftrightarrow 2$ scatterings with soft gauge exchange}
\la{ss:gauge}

In the high-temperature regime, 
the determination of the active neutrino width 
requires a resummed computation~\cite{broken}, which profits
from light-cone sum rules~\cite{sum1,sch,gt,gtm}. 
The leading contribution originates from scatterings
mediated by Bose-enhanced soft gauge bosons. In order to determine this
contribution, the gauge boson propagator needs to be Hard Thermal Loop (HTL)
resummed~\cite{ht1,ht2,ht3,ht4}. Parametrically, HTL effects are
important when $\mW^{ } \lsim gT$, i.e.\ $v\lsim T$. 
As elaborated upon around \eq\nref{broken}, the inclusion of
chemical potentials is complicated in this regime beyond linear
order. Nevertheless, we can show that chemical 
potentials are {\em not} expected to play a role at linear order, 
because of a general symmetry property of the soft 
contribution (see below). 

In terms of $\Phi$ of \eq\nref{splitup}, 
the HTL-resummed result has the form\footnote{%
 Eq.~\nref{HTL} looks like a $1\leftrightarrow 2$ 
 contribution but is really
 a $2\leftrightarrow 2$ contribution, because it originates from the
 Landau damping part of the gauge field propagator, which is itself 
 induced by $2\leftrightarrow 1$ scatterings. 
 }
\be
 \Phi_{ }^\rmii{HTL}(\mathcal{K},...)
 \;  \equiv  \;
 \frac{1}{4}\,
 \Tint{P} \frac{\gamma^{ }_{\mu}
 [ - i (\bsl{\tilde{K}}-\bsl{\tilde{P}}) ]
 \gamma^{ }_{\nu} }
 {(\tilde{k}^{ }_n - \tilde{p}^{ }_n )^2 + (\vec{k-p})^2 }
 \;
 \bigl\langle A^{ }_\mu(\tilde{P}) A^{ }_\nu(-\tilde{P})\bigr\rangle
 \biggr|^{ }_{\tilde{k}^{ }_n \to -i [k^{ }_0 - i 0^+]}
 \hspace*{3mm} \;, \hspace*{3mm} \la{HTL}
\ee
where 
$\tilde{k}^{ }_n \equiv k^{ }_n + i\mu^{ }_3$, 
$\tilde{p}^{ }_n \equiv p^{ }_n + i \mu^{ }_2$, and the 
chemical potentials are related through \eq\nref{mu_rule}. 
In Feynman gauge the gauge propagator can be expressed as
\be
 \bigl\langle A^{ }_\mu(\tilde{P}) A^{ }_\nu(-\tilde{P}) \bigr\rangle
 = 
 \frac{\mathbbm{P}^\rmii{T}_{\mu\nu}(\tilde{P})}
 {\tilde{P}^2 + m^2 + \Pi^{ }_\rmii{T}(\tilde{P})}
 + 
 \frac{\mathbbm{P}^\rmii{E}_{\mu\nu}(\tilde{P})}
 {\tilde{P}^2 + m^2 + \Pi^{ }_\rmii{E}(\tilde{P})}
 + 
 \frac{\tilde{P}^{ }_\mu \tilde{P}^{ }_\nu}
 {\tilde{P}^2(\tilde{P}^2 + m^2)}
 \;,  \la{gauge_prop}
\ee
where $m$ depends on the gauge channel; the self-energies
$\Pi^{ }_\rmii{T,E}$ can be found in appendix~B of ref.~\cite{broken} and their
relevant limiting values in \eqs\nref{limits_PiE} and 
\nref{limits_PiT}; and the projectors read
$
 \mathbbm{P}^\rmii{T}_{\mu\nu}(\tilde{P}) \; \equiv \;  
 \delta^{ }_{\mu i}\delta^{ }_{\nu j}
 ( \delta^{ }_{ij} - {p^{ }_i p^{ }_j} / {p^2} )
$, 
$
 \mathbbm{P}^\rmii{E}_{\mu\nu}(\tilde{P}) \; \equiv \;  
 \delta^{ }_{\mu\nu}
 - { \tilde{P}^{ }_\mu \tilde{P}^{ }_\nu } / {\tilde{P}^2} 
 - \mathbbm{P}^\rmii{T}_{\mu\nu}(\tilde{P})
 \;. 
$
Inserting the projectors into \eq\nref{HTL} 
we obtain
\ba
 \Phi_{ }^\rmii{HTL}(\mathcal{K},...) 
 & = &
 \frac{1}{4}\,
 \Tint{P}
 \biggl\{ 
 \frac{i\bsl{\tilde{K}}}{(\tilde{K}-\tilde{P})^2}
 \biggl[
  \frac{1}{\tilde{P}^2 + m^2}  - 
  \frac{1}{\tilde{P}^2 + m^2 + \Pi^{ }_\rmii{E}(\tilde{P})}  
 \biggr]
 \nn 
 & + &
 \frac{2
 \bigl(\vec{k}\cdot\vec{\gamma} - 
 \frac{\vec{p}\cdot\vec{k}\,\vec{p}\cdot\vec{\gamma}}{p^2} \bigr)}
 {(\tilde{K}-\tilde{P})^2} 
 \biggl[
  \frac{1}{\tilde{P}^2 + m^2 + \Pi^{ }_\rmii{T}(\tilde{P})}  - 
  \frac{1}{\tilde{P}^2 + m^2 + \Pi^{ }_\rmii{E}(\tilde{P})}  
 \biggr]
 \nn 
 & + & 
 \frac{i(D-2)(\bsl{\tilde{K}} - \bsl{\tilde{P}})}{(\tilde{K}-\tilde{P})^2}
 \frac{1}{\tilde{P}^2 + m^2 + \Pi^{ }_\rmii{T}(\tilde{P})}  
 \; 
 \biggr\}^{ }_{\tilde{k}^{ }_n \to -i [k^{ }_0 - i 0^+]}  
 \hspace*{4mm} \;. 
\ea

In order to proceed, 
we write the resummed propagators in a spectral representation: 
\be
 \frac{1}{\tilde{P}^2 + m^2 + \Pi^{ }_i(\tilde{P})}
 \; = \; 
 \int_{-\infty}^{\infty}
 \! \frac{{\rm d}p^{ }_0}{\pi}
 \, \frac{\rho^{ }_i(p^{ }_0,\vec{p})}{p^{ }_0 - i \tilde{p}^{ }_n}
 \;.
\ee
Then the Matsubara sum can be carried out. Subsequently we take the cut
and keep the channel leading to a {\em soft}
contribution from momenta $p^{ }_0,p \ll k, \pi T$.
Setting $\mathcal{K}\to -\mathcal{K}$ as in 
\eq\nref{splitup}, and taking the sign relevant for 
$\bsl{\Sigma}$ in \eq\nref{S_R}, we find 
\ba
 & & \hspace*{-1cm}
 - \im \Phi^\rmii{HTL}_{ }(-\mathcal{K},...) 
  = 
 \fr14 \int_{-\infty}^{k} \! {\rm d}p^{ }_0 
 \int_\vec{p} \!
 \frac{\delta(k - p^{ }_0 - |\vec{k-p}|)}{2|\vec{k-p}|}
 \bigl[ 1 - \nF^{ }(k - p^{ }_0 - \mu^{ }_1)
 + \nB^{ }(p^{ }_0 - \mu^{ }_2)\bigr]
 \nn 
 & \times & \!\!
 \biggl[ 
  \bsl{\mathcal{K}} \, (\rho^{ }_\rmi{free} - \rho^{ }_\rmii{E}) 
 + 2 \, 
 \biggl(\vec{k}\cdot\vec{\gamma} - 
 \frac{\vec{p}\cdot\vec{k}\;\vec{p}\cdot\vec{\gamma}}{p^2}
 \biggr)
 \, \bigl( \rho^{ }_\rmii{T} - \rho^{ }_\rmii{E}\bigr)
 + (D-2) (\bsl{\mathcal{K}} - \bsl{\mathcal{P}})\, \rho^{ }_\rmii{T}
 \biggr]
 \;, \la{Phi_HTL_1} 
\ea
where $\rho^{ }_\rmi{free}$ is the spectral function
corresponding to $1/(\tilde{P}^2 + m^2)$.

Next, let us denote the structures relevant for 
\eqs\nref{final_minus} and \nref{final_plus} by
\be
 \Gamma^\rmii{HTL}_\rmii{$(-)$} \; \equiv \; \Gamma^\rmii{HTL}_{\! u}
 \;, \quad
 \Gamma^\rmii{HTL}_\rmii{$(+)$} \; \equiv \;
 \Gamma^\rmii{HTL}_{\! u} + 2 k \Gamma^\rmii{HTL}_{\!\mathcal{K}} 
 \;. 
\ee
The corresponding contributions to $\Phi$ are denoted
by $\Phi^\rmii{HTL}_{(\tau)}$, $\tau = \pm$. Carrying out the angular integral
in \eq\nref{Phi_HTL_1} and setting $D\to 4$, we get
\ba
 & & \hspace*{-1cm}
 - \im \Phi^\rmii{HTL}_{(\tau)}(-\mathcal{K},...) 
  = 
 \frac{1}{8\pi^2} \int_{-\infty}^{k} \! {\rm d}p^{ }_0 
 \int_{|p^{ }_0|}^{2k-p^{ }_0} \! {\rm d}p \, p \, 
 \bigl[ 1 - \nF^{ }(k - p^{ }_0 - \mu^{ }_1)
 + \nB^{ }(p^{ }_0 - \mu^{ }_2)\bigr]
 \nn 
 & \times & \!\!
 \biggl\{
 \delta^{ }_{\tau,-} \, 
 \biggl[ 
    \frac{p_\perp^2}{p^2} \,
    \bigl( \rho^{ }_\rmii{T} - \rho^{ }_\rmii{E} \bigr)
  + 
    \frac{p^2 - p_0^2}{2k^2} \, \rho^{ }_\rmii{T}
 \biggr]
 \nn 
 & + & \;
 \delta^{ }_{\tau,+} \, 
 \biggl[ 
   \rho^{ }_\rmi{free} - \rho^{ }_\rmii{E}
   - 
    \frac{p_\perp^2}{p^2} \,
    \bigl( \rho^{ }_\rmii{T} - \rho^{ }_\rmii{E} \bigr)
   + 
   \frac{(2k-p^{ }_0)^2-p^2}{2k^2} \, \rho^{ }_\rmii{T}
 \biggr]
 \biggr\}
 \;, \la{Phi_HTL_2} 
\ea
where the energy conservation constraint
 $|\vec{k-p}| + p^{ }_0 = k$
in \eq\nref{Phi_HTL_1}
permitted us to write
\be
 p_\perp^2 \;\equiv\;
 p^2 - \frac{(\vec{k}\cdot\vec{p})^2}{k^2}
 \; = \; 
 (p^2 - p_0^2) 
 \, \biggl[\biggl( 1 - \frac{p^{ }_0}{2k} \biggr)^{\!2} - 
 \biggl( \frac{p}{2k} \biggr)^{\!2}\; \biggr]
 \;. \la{p_perp}
\ee

As a final step, we again focus on the contribution from the soft
domain $p,p^{ }_0 \ll k,\pi T$. Then 
we can drop terms suppressed by $p/k$ or $p^{ }_0/k$
from \eq\nref{Phi_HTL_2}. According to 
\eq\nref{p_perp}, we can subsequently write
$
 \int_{|p^{ }_0|}^{2k-p^{ }_0} \! {\rm d}p \, p  
 \approx 
 \int_{0}^{2k} \! {\rm d}p^{ }_\perp \, p^{ }_\perp
$
and
$p_\parallel^2 \equiv p^2 - p_\perp^2 \approx p_0^2$.  
Furthermore, the leading-order contribution originates
from the Bose-enhanced structure 
$
 \nB^{ }(p^{ }_0 - \mu^{ }_2) \approx T/(p^{ }_0 - \mu^{ }_2)\gg 1
$. 
The oddness of the $p^{ }_0$-integrand implies that $\mu^{ }_2$
only contributes at $\rmO(\mu_2^2)$ and can be omitted. 
Thereby \eq\nref{Phi_HTL_2} becomes
\ba
 & & \hspace*{-1cm}
 - \im \Phi^\rmii{HTL}_{(\tau)}(-\mathcal{K},...) 
  \approx
 \frac{T}{8\pi^2} 
 \int_{0}^{2k} \! {\rm d}p^{ }_\perp \, p^{ }_\perp \, 
 \int_{-\infty}^{\infty} \! \frac{{\rm d}p^{ }_0}{p^{ }_0} 
 \nn 
 & \times & \!\!
 \biggl\{
 \delta^{ }_{\tau,-} \, 
 \biggl[ 
    \frac{p_\perp^2}{p_\perp^2 + p_0^2} \,
    \bigl( \rho^{ }_\rmii{T} - \rho^{ }_\rmii{E} \bigr)
 \biggr]
 + 
 \delta^{ }_{\tau,+} \, 
 \biggl[ 
   \rho^{ }_\rmi{free} - \rho^{ }_\rmii{E} + 2 \rho^{ }_\rmii{T}
   - 
    \frac{p_\perp^2}{p_\perp^2 + p_0^2} \,
    \bigl( \rho^{ }_\rmii{T} - \rho^{ }_\rmii{E} \bigr)
 \biggr]
 \biggr\}
 \;, \la{Phi_HTL_3} 
\ea
where the spatial momentum is 
$\vec{p} \,\approx\, \vec{p}^{ }_\perp + p^{ }_0\, \vec{e}^{ }_\vec{k}$.

The integral over $p^{ }_0$ can now be carried out. 
It is illustrative to first consider 
the term involving $\rho^{ }_\rmi{free}$.
Expressing the spectral function as 
a discontinuity of the ``resolvent''~$\mathcal{R}$,
we are faced with an integral of the type
\ba
 I & \equiv & 
 \int_{-\infty}^{\infty} \! \frac{{\rm d}p^{ }_0}{p^{ }_0}
 \frac{\mathcal{R}(p^{ }_0 + i 0^+,{p}^{ }_\perp,p^{ }_0) - 
  \mathcal{R}(p^{ }_0 - i 0^+,{p}^{ }_\perp,p^{ }_0)}{2i}
 \;, \\
 \mathcal{R}(p^{ }_0,{p}^{ }_\perp,p^{ }_\parallel) 
 & \equiv &
 \frac{1}{{p}_\perp^2 + p_\parallel^2 - p_0^2 + m^2} 
 \;.  
\ea 
Noting that 
\be 
 \mathcal{R}(p^{ }_0 + i 0^+,{p}^{ }_\perp,p^{ }_0) 
 \; = \;
 \frac{1}{{p}_\perp^2 - i p^{ }_0 \, 0^+ + m^2} 
 \la{ftilde}
\ee
is actually regular near the real $p^{ }_0$-axis, 
and defining
$
 \tilde{\mathcal{R}}(p^{ }_0)
  \equiv  
 {1} / ({{p}_\perp^2 + m^2}) 
$,
$I$ can be re-expressed as a complex integral,
\be
 I \; = \; \int_c \frac{\tilde{\mathcal{R}}(p^{ }_0)}{2ip^{ }_0}
 \;, \qquad
 \parbox[c]{190pt}{\begin{picture}(100,30)(0,0)
 \linethickness{0.075mm}
 \put(0,12){c = }
 \put(20,15){\vector(1,0){80}}
 \put(60,0){\vector(0,1){30}}
 \linethickness{0.2mm}
 \put(20,18){\vector(1,0){19.5}}
 \put(39.5,18){\line(1,0){15.5}}
 \put(65,18){\vector(1,0){19.5}}
 \put(84.5,18){\line(1,0){15.5}}
 \put(55,15){\oval(6,6)[r]}
 \put(35.5,12){\line(-1,0){15.5}}
 \put(55,12){\vector(-1,0){19.5}}
 \put(80.5,12){\line(-1,0){15.5}}
 \put(100,12){\vector(-1,0){19.5}}
 \put(65,15){\oval(6,6)[l]}
 \end{picture}%
 \begin{picture}(100,30)(0,0)
 \linethickness{0.075mm}
 \put(3,12){ = }
 \put(20,15){\vector(1,0){80}}
 \put(60,0){\vector(0,1){30}}
 \linethickness{0.2mm}
 \put(40,23){\vector(1,0){1}}
 \put(60,18){\oval(80,10)[t]}
 \put(60,15){\oval(6,4)[r]}
 \put(60,15){\oval(5.5,3.5)[r]}
 \put(80,7){\vector(-1,0){1}}
 \put(60,12){\oval(80,10)[b]}
 \put(60,15){\oval(6,4)[l]}
 \put(60,15){\oval(5.5,3.5)[l]}
 \put(61.0,13.3){\vector(1,0){1}}
 \end{picture}}
 \qquad \;.
\ee
We proceed with the help of the residue theorem. 
There is a contribution from the pole at $p^{ }_0 = 0$, amounting to 
$\pi/(p_\perp^2 + m^2)$. In addition 
there is a contribution from arcs that can be sent to $|p^{ }_0|\to \infty$,  
yielding $-\pi/(p_\perp^2 + m^2)$. 
Summing together, $\rho^{ }_\rmi{free}$ does not contribute. 

Now, let us inspect the other terms in \eq\nref{Phi_HTL_3}. 
For those involving 
${p_\perp^2}/({p_\perp^2 + p_0^2})$, the arcs at 
$|p^{ }_0| \to\infty$ do not contribute 
because of the additional suppression by $\sim 1/p_0^2$. 
On the other hand there is an additional pole at 
$p^{ }_0 = \pm i p^{ }_\perp$, 
but this does not contribute either, because $\Pi^{ }_\rmii{T}$
and $\Pi^{ }_\rmii{E}$ coincide for $p = 0$. Therefore only 
the pole at $p^{ }_0 = 0$ has an effect; this is the content
of the sum rule obtained in refs.~\cite{sum1,sch}. As recalled
in appendix~B, in this limit $\Pi^{ }_\rmii{T}$ vanishes and 
$\Pi^{ }_\rmii{E}$ is replaced by a mass parameter~$\mE^2$.

Finally, for the term $-\rho^{ }_\rmii{E} + 2\rho^{ }_\rmii{T}$
in \eq\nref{Phi_HTL_3}, 
there is a contribution from both the pole at $p_0^{ } = 0$ and 
the arcs at $|p^{ }_0| \to\infty$~\cite{gt,gtm}. 
As elaborated upon
in appendix~B, at the far-away arcs  
$\Pi^{ }_\rmii{E}$ vanishes and 
$\Pi^{ }_\rmii{T}$ is replaced by a mass parameter~$\mE^2/2$.
Combining the terms we obtain
\ba
  - \im \Phi^\rmii{HTL}_{(\tau)}(-\mathcal{K},...) 
 & \approx &  
 \frac{T}{8\pi}
 \int_0^{2k} \!\!\! {\rm d}p^{ }_\perp \, p^{ }_\perp \, 
 \biggl\{ 
  \delta^{ }_{\tau,-} \, 
  \biggl[ \frac{1}{p_\perp^2 + m^2} -
  \frac{1}{p_\perp^2 + m^2 + \mE^2}  \biggr]
 \nn 
 &  & \hspace*{2.1cm} +\, 2 \delta^{ }_{\tau,+} \, 
  \biggl[ \frac{1}{p_\perp^2 + m^2} -
  \frac{1}{p_\perp^2 + m^2 + \mE^2/2} 
  \biggr]
 \biggr\} 
 \;. \la{Phi_HTL_4}
\ea 
For the neutral sector, the temperature-dependent 
weak mixing angle needs to be evaluated in the proper momentum domain. 
Inserting the prefactors from \eq\nref{splitup} and making use of
the angles $\theta,\tilde{\theta},\bar{\theta}$ 
and the thermally modified masses
$\mWt^{ }$, $\mZt^{ }$, $\mQt^{ }$,
$\mWb^{ }$, $\mZb^{ }$, $\mQb^{ }$
defined in appendix~B, we get
\ba
 \Gamma^\rmii{HTL}_{\! u} & \approx & 
 \frac{T}{16\pi} 
 \biggl\{
 2 g_2^2 \ln\biggl( \frac{1 + 4 k^2 / \mW^2 }{ 1 + 4 k^2 / \mWt^2 } \biggr)
 \la{Gamma_HTL_u} \\ 
 && \hspace*{-5mm} +  \,
 (g_1^2 + g_2^2) 
 \biggl[
 \cos^2(\theta - \tilde{\theta}) 
   \ln \biggl( \frac{1 + 4 k^2 / \mZ^2}{1 + 4 k^2 / \mZt^2 } \biggr)
 + 
 \sin^2(\theta - \tilde{\theta}) 
   \ln \biggl( \frac{1 + 4 k^2 / \mZ^2}{1 + 4 k^2 / \mQt^2 } \biggr)
 \biggr]
 \biggr\} 
 \;, 
 \nn 
  \Gamma^\rmii{HTL}_{\! u} + 2 k \Gamma^\rmii{HTL}_{\!\mathcal{K}}
 & \approx & 
 \frac{T}{8\pi} 
 \biggl\{
 2 g_2^2 \ln\biggl( \frac{ 1 + 4 k^2 / \mW^2 }{1 + 4 k^2 / \mWb^2 } \biggr)
 \la{Gamma_HTL_upK} \\ 
 && \hspace*{-5mm} + \,
 (g_1^2 + g_2^2) 
 \biggl[
 \cos^2(\theta - \bar{\theta}) 
   \ln \biggl( \frac{1 + 4 k^2 / \mZ^2 }{ 1 + 4 k^2 / \mZb^2 } \biggr)
 + 
 \sin^2(\theta - \bar{\theta}) 
   \ln \biggl( \frac{1 + 4 k^2 / \mZ^2 }{ 1 + 4 k^2 / \mQb^2 } \biggr)
 \biggr]
 \biggr\} 
 \;. \nonumber
\ea

%
\subsection{Widths at intermediate temperatures:
 Born $1\rightarrow 2$ decays 
 }
\la{ss:Born}

In the intermediate temperature range $\pi T \sim \mW^{ }$, 
no resummations are necessary and 
the inverse neutrino propagator
is given by \eq\nref{Born}. Its imaginary part, 
called the Born rate, can be expressed 
in terms of logarithms and dilogarithms denoted by 
\ba
 && 
 \lnb(p) \; \equiv \; \ln \Bigl( 1 - e^{-p/T} \Bigr)
 \;, \quad
 \lib(p) \; \equiv \; \mbox{Li}^{ }_2 \Bigl(e^{-p/T}\Bigr)
 \;, \la{lnb} \\
 && 
 \lnf(p) \; \equiv \; \ln \Bigl( 1 + e^{-p/T} \Bigr)
 \;, \quad\;
 \lif(p) \;\, \equiv \; \mbox{Li}^{ }_2 \Bigl(-e^{-p/T}\Bigr)
 \;, \la{lnf}
\ea 
which satisfy 
$
 T l'_\rmi{2b} (p) = \lnb(p) 
$,
$
 T l'_\rmi{2f}(p) = \lnf(p) 
$,
$
 T l'_\rmi{1b}(p) = \nB^{ }(p) 
$,
and
$
 T l'_\rmi{1f}(p) = -\nF^{ }(p) 
$.
Parallelling the splitup in \eqs\nref{splitup}
and \nref{bBorn}, we can write
\ba
 \Gamma^\rmi{Born}_{\!u,\mathcal{K}}
 & =& 
 (g_1^2 + g_2^2)\,
   \tilde{\Gamma}_{\!u,\mathcal{K}}^\rmi{Born}
         (\mZ^{ },\munLa,\muZn^{ })
 + 
 2 g_2^2\, \tilde{\Gamma}_{\!u,\mathcal{K}}^\rmi{Born}
                 (\mW^{ },\mueLa,\muW^{ })
 \;. \la{Gamma_upK_Born}
\ea
For $\pi T \sim \mW^{ }$, $v \sim \pi T / g \gg T$, 
so according to \eq\nref{broken2} we can omit $\muZ^{ }$ in 
comparison with $\muQ^{ }$, and set
$
 \munLa \to \mu^{ }_a 
$, 
$
 \muZn \to 0
$, 
$
 \mueLa \to \mu^{ }_a - \muQ^{ } 
$, 
$
 \muW \to \muQ^{ } 
$. 
For the combinations appearing in \eqs\nref{final_minus} 
and \nref{final_plus} we need
(assuming $\mathcal{K}^2 \ll m^2$)
\ba
 \tilde{\Gamma}^\rmi{Born}_{\!u}(m,\mu^{ }_1,\mu^{ }_2) 
 & = &
 \frac{m^2 T}{32\pi k^2}
 \biggl[
  \lnf\biggl( \frac{m^2}{4 k} + \mu^{ }_1  \biggr) 
 - 
  \lnb\biggl( k + \frac{m^2}{4 k} - \mu^{ }_2 \biggr) 
 \biggr]
 \;, \hspace*{7mm} \la{GammaBorn_minus} \\ 
 \bigl( \tilde{\Gamma}^\rmi{Born}_{\!u}
  + 2 k\, \tilde{\Gamma}^\rmi{Born}_{\!\mathcal{K}}
 \bigr) (m,\mu^{ }_1,\mu^{ }_2)
 & = & 
 \frac{T^2}{8\pi k}
 \biggl[
  \lib\biggl( k + \frac{m^2}{4 k} - \mu^{ }_2 \biggr) 
 - 
  \lif\biggl( \frac{m^2}{4 k} + \mu^{ }_1 \biggr) 
 \biggr]
 \;, \la{GammaBorn_plus}
\ea
where it is understood that the results can be expanded
to first order in chemical potentials. 
We note that in the high-temperature limit, when 
$k\sim \pi T$ and $m^2 \ll k T$, the helicity-flipping interaction rate,
$
 \tilde{\Gamma}^\rmi{Born}_{\!u}
  + 2 k\, \tilde{\Gamma}^\rmi{Born}_{\!\mathcal{K}}
$,
is larger than the helicity-conserving one, 
$
 \tilde{\Gamma}^\rmi{Born}_{\!u}
$.
In the low-temperature limit, both become
exponentially suppressed. 

%
\subsection{Widths at low temperatures: 
            Fermi $2\leftrightarrow 2$ scatterings and 
            $1\rightarrow 3$ decays}
\la{ss:Fermi}

\begin{table}[t]

\begin{minipage}[t]{15.5cm}
{ 
\begin{center}
\begin{tabular}{l@{~~~}l@{~~~}l@{~~~}l@{~~~}l}
\hline\hline
 channel & 
 coefficient & 
 $\mu^{ }_1$ & 
 $\mu^{ }_2$ & 
 $\mu^{ }_3$  \\
\hline\hline \\[-3mm]
 WW + quarks & 
 $c^{ }_\rmii{1L} = 2 \Nc^{ }
 (|V^{ }_{\!ud}|^2 + |V^{ }_{\!us}|^2
 + |V^{ }_{\!cd}|^2 + |V^{ }_{\!cs}|^2)$ & 
 $\mueLa$ & 
 $-\mu_{d_\rmii{$L$}}$ & 
 $\mu_{u_\rmii{$L$}}$  \\[2mm]
 WW + leptons & 
 $c^{ }_\rmii{1L} = 2 \sum_{b=1}^{3} $ & 
 $\mueLa$ & 
 $-\mu_{e_{\rmii{$L$}b}}$ & 
 $\mu_{\nu_{\rmii{$L$}b}}$  \\[2mm]
 ZZ + quarks & 
 $c^{ }_\rmii{1L} = \frac{\Nc}{2}
 \bigl(1 - \frac{4 s^2}{3} \bigr)^2\sum_{u,c}$ & 
 $\munLa$ & 
 $-\mu_{u_\rmii{$L$}}$ & 
 $\mu_{u_\rmii{$L$}}$  \\[2mm]
 & 
 $c^{ }_\rmii{1R} = \frac{\Nc}{2}
 \bigl(- \frac{4 s^2}{3} \bigr)^2\sum_{u,c}$ & 
 $\munLa$ & 
 $-\mu_{u_\rmii{$R$}}$ & 
 $\mu_{u_\rmii{$R$}}$  \\[2mm]
 & 
 $c^{ }_\rmii{1L} = \frac{\Nc}{2}
 \bigl(-1 + \frac{2 s^2}{3} \bigr)^2\sum_{d,s,b}$ & 
 $\munLa$ & 
 $-\mu_{d_\rmii{$L$}}$ & 
 $\mu_{d_\rmii{$L$}}$  \\[2mm]
 & 
 $c^{ }_\rmii{1R} = \frac{\Nc}{2}
 \bigl(\frac{2 s^2}{3} \bigr)^2\sum_{d,s,b}$ & 
 $\munLa$ & 
 $-\mu_{d_\rmii{$R$}}$ & 
 $\mu_{d_\rmii{$R$}}$  \\[2mm]
 ZZ + leptons & 
 $c^{ }_\rmii{1L} = \frac{1}{2}
 \bigl(-1 + 2 s^2 \bigr)^2\sum_{b=1}^{3}$ & 
 $\munLa$ & 
 $-\mu_{e_{\rmii{$L$}b}}$ & 
 $\mu_{e_{\rmii{$L$}b}}$  \\[2mm]
 & 
 $c^{ }_\rmii{1R} = \frac{1}{2}
 \bigl(2 s^2 \bigr)^2\sum_{b=1}^{3}$ & 
 $\munLa$ & 
 $-\mu_{e_{\rmii{$R$}b}}$ & 
 $\mu_{e_{\rmii{$R$}b}}$  \\[2mm]
 ZZ + neutrinos & 
 $c^{ }_\rmii{1L} = \frac{1}{2}
 \sum_{b=1}^{3}$ & 
 $\munLa$ & 
 $-\mu_{\nu_{\rmii{$L$}b}}$ & 
 $\mu_{\nu_{\rmii{$L$}b}}$  \\[2mm]
 & 
 $c^{ }_\rmii{2} = -\frac{1}{2}$ & 
 $\munLa$ & 
 $-\munLa$ & 
 $\munLa$  \\[2mm]
 WZ + leptons & 
 $c^{ }_\rmii{2} = 1 - 2 s^2$ & 
 $\mueLa$ & 
 $-\mueLa$ & 
 $\munLa$  \\[2mm]
 & 
 $c^{ }_\rmii{2} = 1 - 2 s^2$ & 
 $\munLa$ & 
 $-\mueLa$ & 
 $\mueLa$  \\[2mm]
\hline\hline
\end{tabular}
\end{center}
}
\end{minipage}

\caption[a]{\small
 The coefficients and chemical potentials 
 that appear in \eq\nref{Delta_Fermi}, 
 with $s^2 \equiv \sin^2(\theta^{ })$. 
 Each ``channel'' is labelled by the gauge bosons
 and fermions participating in the reaction. 
 For clarity we have assigned 
 separate chemical potentials to 
 different chiral states,  
 but when \eq\nref{broken} is satisfied as is necessary for 
 a quasiparticle picture,
 the chemical potentials of 
 chiral partners coincide, cf.\ table~\ref{table:mus}.
}
\label{table:channels}
\end{table}

The low-temperature limits of 
$\Gamma^{ }_{\!u}$ and $\Gamma^{ }_{\!\mathcal{K}}$ originate from 
$2\leftrightarrow 2$ scatterings and $1\rightarrow 3$ decays
among light fermions. To address these, 
we recall that at $\pi T \ll \mW^{ }$, weak gauge bosons can be 
integrated out and the physics described by the Fermi model.  
The four-fermion coupling is proportional to $G^{ }_\rmii{F}$, 
and the rates of $2\leftrightarrow2$ scatterings to $G_\rmii{F}^2$.

In this situation, 
the advanced inverse neutrino propagator 
can be written as 
\be
 \Delta(\mathcal{K} - i u 0^+) 
 = \bsl{\mathcal{K}} + 
 \sum_{\rmi{channels}}
 \bigl\{
   c^{ }_\rmii{1L} T^{ }_\rmii{1L}(\mu^{ }_1,\mu^{ }_2,\mu^{ }_3) 
 +
   c^{ }_\rmii{1R} T^{ }_\rmii{1R}(\mu^{ }_1,\mu^{ }_2,\mu^{ }_3) 
 +
   c^{ }_\rmii{2} T^{ }_\rmii{2}(\mu^{ }_1,\mu^{ }_2,\mu^{ }_3) 
 \bigr\}
 \;, \la{Delta_Fermi} 
\ee
where the coefficients $c^{ }_i$ and the chemical 
potentials $\mu^{ }_i$ are listed in table~\ref{table:channels} 
(given that we are at $T \ll v$, we can again set $\muZ^{ }\to 0$).
The different structures can be compactly 
expressed within the imaginary-time formalism: 
\ba
 T^{ }_\rmii{1L} \!\!& = &\!\! 
 4 G_\rmii{F}^2\, 
 \Tint{\{\tilde{P}^{ }_{\!1}\tilde{P}^{ }_{\!2}\tilde{P}^{ }_{\!3}\}}
 \hspace*{-5mm}
 \deltabar\bigl(\tilde{K} - \Sigma^{ }_i \tilde{P}^{ }_i \bigr)
 \, 
 \aR
 \gamma^{ }_{\mu}
 \frac{i\bsl{\tilde{P}}^{ }_{\!\!\!1}}{\tilde{P}_{\!1}^2}
 \, \gamma^{ }_{\nu}\, \aL \, 
 \tr \biggl[ 
 \frac{i\bsl{\tilde{P}}^{ }_{\!\!\!2}}{\tilde{P}_{\!2}^2}
 \, \gamma^{ }_{\mu}\,
 \frac{i\bsl{\tilde{P}}^{ }_{\!\!\!3}}{\tilde{P}_{\!3}^2}
 \, \gamma^{ }_{\nu}\,
 \aL
 \biggr]^{ }_{\tilde{k}^{ }_n \to -i [k^{ }_0 - i 0^+]}
 \;, \la{T1L} \hspace*{6mm} \\ 
 T^{ }_\rmii{1R} \!\!& = &\!\! 
 4 G_\rmii{F}^2\, 
 \Tint{\{\tilde{P}^{ }_{\!1}\tilde{P}^{ }_{\!2}\tilde{P}^{ }_{\!3}\}}
 \hspace*{-5mm}
 \deltabar\bigl(\tilde{K} - \Sigma^{ }_i \tilde{P}^{ }_i \bigr)
 \, 
 \aR
 \gamma^{ }_{\mu}
 \frac{i\bsl{\tilde{P}}^{ }_{\!\!\!1}}{\tilde{P}_{\!1}^2}
 \, \gamma^{ }_{\nu}\, \aL \, 
 \tr \biggl[ 
 \frac{i\bsl{\tilde{P}}^{ }_{\!\!\!2}}{\tilde{P}_{\!2}^2}
 \, \gamma^{ }_{\mu}\,
 \frac{i\bsl{\tilde{P}}^{ }_{\!\!\!3}}{\tilde{P}_{\!3}^2}
 \, \gamma^{ }_{\nu}\,
 \aR
 \biggr]^{ }_{\tilde{k}^{ }_n \to -i [k^{ }_0 - i 0^+]}
 \;, \la{T1R} \\ 
 T^{ }_\rmii{2} \!\!& = &\!\! 
 4 G_\rmii{F}^2\, 
 \Tint{\{\tilde{P}^{ }_{\!1}\tilde{P}^{ }_{\!2}\tilde{P}^{ }_{\!3}\}}
 \hspace*{-5mm}
 \deltabar\bigl(\tilde{K} - \Sigma^{ }_i \tilde{P}^{ }_i \bigr)
 \, 
 \aR
 \gamma^{ }_{\mu}
 \frac{i\bsl{\tilde{P}}^{ }_{\!\!\!1}}{\tilde{P}_{\!1}^2}
 \, \gamma^{ }_{\nu}\, 
 \frac{i\bsl{\tilde{P}}^{ }_{\!\!\!2}}{\tilde{P}_{\!2}^2}
 \, \gamma^{ }_{\mu}\,
 \frac{i\bsl{\tilde{P}}^{ }_{\!\!\!3}}{\tilde{P}_{\!3}^2}
 \, \gamma^{ }_{\nu}\,
 \aL
 \biggr|^{ }_{\tilde{k}^{ }_n \to -i [k^{ }_0 - i 0^+]}
 \;. \la{T2}
\ea
Here 
$\Tinti{\{...\}}$ indicates a sum-integral over fermionic Matsubara
momenta; 
$\Tinti{P}\,\deltabar(P) \equiv 1$;
$\tilde{P}^{ }_i \equiv (p^{ }_{ni}+ i \mu^{ }_i,\vec{p}^{ }_i)$; 
and Euclidean conventions are used for Dirac matrices. 

After carrying out the Matsubara sums, 
substituting $\tilde{k}^{ }_n \to -i [k^{ }_0 - i 0^+]$,
setting $\mathcal{K}\to -\mathcal{K}$, 
identifying the self-energy
$\bsl{\Sigma}$ according to \eq\nref{S_R}, and taking the imaginary
part, we get
\be
 \im \bsl{\Sigma}(\mathcal{K}) = 
 \sum_{\rmi{channels}}
 \bigl\{
   c^{ }_\rmii{1L} \mathcal{T}^{ }_\rmii{1L}(\mu^{ }_1,\mu^{ }_2,\mu^{ }_3) 
 +
   c^{ }_\rmii{1R} \mathcal{T}^{ }_\rmii{1R}(\mu^{ }_1,\mu^{ }_2,\mu^{ }_3) 
 +
   c^{ }_\rmii{2} \mathcal{T}^{ }_\rmii{2}(\mu^{ }_1,\mu^{ }_2,\mu^{ }_3) 
 \bigr\}
 \;. 
\ee
Here the structures are analytic continuations of 
\eqs\nref{T1L}--\nref{T2}. 
Restricting to those channels that are kinematically allowed in 
the massless limit,\footnote{%
 At very low temperatures, the masses $M_{\!\J}^{ }$ should be 
 kept non-zero, which leads to $1\rightarrow 3$ decays
 through the same expression. In vacuum 
 and with massless final states 
 we find 
 $
  \Gamma^\rmii{Fermi}_{\!u} = 0
 $, 
 $
  \Gamma^\rmii{Fermi}_{\!\mathcal{K}} 
 =  { G_\rmii{F}^2 M_{\!\J}^4}/({192\pi^3})
 \sum_{\rmi{channels}}
 \bigl\{ 
    c^{ }_\rmii{1L} + c^{ }_\rmii{1R} - c^{ }_\rmii{2} 
 \bigr\} 
 $. 
 If only the neutrino channels are open, the sum evaluates to $+2$.
 In this regime the equilibrium distribution 
 $\nF^{ }(\kT^{ })$ should
 also be replaced by 
 $
  \nF^{ }(\sqrt{\kT^2 + M_{\!\J}^2})
 $. 
 } 
we obtain
\ba
 \mathcal{T}^{ }_i & = & 
 2 G_\rmii{F}^2 \, \nF^{-1}\bigl(k^{ }_0 - \Sigma^{ }_i \mu^{ }_i\bigr)
 \int_{\vec{p}^{ }_1\vec{p}^{ }_2\vec{p}^{ }_3}
 \hspace*{-1mm} 
 \frac{
 \mathcal{D}^{ }_i
 }{8 p^{ }_1 p^{ }_2 p^{ }_3}
 \nn 
 & \times & 
 \;\bigl\{\, 
 \deltabar\bigl( 
  \mathcal{P}^{ }_{\!1}
 + \mathcal{P}^{ }_{\!2}
 - \mathcal{P}^{ }_{\!3}
 - \mathcal{K}^{ }_{ }
 \bigr)\,
 \nF^{ }(p^{ }_1 - \mu^{ }_1)\,
 \nF^{ }(p^{ }_2 - \mu^{ }_2)\,
 \bigl[ 1 - \nF^{ }(p^{ }_3 + \mu^{ }_3) \bigr]
 \nn 
 & & + \, 
 \deltabar\bigl( 
  \mathcal{P}^{ }_{\!1}
 + \mathcal{P}^{ }_{\!3}
 - \mathcal{P}^{ }_{\!2}
 - \mathcal{K}^{ }_{ }
 \bigr)\,
 \nF^{ }(p^{ }_1 - \mu^{ }_1)\,
 \bigl[ 1 - \nF^{ }(p^{ }_2 + \mu^{ }_2) \bigr]
 \nF^{ }(p^{ }_3 - \mu^{ }_3)\,
 \nn 
 & & + \, 
 \deltabar\bigl( 
  \mathcal{P}^{ }_{\!2}
 + \mathcal{P}^{ }_{\!3}
 - \mathcal{P}^{ }_{\!1}
 - \mathcal{K}^{ }_{ }
 \bigr)\,
 \bigl[ 1 - \nF^{ }(p^{ }_1 + \mu^{ }_1) \bigr]\,
 \nF^{ }(p^{ }_2 - \mu^{ }_2)\,
 \nF^{ }(p^{ }_3 - \mu^{ }_3)\,
 \nn 
 & &  + \, 
 (\mbox{kinematically forbidden channels})
 \bigr\} 
 \;, \la{channels}
\ea 
where we have gone over to 
Minkowskian conventions, with $p^{ }_i \equiv |\vec{p}^{ }_i|$
and $\mathcal{P}^{ }_i \equiv (p^{ }_i,\vec{p}^{ }_i)$.
The Dirac traces $\mathcal{D}^{ }_i$ appearing in \eq\nref{channels} 
can be easily taken: 
\ba
 \mathcal{D}^{ }_\rmii{1L} & \equiv & 
  \aR \gamma^{\mu}_{ }\,\bsl{\mathcal{P}}^{ }_{\!\!1} \gamma^{\nu}_{ } \aL
  \tr\bigl[
   \bsl{\mathcal{P}}^{ }_{\!\!2} \gamma_{\mu}^{ }
   \bsl{\mathcal{P}}^{ }_{\!\!3} \gamma_{\nu}^{ }
   \aL
  \bigr]
 \; = \; 
 \aR \, 8\, \mathcal{P}^{ }_{\!1}\cdot\mathcal{P}^{ }_{\!3} \, 
 \bsl{\mathcal{P}}^{ }_{\!\!2} \, \aL
 \;, \la{trace1L} \\ 
  \mathcal{D}^{ }_\rmii{1R} & \equiv & 
  \aR \gamma^{\mu}_{ }\,\bsl{\mathcal{P}}^{ }_{\!\!1} \gamma^{\nu}_{ } \aL
  \tr\bigl[
   \bsl{\mathcal{P}}^{ }_{\!\!2} \gamma_{\mu}^{ }
   \bsl{\mathcal{P}}^{ }_{\!\!3} \gamma_{\nu}^{ }
   \aR
  \bigr]
 \; = \;
 \aR \, 8\, \mathcal{P}^{ }_{\!1}\cdot\mathcal{P}^{ }_{\!2} \, 
 \bsl{\mathcal{P}}^{ }_{\!\!3} \, \aL
 \;, \la{trace1R} \\  
 \mathcal{D}^{ }_\rmii{2} & \equiv & 
  \aR \gamma^{\mu}_{ }\,\bsl{\mathcal{P}}^{ }_{\!\!1} \gamma^{\nu}_{ } 
   \bsl{\mathcal{P}}^{ }_{\!\!2} \gamma_{\mu}^{ }
   \bsl{\mathcal{P}}^{ }_{\!\!3} \gamma_{\nu}^{ }
   \aL
 \; = \; 
 - \aR \, 8\, \mathcal{P}^{ }_{\!1}\cdot\mathcal{P}^{ }_{\!3} \, 
 \bsl{\mathcal{P}}^{ }_{\!\!2} \, \aL
 \;. \la{trace2}  
\ea
We refer to the structures containing 
$
 \mathcal{P}^{ }_{\!1}\cdot\mathcal{P}^{ }_{\!3}
$
as ``$t$-channel'' and to those containing
$
 \mathcal{P}^{ }_{\!1}\cdot\mathcal{P}^{ }_{\!2}
$
as ``$s$-channel'' contributions. Through a renaming 
of integration variables, together with a 
permutation of chemical potentials, the three channels 
in \eq\nref{channels} can be transformed into
the appearance of the first channel. 

As a final step, the phase space can be reduced to 
a convergent two-dimensional integral representation. 
For the widths in 
\eqs\nref{final_minus} and \nref{final_plus}, we thereby obtain
\ba 
 \Gamma^\rmi{Fermi}_\rmii{$(-)$} & \equiv & 
 \Gamma^\rmi{Fermi}_{u}
 \;, \quad 
 \Gamma^\rmi{Fermi}_\rmii{$(+)$} \; \equiv \; 
  \Gamma^\rmi{Fermi}_{\!u} 
  +  2 k \Gamma^\rmi{Fermi}_{\!\mathcal{K}}
 \;, \la{Gammau_Fermi} \\[2mm]
 \Gamma^\rmi{Fermi}_{(\tau)}
 & = & 
 \sum_{\rmi{channels}}
 \Bigl\{ 
   \bigl( c^{ }_\rmii{1L} - c^{ }_\rmii{2}\bigr)
   \Bigl[
     \Xi_{(\tau)}^{\,t}(\mu^{ }_1,\mu^{ }_2,\mu^{ }_3) 
  + 
     \Xi_{(\tau)}^{\,t}(\mu^{ }_3,\mu^{ }_2,\mu^{ }_1) 
  + 
     \Xi_{(\tau)}^{\,s}(\mu^{ }_1,\mu^{ }_3,\mu^{ }_2) 
   \Bigr]
 \nn 
 &  & \hspace*{1cm} + \,  
    c^{ }_\rmii{1R}\, 
   \Bigl[
     \Xi_{(\tau)}^{\,t}(\mu^{ }_1,\mu^{ }_3,\mu^{ }_2) 
  + 
     \Xi_{(\tau)}^{\,t}(\mu^{ }_2,\mu^{ }_3,\mu^{ }_1) 
  + 
     \Xi_{(\tau)}^{\,s}(\mu^{ }_1,\mu^{ }_2,\mu^{ }_3) 
   \Bigr]
 \Bigr\} 
 \;. \la{Gammaminus_Fermi}
\ea
Here the $t$ and $s$-channel integrals read
\ba
 & & \hspace*{-1.5cm}
 \Xi_{(\tau)}^{\,t}(\mu^{ }_{\alpha},\mu^{ }_{\beta},\mu^{ }_{\gamma})
 \; = \; 
 \frac{4 G_\rmiii{F}^2}{\pi^3 k^2}
 \int_0^k \! {\rm d}\qp^{ } 
 \int_{-\infty}^{0} \! {\rm d}\qm^{ } \, 
 \Bigl[
   \delta^{ }_{\tau,-}
   \, \qp^2 \qm^2 + 
   \delta^{ }_{\tau,+}
   \, \qp^{ } \qm^{ } (\qp^{ }-k)(k-\qm^{ })    
 \Bigr] 
 \nn 
 \!\!& \times &\!\! 
 \Bigl[ 
  1 - \nF^{ }(k - p^{ }_0 - \mu^{ }_{\beta}) 
    + \nB^{ }(p^{ }_0 - \mu^{ }_{\alpha} - \mu^{ }_{\gamma})
 \Bigr] 
 \nn 
 \!\!& \times &\!\! 
 \Bigl\{ 
  T \bigl[ \lnf(\mu^{ }_{\gamma} -\qm^{ })
  - \lnf(\qp^{ }- \mu^{ }_{\alpha}) \bigr]
 \Bigr\} 
 \;, \la{Xi_t} \\[2mm]
  & & \hspace*{-1.5cm}
 \Xi_{(\tau)}^{\,s}(\mu^{ }_{\alpha},\mu^{ }_{\beta},\mu^{ }_{\gamma})
 \; = \; 
 \frac{4 G_\rmiii{F}^2}{\pi^3 k^2}
 \int_k^{\infty} \! {\rm d}\qp^{ } 
 \int_{0}^{k} \! {\rm d}\qm^{ } \, 
 \Bigl[
   \delta^{ }_{\tau,-}
   \, \qp^2 \qm^2 + 
   \delta^{ }_{\tau,+}
   \, \qp^{ } \qm^{ } (\qp^{ }-k)(k-\qm^{ })    
 \Bigr] 
 \nn 
 \!\!& \times &\!\! 
 \Bigl[ 
      \nF^{ }(p^{ }_0 - k + \mu^{ }_{\gamma}) 
    + \nB^{ }(p^{ }_0 - \mu^{ }_{\alpha} - \mu^{ }_{\beta})
 \Bigr] 
 \nn 
 \!\!& \times &\!\! 
 \Bigl\{ 
  p + 
  T \bigl[
  \lnf(\qp^{ } - \mu^{ }_{\alpha})
  + \lnf(\qp^{ } - \mu^{ }_{\beta})
  - \lnf(\qm^{ }- \mu^{ }_{\alpha}) 
  - \lnf(\qm^{ }- \mu^{ }_{\beta}) 
 \bigr]
 \Bigr\} 
 \;, \la{Xi_s}
\ea
where 
$\lnf$ is defined in \eq\nref{lnf}
and $p^{ }_{\pm} \equiv (p^{ }_0 \pm p)/2$.
The integrands are supposed to be  
expanded to leading order 
in chemical potentials; 
the coefficients appearing after this expansion are
collected in appendix~C.
At zeroth order in chemical potentials, \eq\nref{Fermi_simplified} reproduces 
\eqs(5.34-37) of ref.~\cite{broken}.

%
\section{Determination of rate coefficients for the direct contribution}
\la{se:direct}

Let us turn to the direct contribution, which adds up to 
the indirect contribution according to \eq\nref{dir_indir}. 
At low temperatures, 
the largest indirect contribution is helicity-conserving, 
cf.\ \eq\nref{final_minus}, 
with the helicity-flipping channel in \eq\nref{final_plus}
lacking the possibility of resonant enhancement. 
For the direct contribution the roles are interchanged.

In the ultrarelativistic regime $m \ll \pi T$, $1\leftrightarrow 2$
reactions are phase-space suppressed. If $m\sim gT$, this implies 
that $1\leftrightarrow 2$ rates are of the same order as unsuppressed
$2\leftrightarrow 2$ rates. 
The $1\leftrightarrow 2$ processes
are also substantially modified by soft higher-order scatterings, i.e.\ 
by $1+n\leftrightarrow 2+n$ processes with $n\ge 1$, 
which therefore need to be 
summed to all orders, via a procedure known as 
Landau-Pomeranchuk-Migdal (LPM) resummation~\cite{bb1}. 
At low temperatures, 
when $m \sim \pi T$, the phase-space suppression is not present, and it
is sufficient to consider Born level $1\rightarrow 2$ decays. 
In the following we consider 
$2\leftrightarrow 2$,  
resummed $1+n \leftrightarrow 2+n$, 
and Born $1\leftrightarrow 2$ processes in turn. 

%
\subsection{High temperatures: 
$2\leftrightarrow 2$ scatterings with lepton or scalar exchange}
\la{ss:direct_highT}

The direct contribution from $2\leftrightarrow 2$ 
scatterings was originally determined
in ref.~\cite{bb2}, and subsequently 
resolved into helicity channels and generalized
to include chemical potentials relevant 
for the symmetric phase in ref.~\cite{cptheory}.
Two separate resummations were needed in the presence
of chemical potentials. 
In the broken phase, the chemical potentials and masses need 
to be re-adjusted, so that the results
of ref.~\cite{cptheory} change moderately. 

\begin{table}[t]

\begin{minipage}[t]{15.5cm}
{ 
\begin{center}
\begin{tabular}{l@{~~~}l@{~~~}l@{~~~}l@{~~~}l}
\hline\hline
 channel & 
 coefficient & 
 $\mu^{ }_1$ & 
 $\mu^{ }_2$ & 
 $\mu^{ }_3$  \\
\hline\hline \\[-3mm]
 $Z\, \phi^{ }_0 \to \bar{\nu}^{ }_\rmii{$L$} \nu^{ }_\rmii{$R$} $  & 
 $ c^{ }_{t1} = ({g_1^2 + g_2^2})/{2 }$ & 
 $\muZn^{ }$ & 
 $\muphin$ & 
 $\munLa$  \\[1mm]
 $Z' \phi^{ }_+ \to \bar{e}^{ }_\rmii{$L$} \nu^{ }_\rmii{$R$} $  & 
 $ c^{ }_{t1} = ({g_1^2 + g_2^2})/{2 }$ & 
 $\muZn^{ }$ & 
 $\muphip$ & 
 $\mueLa$  \\[1mm]
 $W^+ \phi^{ }_0 \to \bar{e}^{ }_\rmii{$L$} \nu^{ }_\rmii{$R$} $  & 
 $ c^{ }_{t1} = g_2^2$ & 
 $\muW^{ }$ & 
 $\muphin$ & 
 $\mueLa$  \\[1mm]
 $W^- \phi^{ }_+ \to \bar{\nu}^{ }_\rmii{$L$} \nu^{ }_\rmii{$R$} $  & 
 $ c^{ }_{t1} = g_2^2 $ & 
 $-\muW^{ }$ & 
 $\muphip$ & 
 $\munLa$  \\[1mm]  \hline
 $Z\, {\nu}^{ }_\rmii{$L$} \to \phi^{*}_0\, \nu^{ }_\rmii{$R$} $  & 
 $ c^{ }_{s1} = ({g_1^2 + g_2^2})/{2 }$ & 
 $\muZn^{ }$ & 
 $\munLa$ & 
 $\muphin$  \\[1mm]
 $Z' {e}^{ }_\rmii{$L$} \to \phi^{*}_+ \nu^{ }_\rmii{$R$} $  & 
 $ c^{ }_{s1} = ({g_1^2 + g_2^2})/{2 }$ & 
 $\muZn^{ }$ & 
 $\mueLa$ & 
 $\muphip$  \\[1mm]
 $W^+ {e}^{ }_\rmii{$L$} \to \phi^{*}_0\, \nu^{ }_\rmii{$R$} $  & 
 $ c^{ }_{s1} = g_2^2$ & 
 $\muW^{ }$ & 
 $\mueLa$ & 
 $\muphin$  \\[1mm]
 $W^- {\nu}^{ }_\rmii{$L$} \to \phi^{*}_+ \nu^{ }_\rmii{$R$} $  & 
 $ c^{ }_{s1} = g_2^2$ & 
 $-\muW^{ }$ & 
 $\munLa$ & 
 $\muphip$  \\[1mm]  \hline
 $\phi^{ }_0\, {\nu}^{ }_\rmii{$L$} \to Z\, \nu^{ }_\rmii{$R$} $  & 
 $ c^{ }_{u1} = ({g_1^2 + g_2^2})/{2 }$ & 
 $\muphin$ & 
 $\munLa$ & 
 $-\muZn^{ }$  \\[1mm]
 $\phi^{ }_+ {e}^{ }_\rmii{$L$} \to Z' \nu^{ }_\rmii{$R$} $  & 
 $ c^{ }_{u1} = ({g_1^2 + g_2^2})/{2 }$ & 
 $\muphip$ & 
 $\mueLa$ & 
 $-\muZn^{ }$  \\[1mm]
 $\phi^{ }_0\, {e}^{ }_\rmii{$L$} \to W^- \nu^{ }_\rmii{$R$} $  & 
 $ c^{ }_{u1} = g_2^2$ & 
 $\muphin$ & 
 $\mueLa$ & 
 $\muW^{ }$  \\[1mm]
 $\phi^{ }_+ {\nu}^{ }_\rmii{$L$} \to W^+ \nu^{ }_\rmii{$R$} $  & 
 $ c^{ }_{u1} = g_2^2$ & 
 $\muphip$ & 
 $\munLa$ & 
 $-\muW^{ }$  \\[1mm] \hline
 $\bar{t}^{ }_\rmii{$L$} t^{ }_\rmii{$R$} \to 
 \bar{\nu}^{ }_\rmii{$L$}\, \nu^{ }_\rmii{$R$} $  & 
 $ c^{ }_{s0} = h_t^2\Nc^{ }$ & 
 $-\mu_{t_\rmii{$L$}}$ & 
 $\mu_{t_\rmii{$R$}}$ & 
 $\munLa$  \\[1mm]
 $\bar{b}^{ }_\rmii{$L$} t^{ }_\rmii{$R$} \to 
 \bar{e}^{ }_\rmii{$L$}\, \nu^{ }_\rmii{$R$} $  & 
 $ c^{ }_{s0} = h_t^2\Nc^{ }$ & 
 $-\mu_{b_\rmii{$L$}}$ & 
 $\mu_{t_\rmii{$R$}}$ & 
 $\mueLa$  \\[1mm] \hline
 $\bar{t}^{ }_\rmii{$L$} {\nu}^{ }_\rmii{$L$}  \to 
 \bar{t}^{ }_\rmii{$R$}\, \nu^{ }_\rmii{$R$} $  & 
 $ c^{ }_{t0} = h_t^2\Nc^{ }$ & 
 $-\mu_{t_\rmii{$L$}}$ & 
 $\munLa$ & 
 $\mu_{t_\rmii{$R$}}$  \\[1mm]
 $\bar{b}^{ }_\rmii{$L$} e^{ }_\rmii{$L$} \to 
 \bar{t}^{ }_\rmii{$R$}\, \nu^{ }_\rmii{$R$} $  & 
 $ c^{ }_{t0} = h_t^2\Nc^{ }$ & 
 $-\mu_{b_\rmii{$L$}}$ & 
 $\mueLa$ & 
 $\mu_{t_\rmii{$R$}}$  \\[1mm] \hline
 ${t}^{ }_\rmii{$R$} {\nu}^{ }_\rmii{$L$}  \to 
 {t}^{ }_\rmii{$L$}\, \nu^{ }_\rmii{$R$} $  & 
 $ c^{ }_{t0} = h_t^2\Nc^{ }$ & 
 $\mu_{t_\rmii{$R$}}$ &
 $\munLa$ & 
 $-\mu_{t_\rmii{$L$}}$  \\[1mm]
 ${t}^{ }_\rmii{$R$} e^{ }_\rmii{$L$} \to 
 {b}^{ }_\rmii{$L$}\, \nu^{ }_\rmii{$R$} $  & 
 $ c^{ }_{t0} = h_t^2\Nc^{ }$ & 
 $\mu_{t_\rmii{$R$}}$ & 
 $\mueLa$ & 
 $-\mu_{b_\rmii{$L$}}$ \\[1mm]
\hline\hline
\end{tabular}
\end{center}
}
\end{minipage}

\caption[a]{\small
 The channels, coefficients and chemical potentials
 (cf.\ table~\ref{table:mus})
 that appear in \eq\nref{direct_2to2_hard}. 
 The field $Z'$, which is a linear combination of the physical
 $Z$ and photon fields, is defined below \eq\nref{Zp}. 
 When \eq\nref{broken} is satisfied, the chemical potentials of 
 chiral partner states coincide.
}
\label{table:channels_direct}
\end{table}

The $2\leftrightarrow 2$ contribution originates
from scatterings with 
hard momenta $p^{ }_i \sim \pi T$ and 
is not phase-space suppressed. 
Therefore it can be evaluated in the massless limit, 
i.e.\ restricting to a term of the type $\beta^{ }_u\,(2k)$ in 
the language of \eq\nref{proj_plus}. 
The numerator of \eq\nref{Phi} becomes
\be
 \bar{u}^{ }_{\vec{k}\tau\J} 
 \, \aL \, \rho^{2\leftrightarrow 2,\rmi{direct}}_a
   (\mathcal{K}^{ }_\J) \, \aR \, 
 u^{ }_{\vec{k}\tau\I} 
  \approx 
 \delta^{ }_{\tau,+}
 \tr \{
  \bsl{\mathcal{K}}^{ }_{\!\J}  
  \, \aL \,  \rho^{2\leftrightarrow 2,\rmi{direct}}_a
    (\mathcal{K}^{ }_\J) \,  \aR \,
  \}
  \;.  
\ee
Consequently we can write the contribution from hard momenta as 
\be
 \Omega^{2\leftrightarrow 2,\rmi{direct,hard}}_{(a+)\I\J} = 
 \sum_\rmi{channels} c^{ }_{i} \, \Xi_\rmii{$(+)$}^{\,i}(\{\mu^{ }_i\})
 \;, \la{direct_2to2_hard}
\ee
where the coefficients $c^{ }_i$ and
the associated chemical potentials 
are listed in table~\ref{table:channels_direct}. 
The phase space integrals have forms analogous to \eq\nref{channels}, 
and are collected in appendix~D. 

The contribution from hard $2\leftrightarrow 2$ scatterings, 
\eq\nref{direct_2to2_hard}, needs to be resummed in two ways in order
to render it IR finite. This can be implemented by subtracting
the problematic terms, and subsequently adding
them in a resummed form: 
\ba
  \Omega^{2\leftrightarrow 2,\rmi{direct}}_{(a+)\I\J}
 & \;\equiv\; & 
  \Omega^{2\leftrightarrow 2,\rmi{direct,hard}}_{(a+)\I\J} 
 \; - \; 
  \Omega^{2\leftrightarrow 2,\rmi{direct,subtrL}}_{(a+)\I\J} 
 \; + \;
  \Omega^{2\leftrightarrow 2,\rmi{direct,softL}}_{(a+)\I\J} 
 \nn[2mm] 
 & \;-\; &  
  \Omega^{2\leftrightarrow 2,\rmi{direct,subtrH}}_{(a+)\I\J} 
 \; + \;  
  \Omega^{2\leftrightarrow 2,\rmi{direct,softH}}_{(a+)\I\J} 
 \;. \la{direct_2to2_all}
\ea
Here ``L'' and ``H'' refer to scatterings mediated by soft lepton exchange 
and taking place off soft Higgs bosons, respectively. 

Considering first the lepton exchange contribution, we define 
a thermal lepton mass as~\cite{weldon}
\be
 m_\ell^2 = \frac{(g_1^2 + 3 g_2^2)T^2}{16} + \rmO(\mu_i^2)
 \;. \la{mell}
\ee
The IR-sensitive contribution originates from the $t$ and $u$-channel  
terms, 
$
 c^{ }_{t1} \Xi_{(+)}^{t1} + c^{ }_{u1} \Xi_{(+)}^{u1}
$. 
The logarithmic divergence from 
small momenta can be subtracted with
\ba
  \Omega^{2\leftrightarrow 2,\rmi{direct,subtrL}}_{(a+)\I\J} 
 & \equiv & 
  \frac{m_\ell^2}{8\pi k}
  \int_0^{k} \! {\rm d}\qp \!
  \int_{-\infty}^{0} \!\!\! {\rm d}\qm 
  \, \biggl[\, 
  \frac{
   \nB^{ }(k-\muphin) + \nF^{ }(\munLa^{ })
  }
  {p^2}
 \nn 
 &  & \hspace*{3.2cm}
 + \,   
  \frac{
   \nB^{ }(k-\muphip) + \nF^{ }(\mueLa^{ })
  }
  {p^2} \, \biggr]
 \;, \quad
 p^{ }_{\pm} \;\equiv\; \frac{ p^{ }_0 \pm p }{ 2 }
 \;. \hspace*{7mm}
\ea
The resummed term, obtained by using 
a HTL propagator for the soft lepton, reads
\be
 \Omega^{2\leftrightarrow 2,\rmi{direct,softL}}_{(a+)\I\J} 
  = 
 \frac{m_\ell^2}{16\pi k}
 \Bigl[ 
   \nB^{ }(k - \muphin^{ }) + \nF^{ }(\munLa^{ })
  + \nB^{ }(k - \muphip^{ }) + \nF^{ }(\mueLa^{ })
 \Bigr] 
 \ln \biggl( 1 + \frac{4k^2}{m_\ell^2} \biggr) 
 \;. 
\ee 

Turning to scatterings off soft Higgs bosons, 
the problem arises from expanding 
$\nB^{ }(k - p^{ }_0 - \mu^{ }_2)$ in 
\eq\nref{channels_direct_t1}, 
$\nB^{ }(k - p^{ }_0 - \mu^{ }_1)$ in 
\eq\nref{channels_direct_u1}, and
$\nB^{ }(p^{ }_0 - k + \mu^{ }_3)$ in 
\eq\nref{channels_direct_s1}, 
to first order in chemical potentials, 
yielding $\pm \mu^{ }_i/(k - p^{ }_0)^2$. 
Then there is a logarithmic
divergence from momenta $p^{ }_0 \approx k$.
Inserting the coefficients from table~\ref{table:channels_direct}, 
the problematic terms can be subtracted with 
\ba
 & & \hspace*{-1.5cm}
 \Omega^{2\leftrightarrow 2,\rmi{direct,subtrH}}_{(a+)\I\J} 
 \; \equiv \; 
 \frac{(g_1^2 + 3 g_2^2)(\muphin+\muphip)T}{4(4\pi)^3 k^2}\; \biggl\{ 
 \nn 
 & & 
 \int_0^k \! {\rm d}p^{ }_0\int_{p^{ }_0}^{2k - p^{ }_0} \!\! {\rm d}p \; 
 \frac{
  T \bigl[ \lnf(k) - \ln(-\frac{\qm}{T}) \bigr] 
  + \frac{T^2}{k} 
   \bigl[
      \lib(k) - \lif(k) + \frac{\pi^2}{4}  
   \bigr] 
 }{(k - p^{ }_0)^2}
 \nn 
 & - & 
 \int_k^{2k} \! {\rm d}p^{ }_0\int^{p^{ }_0}_{2k - p^{ }_0} \!\! {\rm d}p \; 
 \frac{
 \frac{k}{2}
 + T \bigl[ \lnf(k) - \ln(\frac{\qm}{T}) \bigr] 
 + \frac{T^2}{k} 
   \bigl[
      \lib(k) - \lif(k) - \frac{\pi^2}{4}  
   \bigr] 
 }{(p^{ }_0 - k)^2}
 \; \biggr\} 
 \;. \hspace*{5mm}
\ea

The resummed result
is obtained by putting the integration domains together, whereby most
terms cancel, and integrating the remainder over a domain 
regularized by a scalar mass. 
At this point, we recall the discussion around \eqs\nref{SEZ}, 
\nref{propZ}, namely that neutral scalars cannot be treated
as being in chemical equilibrium if $\muZ^{ }\neq 0$. Therefore we 
borrow an argument from the parametric regime $v \gg T$,  
and impose \eq\nref{broken}. Then 
$ 
 \muphin^{ }+\muphip^{ } = \muQ^{ } + s^2 \muZ^{ } \simeq \muQ^{ } 
$.  
The resummed contribution, originating from charged scalars, reads
\ba
 \Omega^{2\leftrightarrow 2,\rmi{direct,softH}}_{(a+)\I\J} 
 & \simeq &
 \frac{(g_1^2 + 3 g_2^2) \muQ^{ } T }
 {4(4\pi)^3 k}
 \biggl( \frac{\pi^2 T^2}{k^2} -1 \biggr)
 \nn 
 & \times & \theta(k - \mW^{ })\,
 \biggl[\; 
  \ln \biggl( \frac{k + \sqrt{k^2 - \mW^2}}{\mW^{} } \biggr)
 - 
  \frac{\sqrt{k^2 - \mW^2}}{k}
 \;\biggr]
 \;. \hspace*{5mm} \la{resum_S}
\ea
The result for the symmetric phase is recovered by setting
$\muQ^{ } \to \muY^{ }$ and $\mW^{ }\to m^{ }_{\phi}$. In fact, 
apart from the values of running couplings, \eq\nref{resum_S}
represents the only difference of the symmetric and broken phase
values of the direct $2\leftrightarrow 2$ contribution. 

When $\pi T \ll \mW^{ }$, the $2\leftrightarrow 2$ contributions
determined by using massless propagators need to switched off. We have 
done this by multiplying 
$
 \Omega^{2\leftrightarrow 2,\rmi{direct}}_{(a+)\I\J} 
$ 
by a phenomenological factor $\kappa(\mW^{ })$, 
defined as  
\be
 \kappa(\mW) 
 \; \equiv \; \frac{3}{\pi^2 T^3}
 \int_0^\infty \! {\rm d}p\, p^2
 \nB^{ }(\sqrt{p^2 + \mW^2})
 \bigl[ 1 + \nB^{ }(\sqrt{p^2 + \mW^2}) \bigr]
 \;. \la{kappa}
\ee

%
\subsection{High temperatures:
ultrarelativistic $1+n\leftrightarrow 2+n$ scatterings and decays}

The treatment of direct $1+n\leftrightarrow 2+n$ scatterings requires
LPM resummation, a procedure that was first
worked out for right-handed neutrinos in ref.~\cite{bb1}. Some chemical
potentials were included in ref.~\cite{n3}. 
These results
were resolved into helicity channels and generalized to include all chemical 
potentials relevant for the symmetric phase
in ref.~\cite{cptheory}. 
In the broken phase, the assignment of
chemical potentials and masses needs to be reconsidered. 

As discussed in ref.~\cite{broken}, the LPM contribution originates 
from four components of a wave function, describing different 
annihilation channels. We express this as 
\be
 \Omega^\rmi{LPM,direct}_{(a\tau)\I\J}
 = 
 \im\Bigl\{\,  
 \Psi^{\rmii{LPM}(\rmii{$H$})}_{(\tau)\I\J}(\munLa^{ },\muphin)
 + 
 \Psi^{\rmii{LPM}(\rmii{$Z$})}_{(\tau)\I\J}(\munLa^{ },\muphin) 
 +  
 2\,  \Psi^{\rmii{LPM}(\rmii{$W$})}_{(\tau)\I\J} (\mueLa^{ },\muphip)
 \, \Bigr\}
 \;, \la{channels_LPM_direct}
\ee
where the superscript $\alpha\in\{H,Z,W\}$ enumerates the components. 
According to table~\ref{table:mus} and the definitions
in \eq\nref{broken2}, the chemical potentials read
$
 \munLa = \mu^{ }_a - \muZ^{ }/2
$, 
$
 \muphin = \muZ^{ }/2
$, 
$
 \mueLa = \mu^{ }_a - \muQ^{ } + \muZ^{ }(\frac{1}{2} - s^2)
$, 
$
 \muphip = \muQ^{ } - \muZ^{ }(\frac{1}{2} - s^2)
$. 
Because of issues discussed above \eq\nref{resum_S}, we impose
\eq\nref{broken}, omitting contributions from $\muZ^{ }$.
The resummed terms read
\ba
 \im \Psi^{\rmii{LPM}(\alpha)}_{(\tau)\I\J}(\mu^{ }_1,\mu^{ }_2)
 & = & 
  \frac{1}{16 \pi} 
 \int_{-\infty}^{\infty} \!\!\!\! {\rm d}\omega^{ }_1 \, 
 \int_{-\infty}^{\infty} \!\!\!\! {\rm d}\omega^{ }_2 \;
 \delta(k - \omega^{ }_1 - \omega^{ }_2) \, 
 \bigl[ 1 - \nF^{ }(\omega^{ }_1 - \mu^{ }_1)
          + \nB^{ }(\omega^{ }_2 - \mu^{ }_2) \bigr] 
 \nn  
 & & \hspace*{-2cm} \times 
  \, \frac{1}{\omega^{ }_2} 
 \lim_{\vec{y}_\perp \to \vec{0}}
 \biggl\{
  \frac{M^{ }_\I M^{ }_\J\, \delta^{ }_{\tau,-} }{k^2}
  \im\, \bigl[g_{ }^{(\alpha)} (\vec{y}_\perp )\bigr]
   \;  +  \;
  \frac{ \delta^{ }_{\tau,+} }{\omega_1^2}
  \im\, \bigl[\nabla_\perp\cdot \vec{f}_{ }^{(\alpha)}(\vec{y}_\perp )\bigr] 
 \biggr\}
 \;. \hspace*{3mm} 
\ea
The $s$ and $p$-wave functions $g_{ }^{(\alpha)}$ and $\vec{f}_{ }^{(\alpha)}$
satisfy the matrix equations
\ba
 && (\hat{H}^{ }_{\!\J} - i 0^+)\, g^{ }(\vec{y}_\perp) \, = \, 
  \delta^{(2)}(\vec{y}_\perp) \;, \quad 
 (\hat{H}^{ }_{\!\J} - i 0^+)\, \vec{f}^{ }(\vec{y}_\perp) \, = \, 
  -\nabla^{ }_\perp \delta^{(2)}(\vec{y}_\perp) 
 \;, \\[3mm]
 && 
 \hat{H}^{ }_{\!\J} \; \equiv \; - \frac{M^2_{\!\J}}{2 k}
 + 
   \frac{m_{\ell}^2 - \nabla_\perp^2}{2\omega^{ }_1}
 + 
   \frac{\mbox{diag}(\mH^2,\mZ^2,\mW^2,\mW^2) - \nabla_\perp^2}{2\omega^{ }_2}
 \; - \; 
    i \, \Gamma^{ }_{4\times 4}(\vec{y}_\perp) 
 \;. 
\ea
Here $m_\ell^{ }$ is from \eq\nref{mell}, 
whereas the matrix $\Gamma^{ }_{4\times 4}$ is given in \eq(3.20) 
of ref.~\cite{broken}.

%
\subsection{Intermediate temperatures:
 Born $1\rightarrow 2$ decays 
 }
\la{ss:Born_direct}

For $\pi T \lsim \mW^{ }$, the relevant direct processes are Born-level
decays of Higgs, $Z^0_{ }$ and $W^{\pm}_{ }$ bosons. 
Adopting our previous 
trick of analytically continuing into an advanced propagator
and subsequently inverting the sign of the four-momentum, 
the result can be written in a form 
analogous to \eqs\nref{splitup}, \nref{Born}
and \nref{channels_LPM_direct},\footnote{%
 The real part of $\Psi^\rmii{Born}_\rmi{ }$ also plays a role, 
 leading to a ``dispersive'' correction
 in \eq\nref{evol_rho}. In terms of the function
 $\mathcal{E}$ in \eq\nref{b_full}, 
 this amounts to the 
 terms $\propto T^2$ in \eqs\nref{H0} and \nref{D0}, with 
 \be
  \beta^{\pm}_{(a)}
 = 
 -\frac{2k}{T^2} 
 \biggl[ 
 \mathcal{E}(\mH^{ },\munLa^{ },\muphin)
 + 
 \mathcal{E}(\mZ^{ },\munLa^{ },\muphin)
   +   
 2 \mathcal{E}(\mW^{ },\mueLa^{ },\muphip)
 \biggr]^{ }_{\mu^{\pm}_{ }}
 \;, \la{dispersive2}
 \ee
 where 
 $
 \mu^{\pm}_{ }
 $ 
 indicates a symmetrization/antisymmetrization
 with respect to chemical potentials. 
 Eq.~\nref{dispersive2} originates from a helicity-flipping
 process like the factors $\frac{1}{2}$
 in \eq\nref{dispersive}. 
 Similarly to \eq\nref{b_indir},
 we can expand
 $ 
  \beta^+_{(a)} = \beta^\rmii{$(0)$}_{ }
 $ 
 and 
 $ 
  \beta^-_{(a)} =  
   \bar{\mu}^{ }_a \, \beta^{\rmii{$($}a\rmii{$)$}}_{ }
  + \sum_i \bar{\mu}^{ }_i \, \beta^{\rmii{$($}i\rmii{$)$}}_{ }
 $.  
 For $m \ll k,\pi T$,
 these evaluate to 
 $\beta^+_{(a)} \approx 1/4$, 
 $\beta^{-}_{(a)} \approx 0$. 
 }
\ba
 \rho_a^\rmi{Born,direct}(\mathcal{K})
 & = &
 \im \Bigl\{ \,  
     \Psi^{\rmi{Born}(\rmii{$H$})}_\rmi{ }(-\mathcal{K},\munLa^{ },\muphin)
 + 
     \Psi^{\rmi{Born}(\rmii{$Z$})}_\rmi{ }(-\mathcal{K},\munLa^{ },\muphin)
 \nn 
 &  & \hspace*{0.3cm} + \,   
 2   \Psi^{\rmi{Born}(\rmii{$W$})}_\rmi{ }(-\mathcal{K},\mueLa^{ },\muphip)
 \, \Bigr\} 
 \;. \la{channels_Born_direct}
\ea
Here,  denoting 
$\epsilon^{ }_1 \equiv |\vec{k-p}|$,  
$\epsilon^{ }_2 \equiv \sqrt{p^2 + m_\alpha^2}$ and
$\mu^{ }_3 \equiv \mu^{ }_1 + \mu^{ }_2$, 
the basic structure reads
\be
 \Psi_\rmi{ }^{\rmi{Born}(\alpha)}(\mathcal{K},\mu_1^{ },\mu^{ }_2)  \equiv  
 \left. 
 - \frac{1}{2}
 \Tint{P} \frac{i (\bsl{\tilde{K}}-\bsl{\tilde{P}})  }
 {[(\tilde{k}^{ }_n - \tilde{p}^{ }_n)^2 + \epsilon_1^2 ]
 (\tilde{p}_n^{\,2} + \epsilon_2^2)}
 \right|^{ }_{\tilde{k}^{ }_n \to -i [k^{ }_0 - i 0^+]}
 \hspace*{-1.5cm} \;. \la{Born_direct}
\ee
Carrying out the Matsubara sum, taking the imaginary part, and restricting
to the kinematics $\mathcal{K}^2 < m^2$ relevant for low temperatures, we find
\be
 \im\Psi_\rmi{ }^{\rmi{Born}(\alpha)}(-\mathcal{K},...)
 = 
 \int_{\vec{p}}
 \!\frac{\pi\,\delta(k^{ }_0 + \epsilon^{ }_1 - \epsilon^{ }_2)
 }{8\epsilon^{ }_1 \epsilon^{ }_2}
 \bigl[ \epsilon^{ }_1 \gamma^0_{ } 
 + (\vec{k+p})\cdot\vec{\gamma} \bigr]
 \bigl[
   \nF^{ }(\epsilon^{ }_1 + \mu^{ }_1)
 + \nB^{ }(\epsilon^{ }_2 - \mu^{ }_2)
 \bigr]
 \;,
\ee
where now $\epsilon^{ }_1 = |\vec{k+p}|$. After performing the angular
integral, the result can be decomposed as  
$ 
 \beta^{ }_{\mathcal{K}}\, \bsl{\mathcal{K}}^{ }_{ }
  + \beta^{ }_{u} \msl{u} 
$, 
so that matrix elements can be taken according to 
\eqs\nref{proj_minus} and \nref{proj_plus}. 
Thereby we obtain for $M \ll k$
expressions similar to
\eqs\nref{GammaBorn_minus} and \nref{GammaBorn_plus}, 
except that the roles of the helicity channels have swapped places: 
\ba
 \Omega^\rmi{Born,direct}_{(a-)\I\J}
 & = & 
 \sum_\rmi{channels}
 \frac{M^{ }_{\I} M^{ }_{\J} T^2}{32\pi k^3}
 \biggl[
  \lib\biggl( k + \frac{m^2}{4 k} - \mu^{ }_2 \biggr) 
 - 
  \lif\biggl( \frac{m^2}{4 k} + \mu^{ }_1 \biggr) 
 \biggr]
 \;, \la{PhiBorn_minus}
 \\ 
 \Omega^\rmi{Born,direct}_{(a+)\I\J}
 & = &
 \sum_\rmi{channels}
 \frac{m^2 T}{32\pi k^2}
 \biggl[
  \lnf\biggl( \frac{m^2}{4 k} + \mu^{ }_1  \biggr) 
 - 
  \lnb\biggl( k + \frac{m^2}{4 k} - \mu^{ }_2 \biggr) 
 \biggr]
 \;. \hspace*{7mm} \la{PhiBorn_plus}
\ea
Here the channels have the masses and chemical potentials
given in \eq\nref{channels_Born_direct}, and polylogarithms
are defined in \eqs\nref{lnb}--\nref{lnf}. 
Given that $\pi T \lsim \mW^{ }$ implies
$v \gg T$, we make use of \eq\nref{broken} and set 
$\muZ^{ }\to 0$ in the chemical potentials. 
If $m^2 < \mathcal{K}^2$, we omit this contribution. 

%
\subsection{Summary: putting everything together}
\la{ss:summary}

In \ses\ref{ss:direct_highT}--\ref{ss:Born_direct} we have discussed
the different direct contributions to the rate 
coefficients. Let us now specify how these are put together and
then combined with the indirect ones. 

The full direct contribution to the rate coefficients
in the broken phase can be expressed as 
\be
 \Omega^\rmi{direct}_{(a\tau)\I\J}  \;=\;  
 \Omega^{2\leftrightarrow 2,\rmi{direct}}_{(a\tau)\I\J} 
   \;+\; 
 \mathcal{I}
 \Bigl\{ 
    \Omega^\rmi{LPM,direct}_{(a\tau)\I\J}
    \,,\; 
    \Omega^\rmi{Born,direct}_{(a\tau)\I\J}
 \Bigr\} 
 \;. 
\ee
Here the $2\leftrightarrow 2$ part is from \eqs\nref{direct_2to2_all}
and \nref{kappa}. The function $\mathcal{I}$ represents an interpolation
between the two different $1+n \leftrightarrow 2+n$ computations, 
cf.\ \eqs\nref{channels_LPM_direct},
\nref{PhiBorn_minus} and
\nref{PhiBorn_plus}, 
in analogy with the procedure
discussed in ref.~\cite{broken}.\footnote{%
 More precisely, the LPM contribution is overtaken by the Born
 contribution at the smallest $k/T$ and $T/$GeV when the Born
 contribution is smaller than the LPM one. The reason is that 
 in these regimes the practical determination of 
 the LPM contribution, 
 making use of ultrarelativistic kinematics, 
 becomes unreliable and overestimates the correct result. 
 } 
The chemical potential dependence
is expanded to linear order, 
\be
 \Omega^\rmi{direct}_{(a\tau)\I\J}  \;\equiv\;  
 Q^\rmi{direct}_{(a\tau)\I\J} 
 + \bar{\mu}^{ }_a\, R^\rmi{direct}_{(a\tau)\I\J} 
 + {\textstyle\sum_i}\, \bar{\mu}^{ }_i\, 
 S^{(i)\rmi{direct}}_{(a\tau)\I\J}
 + \rmO(\bar{\mu}^2)
 \;, \quad
 \bmu^{ }_i \in \{ \bmuQ^{ }, \bmuZ^{ },
 \Sigma^{ }_b\, \bar{\mu}^{ }_b, \bmuB^{ } \} 
 \;. \la{QRS_direct}
\ee
This defines the coefficients 
$
  Q^\rmi{direct}_{(a\tau)}
$, 
$
  R^\rmi{direct}_{(a\tau)}
$
and 
$
  S^{(i)\rmi{direct}}_{(a\tau)}
$
that are subsequently summed together with the indirect contributions. 

As far as the indirect contributions go 
(cf.\ \ses\ref{se:indirect} and \ref{se:rates}), 
we invoke an expansion similar to \eq\nref{QRS_direct}, 
but this time for the coefficients appearing in the 
active neutrino self-energy, 
cf.\ \eq\nref{Sigma_form}: 
\ba
 b & = & 
 b^\rmii{$(0)$}_{ } 
 + \bar{\mu}^{ }_a \, b^{\rmii{$($}a\rmii{$)$}}_{ }
 + {\textstyle\sum_i}\, \bar{\mu}^{ }_i\, b^{(i)}_{ }
 + \rmO(\bmu^2)
 \;, \la{b_indir} \\[2mm] 
 \Gamma^{ }_{\!u} & = & 
 \Gamma_\rmii{$(-)$}^\rmii{$(0)$}
 + \bar{\mu}^{ }_a \, \Gamma_\rmii{$(-)$}^{\rmii{$($}a\rmii{$)$}}
 + {\textstyle\sum_i}\, \bar{\mu}^{ }_i\, 
   \Gamma_\rmii{$(-)$}^{(i)}
 + \rmO(\bmu^2)
 \;, \la{G_indir_m} \\[2mm]
 \Gamma^{ }_{\!u} + 2 k \Gamma^{ }_{\!\mathcal{K}} & = & 
 \Gamma_\rmii{$(+)$}^\rmii{$(0)$}
 + \bar{\mu}^{ }_a \, \Gamma_\rmii{$(+)$}^{\rmii{$($}a\rmii{$)$}}
 + {\textstyle\sum_i}\, \bar{\mu}^{ }_i\, 
   \Gamma_\rmii{$(+)$}^{(i)}
 + \rmO(\bmu^2)
 \;. \la{G_indir_p}
\ea
Here $b$ is given by \eqs\nref{bBorn} and \nref{b_full}, whereas
the other parts are obtained through an interpolation of the type 
discussed in ref.~\cite{broken},\footnote{%
  More precisely, the interpolation $\mathcal{I}$
  makes use of the Fermi contribution at $T \le \mW^{ }/\pi$ 
  and then freezes its value (in units of $T$). The HTL contribution
  overtakes the Fermi one, 
  once it exceeds the frozen value. In the rare case that
  the rapidly growing 
  Fermi contribution is still smaller than the HTL one at 
  $T = \mW^{ }/\pi$, we continue to follow it until the two cross, 
  and go over to the HTL one at higher temperatures.  
  }
based on the ingredients in 
\eqs\nref{Gamma_upK_Born}--\nref{GammaBorn_plus},
\nref{Gamma_HTL_u}--\nref{Gamma_HTL_upK}, and
\nref{Gammau_Fermi}--\nref{Gammaminus_Fermi}, respectively:
\be
 \Gamma_{(\tau)}^{\rmii{$($}0,a,i\rmii{$)$}}
   \;=\; 
 \Gamma_{(\tau)}^{\rmii{$($}0,a,i\rmii{$)$}\rmi{Born}}
 \;
  +
 \;
 \mathcal{I}
 \Bigl\{ 
  \Gamma_{(\tau)}^{\rmii{$($}0,a,i\rmii{$)$HTL}}
 \,,\;
  \Gamma_{(\tau)}^{\rmii{$($}0,a,i\rmii{$)$}\rmi{Fermi}}   
 \Bigr\} 
 \;. 
\ee
Here ``Born'' accounts for $1\rightarrow 2$ decays, whereas 
``HTL'' and ``Fermi'' are $2\leftrightarrow 2$ processes. 

With the ingredients in \eqs\nref{b_indir}--\nref{G_indir_p}, 
$
  \Omega^\rmi{indirect}_{(a\tau)\I\J}
$
is obtained from \eqs\nref{final_minus} and \nref{final_plus}, 
and we can construct the rate coefficients 
according to \eq\nref{RS}: 
\ba
   Q^\rmi{indirect}_{(a\tau)\I\J}
 & \equiv & 
 \fr12 \Bigl[ 
   \left. \Omega^\rmi{indirect}_{(a\tau)\I\J} \right|^{ }_{\mu}
 + \left. \Omega^\rmi{indirect}_{(a\tau)\I\J} \right|^{ }_{-\mu} \, 
 \Bigr]
 \;, \la{Q_indirect} \\
 \bar{\mu}^{ }_a R^\rmi{indirect}_{(a\tau)\I\J}
 + {\textstyle\sum_i}\,\bar{\mu}^{ }_i\, S^{(i)\rmi{indirect}}_{(a\tau)\I\J}
 & \equiv & 
 \fr12 \Bigl[ 
   \left. \Omega^\rmi{indirect}_{(a\tau)\I\J} \right|^{ }_{\mu}
 - \left. \Omega^\rmi{indirect}_{(a\tau)\I\J} \right|^{ }_{-\mu} \, 
 \Bigr] 
 \;. \la{RS_indirect}
\ea
Note that no Taylor expansion in $\mu$ is invoked here, given that 
\eq\nref{final_minus} may contain a resonance.
Afterwards, the full 
$Q^{ }_{ }$, 
$R^{ }_{ }$, 
and 
$S^\rmii{$(i)$}_{ }$ 
are obtained 
according to \eq\nref{dir_indir}.

%
\section{Approximate solution and overall parametric dependences}
\la{se:scan}

In order to gain insight on the behaviour of the equations 
specified in \se\ref{se:overview}, we first consider an approximate
solution, similar to that followed in most of the literature. 
The idea is to assume that all components of the density matrix are
in kinetic equilibrium, with 
$
 \rho^{\pm}_{ }(\kT^{ })
 \equiv 
 \hat{\rho}^{\pm}_{ }(x)\, \nF^{ }(\kT^{ })
$,
$x = \ln (T^{ }_\rmi{max}/T)$. 
If we subsequently integrate both sides of 
\eq\nref{evol_rho} over $\kT^{ }$, we end up with a coupled 
set of equations for the lepton asymmetries and the variables 
$\hat{\rho}^{\pm}_{ }$, parametrized by integrals of the
rate coefficients, weighted by $\nF^{ }(\kT^{ })$
or $\nF^{ }(\kT^{ }) [ 1 - \nF^{ }(\kT^{ })]$. The latter
integrals can be carried out once and for all. As inputs for
this we employ the values of $Q,R,S,\beta,\kappa,\delta$ Taylor-expanded
to linear order in chemical potentials. 

It is appropriate to remark that, 
based on the analysis in ref.~\cite{cpnumerics},  
it is not clear {\it a priori} whether  
a momentum-averaged solution
can be accurate. First of all, 
the density matrices
found in ref.~\cite{cpnumerics} have the characteristic 
feature that they kinetically equilibrate very fast at small momenta, 
and remain close to their vanishing initial values 
at large momenta (a similar finding had been made 
in ref.~\cite{kinetic}). 
Consequently, 
$\rho^{\pm}_{12}(\kT^{ }) / \nF^{ }(\kT^{ })$ are peaked at  
around $\kT^{ } \sim 0.5 T$ in \fig{3} of ref.~\cite{cpnumerics}, 
rather than being constant. 
Second, the different momentum modes add up incoherently 
in the source terms for the lepton asymmetries, so that $Y^{ }_a - \YB^{ }/3$ 
show much less oscillations than the momentum-averaged recipe suggests. 
In spite of these differences, we find that the
momentum-averaged recipe performs reasonably well, with 
errors $\lsim 50$\% in many cases, even if differences 
of $\rmO(10)$ can also be found 
(cf.\ \figs\ref{fig:Ya_YB_circle} and \ref{fig:Ya_YB_square}). 

For the numerical solution itself, 
we remark that the system contains a ``charge'' that is almost
conserved at high temperatures, 
sometimes referred to as the ``fermion number'', 
and defined as the sum of the 
helicity asymmetries of right-handed neutrinos and 
the lepton asymmetries of the Standard Model particles. 
It is important to make sure that the integration algorithm
respects this symmetry, as otherwise a non-zero fermion number 
generated inadvertently by numerical inaccuracy may have
a large effect on late-time lepton asymmetries. 
On the other hand, physical
fermion-number violating interactions do originate from the 
helicity-conserving coefficients 
$Q^{ }_{(a-)}$, 
$R^{ }_{(a-)}$, 
$S^{ }_{(a-)}$~\cite{cptheory}, 
and they do become appreciable 
in the broken phase (cf.\ \eq\nref{final_minus}).

As far as the parameter values go, 
the neutrino Yukawa couplings are fixed as specified 
in ref.~\cite{cpnumerics}, by making use of active neutrino properties
from ref.~\cite{npheno} as well as the Casas-Ibarra 
parametrization from ref.~\cite{ci} (which can be generalized beyond
the see-saw limit~\cite{n0}). Choosing 
the right-handed neutrino mass to be $M\sim 1$~GeV, 
and noting that complex phases have effects of $\rmO(1)$,  
the results depend substantially on just two quantities, 
the mass splitting $\Delta M$ and the Casas-Ibarra parameter $\im z$.
The goal now is to map the viable parameter space in this plane.
The viability concerns both the baryon asymmetry, 
$
  \YB^{ } 
  = 0.87(1) \times 10^{-10} 
$~\cite{planck}, 
and low-scale lepton asymmetries, which we monitor through 
$\mathop{\mbox{max}}(\{|Y^{ }_a|\})$ evaluated 
at $T = 5$~GeV. 
Specifically we fix, following refs.~\cite{n3,cpnumerics}, 
the non-critical parameters to the benchmark point
\ba
 M^{ }_{1(2)} & = & M \mp \frac{\Delta M}{2} \;, \quad
 M \; = \; 0.7732\,\mbox{GeV}\;,  \quad
 \mbox{``inverted hierarchy''}\;,  
 \la{benchmark1} \\ 
 \re z & = & 2.444 \;, \quad 
 \phi^{ }_1 \; = \; - 1.857 \;, \quad
 \delta \; = \; - 2.199 \;. \la{benchmark2}
\ea

\begin{figure}[t]
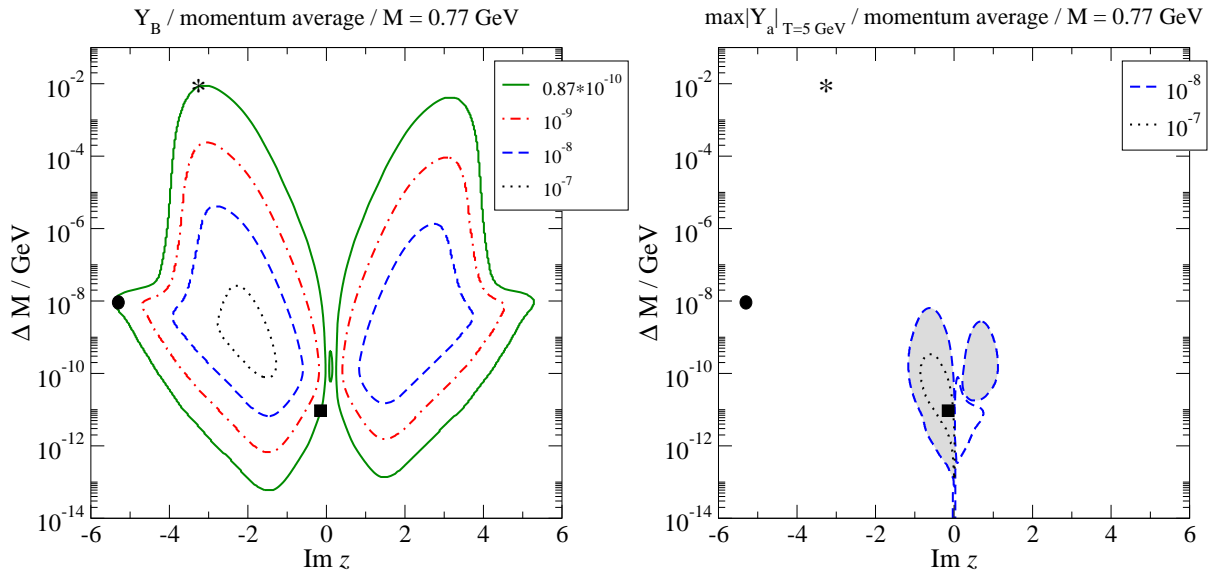


\hspace*{-0.1cm}
\centerline{%
 \epsfysize=7.5cm\epsfbox{contour_YB_highT.eps}%
 \hspace{0.1cm}%
 \epsfysize=7.5cm\epsfbox{contour_Ya_lowT.eps}
}

\caption[a]{\small
 Left: 
 contours of the total baryon yield $\YB^{ }$ 
 in the plane of $\im z$ and $\Delta M$, 
 with other parameters fixed according to \eqs\nref{benchmark1} 
 and \nref{benchmark2}.
 The smallest value considered is 
 $
  \YB^{ } = 0.87(1) \times 10^{-10} 
 $~\cite{planck}.
 Two benchmarks (filled circle and square) 
 are studied in more detail in \ses\ref{se:large_h} and 
 \ref{se:small_h}, respectively, whereas the point indicated with
 a star was studied in ref.~\cite{cpnumerics}. 
 Right: 
 analogous results for the maximal lepton asymmetry 
 $\mathop{\rm max}\{|Y^{ }_a|\}$, 
 evaluated at $T = 5$~GeV. 
 Within the shaded domains
 $Y^{ }_a,\YB^{ }$ are negative, within the unshaded positive.
}

\la{fig:contour_Ya_YB}
\end{figure}

Results obtained from the numerical solution of this system 
are shown in \fig\ref{fig:contour_Ya_YB}. We observe that largest 
values of $|\im z|$ are obtained for $\Delta M /M \sim 10^{-8}$. 
Because of the largest neutrino Yukawa couplings and 
consequently the largest mixing angle with active neutrinos,  
this situation, 
studied in more detail in \se\ref{se:large_h},  
is ideal for the experimental
search for right-handed neutrinos. On the other hand late-time 
lepton asymmetries can be considerably larger than the 
baryon asymmetry, but are obtained preferably with small 
values of $\im z$ and a more extreme degeneracy around
$\Delta M /M \sim 10^{-11}$, so that leptogenesis takes place
as late as possible. Such a situation is studied in more detail 
in \se\ref{se:small_h}.

%
\section{Accurate solution for large neutrino Yukawa couplings}
\la{se:large_h}

\begin{figure}[t]
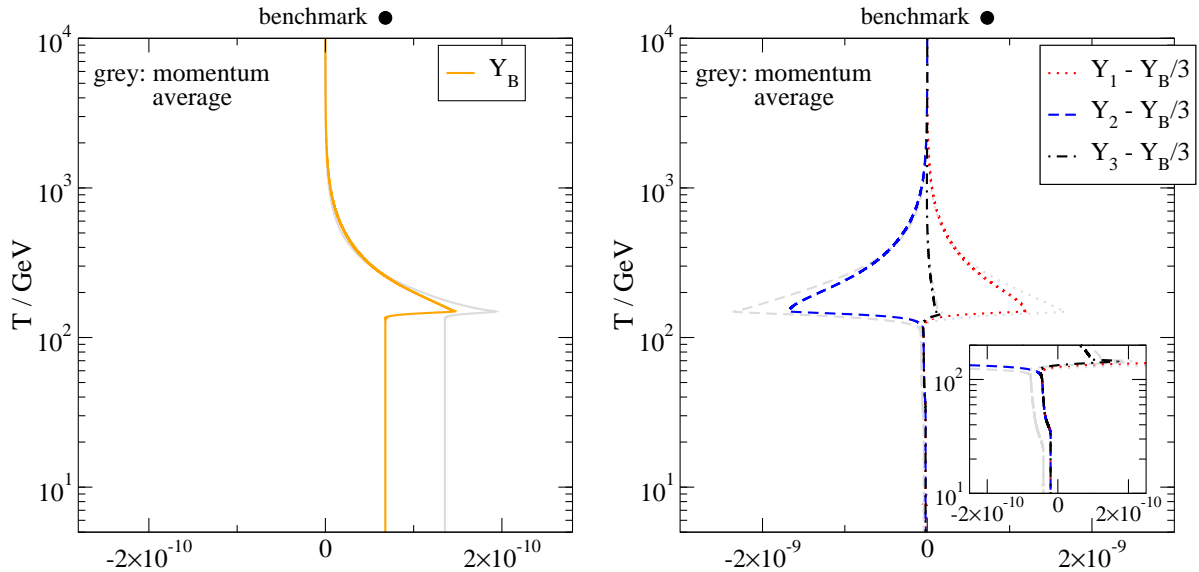


\hspace*{-0.1cm}
\centerline{%
 \epsfysize=7.5cm\epsfbox{YB_circle.eps}%
 \hspace{0.5cm}%
 \epsfysize=7.5cm\epsfbox{Ya_circle.eps}
}

\caption[a]{\small
 Left: 
 baryon yield
 as a function of $T/\mbox{GeV}$
 for the benchmark point defined in \se\ref{se:large_h}.
 Grey lines show the outcome if we resort to
 momentum averaging, like in \fig\ref{fig:contour_Ya_YB}.
 Momentum averaging 
 overestimates $\YB^{ }$ by a factor $\sim 2$. 
 Right: 
 the corresponding 
 $Y^{ }_a - {\YB^{ }}/{3}$. 
}

\la{fig:Ya_YB_circle}
\end{figure}

Consider the filled circle from \fig\ref{fig:contour_Ya_YB},
corresponding to 
$\Delta M \equiv 10^{-8}$~GeV, 
$\im z \equiv -5.3$.  
The magnitudes of the neutrino Yukawa couplings are conveniently 
characterized by the square roots of the eigenvalues 
of the matrix $h \, h^\dagger$, which read 
$
 0.7 \times 10^{-5}
$
and
$
 0.2 \times 10^{-9}
$.
Baryon asymmetry production peaks at temperatures just above
the freeze-out one, $T \sim 130$~GeV, so that little 
washout has time to take place while sphaleron transitions
are active, even if the washout rate is large. 
Lepton asymmetries are, however, efficiently washed out once sphaleron 
processes have decoupled.  

Let us mention that the
numerical integration of the basic equations is somewhat
demanding in this case. 
In the language of \eqs\nref{dp} and \nref{dm}, the dimensionless
rate coefficients are 
${\textstyle\sum_a}
   \re (h^{ }_{\I a} h^{*}_{\J a})   
   \,  
   \widehat Q^{+}_{\I\J}
 \sim 
 {\textstyle\sum_a} 
  \im (h^{ }_{\I a} h^{*}_{\J a}) 
   \, 
   \widehat Q^{-}_{\I\J}
  \sim 
 10^3  
$
at $T\sim 160$~GeV.  
Therefore a very fast equilibration process is taking place, 
and needs to be tracked with high accuracy, in a regime in which 
the rate coefficients vary rapidly~\cite{cptheory}. 
We have written two independent routines for the integration, 
utilizing different languages and platforms, and verified that
the results agree in general down to the $1 ... 2$\% level
(this applies also to wiggly features such as those observed 
in \fig\ref{fig:Ya_YB_square}(right)). 

The results from the numerical integration are shown 
in \fig\ref{fig:Ya_YB_circle}, where they are also compared
with the momentum-averaged treatment of \se\ref{se:scan}. The basic
feature of this benchmark is that baryon asymmetry freezes
out close to when lepton asymmetries are maximal. After the 
freeze-out, lepton asymmetries are rapidly erased. 
These qualitative features are correctly 
reproduced by the momentum-averaged approximation, even if  
momentum averaging is seen to 
overestimate the correct result by a factor $\sim 2$.

%
\section{Accurate solution for small neutrino Yukawa couplings}
\la{se:small_h}

\begin{figure}[t]
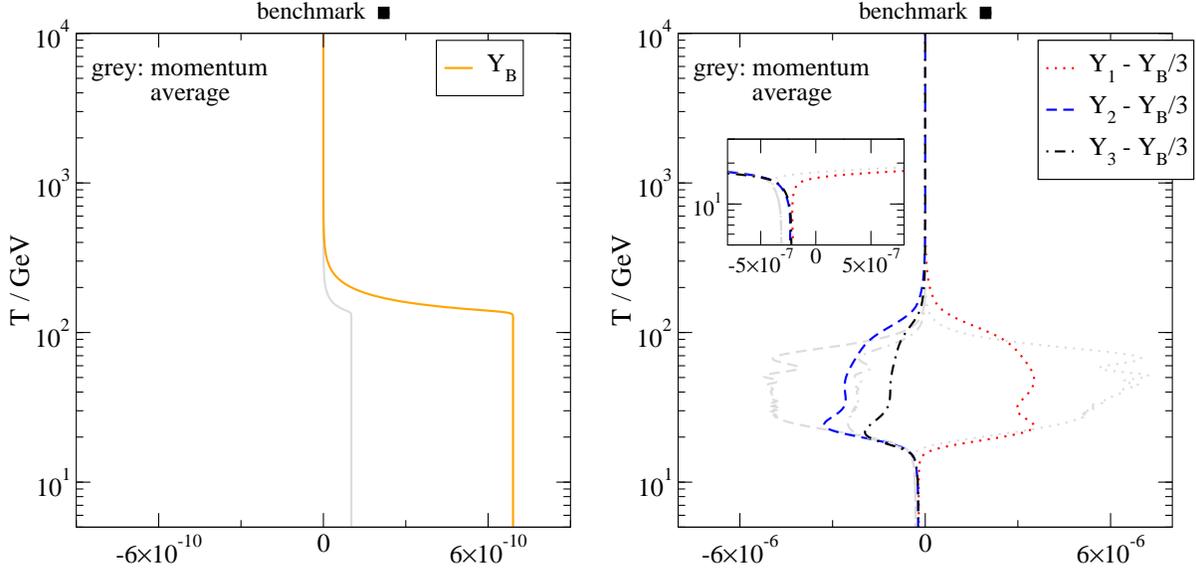


\hspace*{-0.1cm}
\centerline{%
 \epsfysize=7.5cm\epsfbox{YB_square.eps}%
 \hspace{0.5cm}%
 \epsfysize=7.5cm\epsfbox{Ya_square.eps}
}

\caption[a]{\small
 Left: 
 baryon yield 
 as a function of $T/\mbox{GeV}$
 for the benchmark point defined in \se\ref{se:small_h}.
 Grey lines show the outcome if we resort to
 momentum averaging, like in \fig\ref{fig:contour_Ya_YB}.
 Right: 
 the corresponding 
 $Y^{ }_a - {\YB^{ }}/{3}$. 
 Momentum averaging 
 underestimates $\YB^{ }$ by a factor~$\sim7$ but overestimates
 $Y^{ }_a - {\YB^{ }}/{3}$. 
}

\la{fig:Ya_YB_square}  
\end{figure}

Finally we consider the filled square from \fig\ref{fig:contour_Ya_YB},
corresponding to 
$\Delta M \equiv 10^{-11}$~GeV, 
$\im z \equiv -0.15$.  
In this case the square roots of the eigenvalues 
of the matrix $h \, h^\dagger$ are 
$
 4.1 \times 10^{-8}
$
and
$
 3.0 \times 10^{-8}
$.
Most of the lepton asymmetry generation takes place after sphaleron
processes have ceased to be active, i.e.\ at $T < 130$~GeV.

Our numerical solution is shown 
in \fig\ref{fig:Ya_YB_square}, where we have also compared
with the momentum averaged treatment (grey lines). 
Baryon asymmetry is seen to freeze out
already during an early stage of lepton asymmetry generation
(left panel).
In this particular case the momentum-averaged 
treatment is seen to underestimate the full result by 
a factor~$\sim 7$. 

The most remarkable feature of our solution concerns 
the lepton asymmetries, which are shown in the right panel of 
\fig\ref{fig:Ya_YB_square}. We observe 
that lepton asymmetries obtain a constant value below
$T \sim 15$~GeV, which is furthermore the same
in all flavours. This is the case 
for low-temperature lepton asymmetries in general. 
The existence of such a state
was proposed in ref.~\cite{eijima}, 
whose \eq(61) can be derived from
our \eq\nref{stationary} by summing over both active and sterile 
flavours, integrating over momenta, and approximating susceptibilities. 

\begin{figure}[t]

\hspace*{-0.1cm}
\centerline{%
 \epsfysize=7.5cm\epsfbox{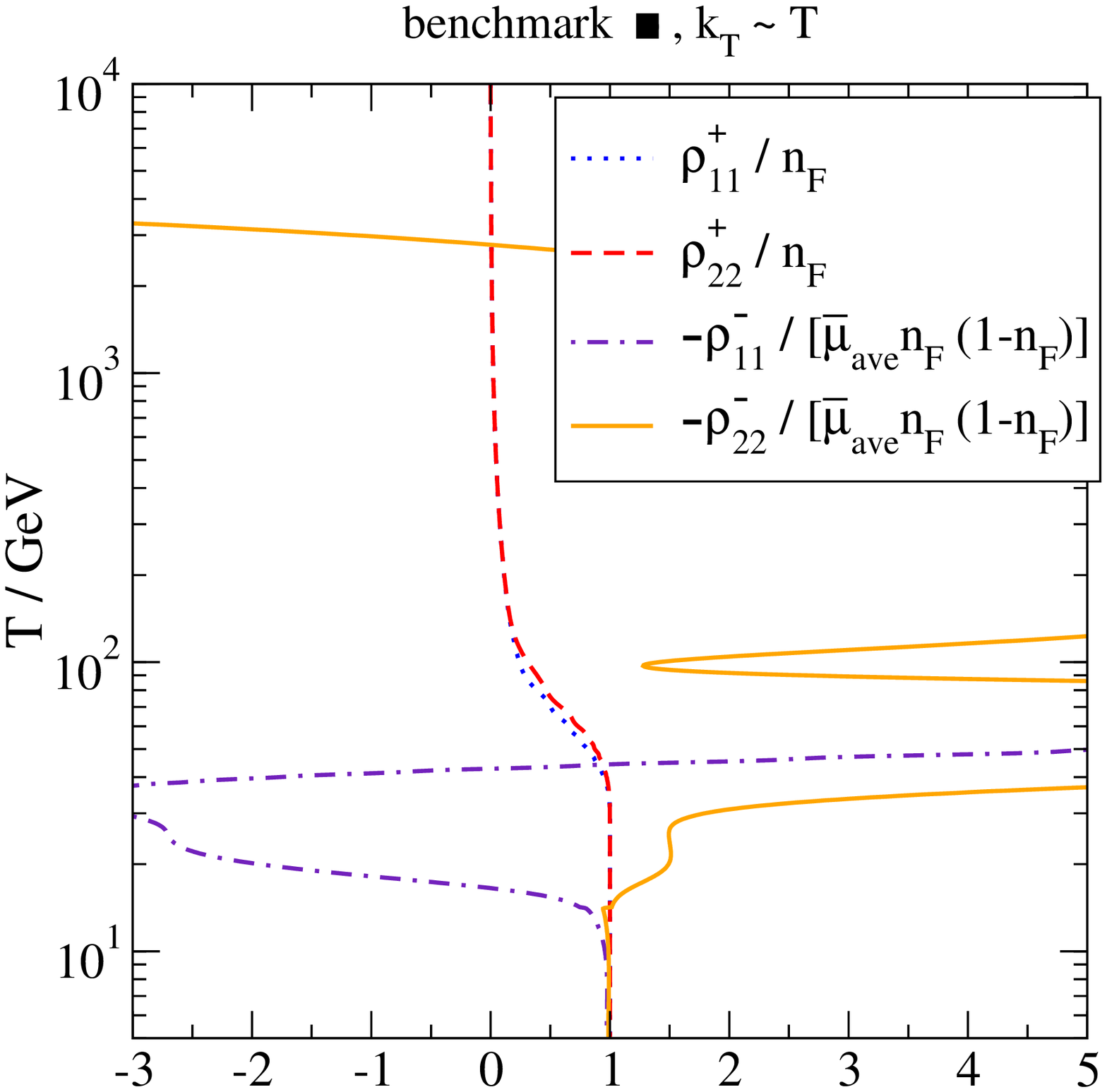}%
 \hspace{0.5cm}%
 \epsfysize=7.5cm\epsfbox{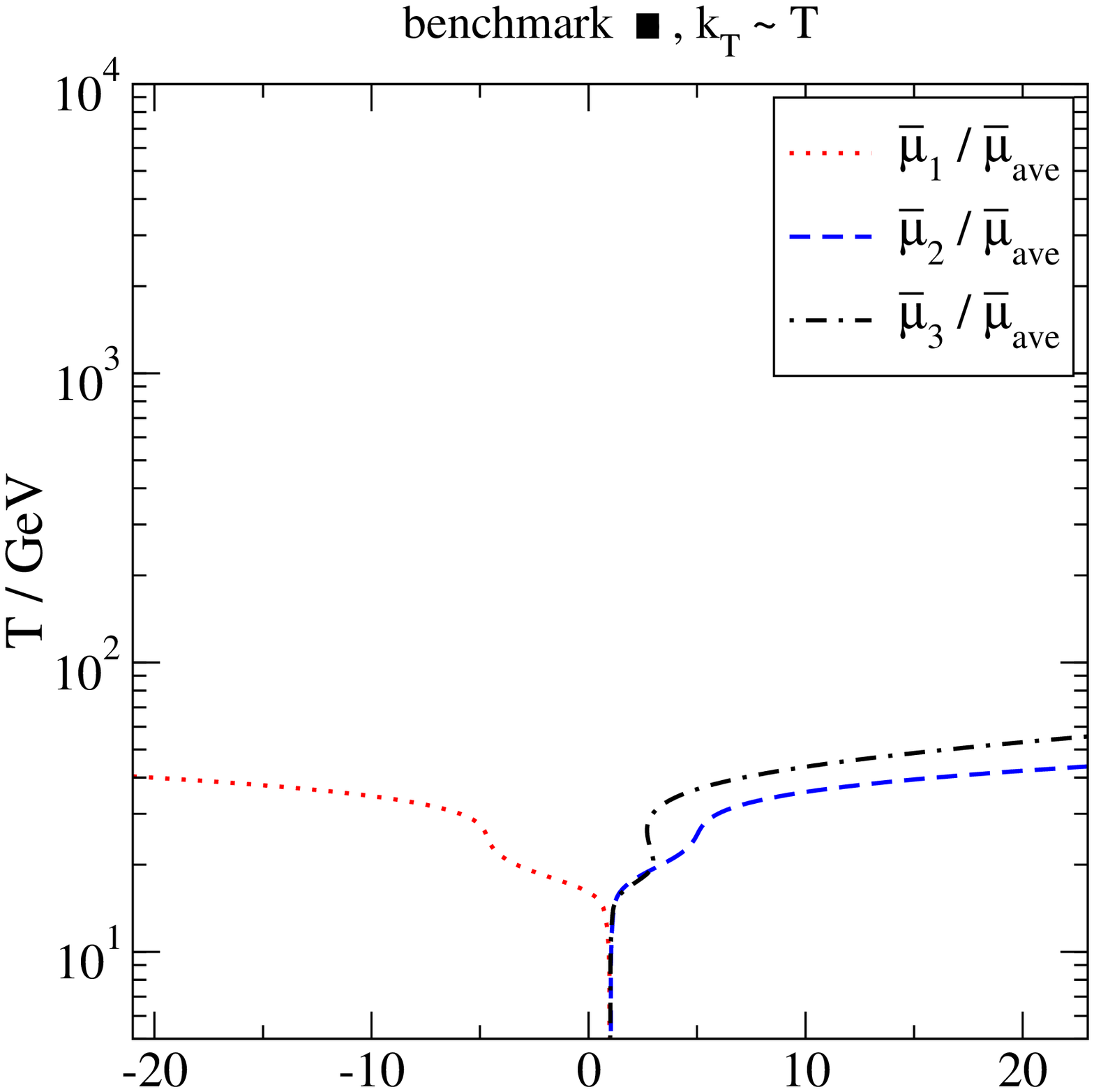}
}

\caption[a]{\small
 Left: an illustration of the equilibration of the diagonal 
 components of the density matrix. For the normalization of
 the helicity asymmetries, we have defined
 $\bmu^{ }_\rmi{ave} \equiv \frac{1}{3} \sum_a \bar{\mu}^{ }_a$. 
 The comoving momentum was chosen as $\kT^{ }\sim T$ at $T\sim 100$~GeV.
 Right: an illustration of the equilibration of the lepton 
 asymmetries in the different flavours, expressed in terms of
 the lepton chemical potentials $\bar{\mu}^{ }_a$. 
}

\la{fig:equilibration}  
\end{figure}

The reason for this behaviour can be understood as follows. 
Consider a state in which the helicity-symmetric density matrix has
equilibrated, $\rho^{+}_{ }= \mathop{\mbox{diag}}(\nF^{ },\nF^{ })$, 
and the helicity-asymmetry is diagonal, 
$\rho^{-}_{ }= \mathop{\mbox{diag}}(\rho^{-}_{11},\rho^{-}_{22})$. 
In \eq\nref{evol_Ya}, only the first and last term play a role.
In \eq\nref{evol_rho}, only the third
and fifth term play a role. At low temperatures, the rate coefficients
are dominated by the helicity-conserving components, so that 
$
     \widehat Q^{+}_{(a)\{\I\J\}} 
  \approx 
     \fr12\, \widehat Q^{ }_{(a-)\{\I\J\}} 
$,
$
     \widehat Q^{-}_{(a)\{\I\J\}} 
  \approx 
     -\fr12\, \widehat Q^{ }_{(a-)\{\I\J\}} 
$. 
Then the right-hand side of 
\eq\nref{evol_Ya} vanishes if\hspace*{0.2mm}\footnote{%
 We stress that the rates themselves 
 (helicity-conserving transitions between 
 lepton asymmetries and helicity asymmetries) 
 may remain appreciable, but the processes and inverse
 processes cancel against each other. Note that the total 
 ``fermion number'' which is conserved in the 
 helicity-flipping processes dominating at
 high temperatures, is not conserved here
 (unless the conversion rates 
 $ | h^{ }_{\I a} |^2  \widehat Q^{ }_{(a-)\I\I} $
 vanish).  
 }  
\be
 \forall a: \quad
 {\textstyle\sum_{\I}} | h^{ }_{\I a} |^2 
 \bar{\mu}^{ }_a \,
 \nF^{ }(1 - \nF^{ }) \, 
 \widehat Q^{ }_{(a-)\I\I}
 \; = \; 
 - {\textstyle\sum_{\I}} | h^{ }_{\I a} |^2 
 \rho^{-}_{\I\I} \,
 \widehat Q^{ }_{(a-)\I\I} 
 \;, \la{req1}
\ee
whereas the right-hand side of 
\eq\nref{evol_rho} vanishes if
\be
 \forall I: \quad
 {\textstyle\sum_{a}} | h^{ }_{\I a} |^2 
 \bar{\mu}^{ }_a \,
 \nF^{ }(1 - \nF^{ }) \, 
 \widehat Q^{ }_{(a-)\I\I}
 \; = \; 
 - {\textstyle\sum_{a}} | h^{ }_{\I a} |^2 
 \rho^{-}_{\I\I} \,
 \widehat Q^{ }_{(a-)\I\I} 
 \;. \la{req2}
\ee
Eqs.~\nref{req1} and \nref{req2} can be satisfied simultaneously
if 
\be
 \forall a,I: \quad 
 \bar{\mu}^{ }_a\, \nF^{ } (1-\nF^{ }) = - \rho^{-}_{\I\I}
 \;. \la{stationary} 
\ee
Alternatively, this can be expressed as 
$
 [\nF^{ }(\kT^{ }+\mu^{ }_a) - \nF^{ }(\kT^{ }-\mu^{ }_a)]/2
 = 
 [\rho^{ }_{(+)\I\I} - \rho^{ }_{(-)\I\I}]/2 
$.
Eq.~\nref{stationary} implies that 
$
 \bar{\mu}^{ }_1 = \bar{\mu}^{ }_2 = \bar{\mu}^{ }_3 
$
and 
$
 \rho^{-}_{11} = \rho^{-}_{22} 
$. 

Numerically, we find that the process towards the stationary
state starts with the equilibration of $\rho^{+}_{\I\I}$. 
Later on this is followed by $\rho^{-}_{\I\I}$ and the lepton asymmetries, 
cf.\ \fig\ref{fig:equilibration}. 
Afterwards the system remains in this
state at least as long as $\pi T \gsim M$.  

%
\section{Conclusions}
\la{se:concl}

The purpose of this paper has been to carry out  
a precise study of two carefully tuned benchmark points of 
GeV-scale resonant leptogenesis. By precision we mean that
the rate equations and most rate coefficients have been consistently 
determined to complete leading order in Standard Model couplings
in the parametric regime $g T \ll k,\pi T$, 
where $g^2 = 4\pi\alpha^{ }_w$
and $k$ is the right-handed neutrino co-moving momentum. 
Due to soft thermal effects,
a resummation of the naive loop expansion was necessary for  
achieving this goal. Based
on an analysis of lepton number susceptibilities (i.e.\ relations
of chemical potentials and lepton asymmetries), which play a role
in our master equations and for which higher-order
corrections have been determined~\cite{kubo,sangel}, 
we expect the theoretical uncertainty 
to be on the $\sim 20$\% level. 
There is one ingredient which was not fully
resolved yet, namely the effect of the ``chiral'' chemical
potential $\muZ^{ }$ on the ``direct'' rate coefficients
(cf.\ \se\ref{se:direct})
in the intermediate domain $v\sim T$, however we expect
the numerical influence from here to be on the $\sim 1\%$ level.
In addition there are non-perturbative uncertainties which are 
difficult to quantify at present, such as that the non-perturbative
crossover is at $T\sim 160$~GeV whereas within our perturbative
treatment the Higgs phenomenon sets in at $T \sim 150$~GeV. 

As main ingredients of our analysis, we track
both helicity states of right-handed neutrinos; consider 
both the symmetric and broken phase 
of the electroweak theory; 
allow for kinetic non-equilibrium; and
include a large set of chemical potentials 
(including gauge field ``tadpoles'').
To contrast this with extensive recent
parameter scans, kinetic non-equilibrium, 
helicity-conserving rates, 
a smoothly evolving sphaleron rate, 
hypercharge chemical potential, 
as well as all indirect contributions relevant for the broken phase, 
were omitted in ref.~\cite{n3}.
In ref.~\cite{new3}, kinetic non-equilibrium, 
the term $2 k \Gamma^{ }_{\!\mathcal{K}}$ 
in the helicity-flipping indirect contribution
(cf.\ \eq\nref{final_plus}), 
the running of Standard Model couplings,  
as well as the chemical potential dependences of the rates 
$B^\pm$, $D^\pm$ (cf.\ \eqs\nref{bp}, \nref{bm}, \nref{dp}, \nref{dm}) 
and of the mass corrections $\beta,\kappa,\delta$,
were omitted. On the $\mu$-dependence of $B^\pm$, $D^\pm$ 
we remark that even if such effects 
are formally of second order in deviations from equilibrium, 
it may be prudent to include them, given 
that $\rho^{\pm}_{ }$ can deviate from equilibrium 
by $\rmO(1)$. Nevertheless, the results in our
\fig\ref{fig:contour_Ya_YB}(left) agree 
semi-quantitatively with ref.~\cite{new3}. 

The first of our benchmarks
(cf.\ 
the filled circle in \fig\ref{fig:contour_Ya_YB}, 
and 
\se\ref{se:large_h}) 
concentrated 
on large neutrino Yukawa couplings, 
whereas the second
(cf.\ 
the filled square in \fig\ref{fig:contour_Ya_YB},
and
\se\ref{se:small_h}) 
focussed on small ones.
On the methodological side, our main finding was that
kinetic equilibrium, even if not justifiable theoretically,
is often a reasonable approximation, 
even if differences of $\rmO(10)$ can be found 
(cf.\ \figs\ref{fig:Ya_YB_circle} and \ref{fig:Ya_YB_square}). 
Assuming kinetic equilibrium is attractive in that it accelerates
numerics and therefore permits for overall parameter scans.

Apart from kinetic non-equilibrium, another  
ingredient worth elaborating upon are the mass corrections, parametrized by 
$\beta^{\pm}_{},\kappa^{\pm}_{ },\delta^{\pm}_{ }$
in \eqs\nref{H0} and \nref{D0}. Like the rate coefficients, these
can originate either from 
``indirect'' processes (the terms $\propto v^2$)
or from 
``direct'' processes (the terms $\propto T^2$). We find that implementing 
precisely the mass corrections has a very important $\rmO(10)$ suppressive
effect on late-time lepton asymmetries (less so on $\YB^{ }$). 

On the physics side, 
our main conclusion concerns the strong interplay 
between helicity  and lepton asymmetries. 
Following an earlier hint~\cite{eijima}, 
we have demonstrated that, after undergoing complicated 
dynamics, the system settles into a stationary
state, or ``fixed point'', at low temperatures 
(cf.\ \figs\ref{fig:Ya_YB_square}, \ref{fig:equilibration}), 
in which there is flavour equilibrium
both in the active and sterile sectors (cf.\ \eq\nref{stationary}). 
The temperature at which this happens 
lies typically in the range $T \sim 15 ... 50$~GeV. The remnant
lepton asymmetries can reach values 
$|Y^{ }_a| > 10^{-7} \gg |\YB^{ }|$.

The significance of this finding originates from its connection
to dark matter physics~\cite{singlet}. 
Thanks to flavour
equilibrium, values $|Y^{ }_a| \gsim 10^{-5}$ would be  
large enough to permit for resonant keV-scale sterile neutrino
dark matter production at $T \sim 0.1 ... 1.0$~GeV~\cite{dmpheno},  
proceeding via the Shi-Fuller mechanism~\cite{sf}. 
The existence of a stationary state suggests that the leptogenesis
and dark matter processes nicely factorize from each other.

Our finding should motivate further work in this direction. 
Even if our results got to $|Y^{ }_a| > 10^{-7}$, 
they fell short of $|Y^{ }_a| \sim 10^{-5}$ for 
the parameters in \eqs\nref{benchmark1}, \nref{benchmark2} 
(cf.\ \figs\ref{fig:contour_Ya_YB} and \ref{fig:Ya_YB_square}). 
This justifies broader parameter scans, 
as well as further refinements of the theoretical framework. 
For instance, at very low temperatures $T \ll M/\pi$, 
an additional contribution to lepton asymmetries could originate from 
the non-equilibrium decays of the right-handed neutrinos~\cite{singlet,late}.
Considering such contributions leads to the need to include many 
new mass effects (such as from $m^{ }_{\tau}$, $m^{ }_c$). 
Finally, it should be clarified whether 
the sterile neutrino helicity asymmetries that we observed
(cf.\ \eq\nref{stationary}) could constitute 
``reservoirs'', which might facilitate dark matter
production, thereby rendering the Shi-Fuller mechanism viable even 
if Standard Model lepton asymmetries 
remain somewhat below  $|Y^{ }_a| \sim 10^{-5}$. 
Unlike the 
existence of the stationary state itself, this seems to be 
a dynamical question, whose resolution depends on the values 
of the conversion rates 
$ | h^{ }_{\I a} |^2  \widehat Q^{ }_{(a-)\I\I} $
(cf.\ \eqs\nref{req1}, \nref{req2}). 

%
\section*{Acknowledgements}

We thank S.~Eijima, J.~L\'opez-Pav\'on, 
M.~Shaposhnikov and I.~Timiryasov for helpful suggestions and discussions. 
This work was partly supported by the Swiss National Science Foundation
(SNF) under grant 200020-168988.

%
\appendix
\renewcommand{\thesubsection}{\Alph{section}.\arabic{subsection}}
\renewcommand{\theequation}{\Alph{section}.\arabic{equation}}

%
\section{Relations between chemical potentials and asymmetries}

We specify here the expressions resulting from 
\eq\nref{pressure} for three cases: 
when restricting to \eq\nref{symmetric}
relevant for the symmetric phase, or to \eq\nref{broken} relevant
for the deep Higgs phase, or when we are in 
the intermediate regime $v\sim T$, when neither of these limits is viable. 
At high temperatures, when $v \ll T$
and $\muA^{ } = 0$, we find
\ba
 && \hspace*{-1.5cm}
 p(T,\mu) - p(T,0) \Big|^{ }_\rmi{$v \ll T$}
 \nn 
 & \approx & 
 \sum_a
 \chi^{ }_\rmii{F}(m_{\nu_a}) 
 \, \biggl[ 
  \frac{\mu_a^2}{2}
 - \frac{\muY^{ }\mu^{ }_a}{2}
 + \frac{\muY^2}{8}
 \biggr]
  +  
 \sum_a
 \chi^{ }_\rmii{F}(m_{e_a}) 
 \, \biggl[ 
   \mu_a^2
 - \frac{3\muY^{ }\mu^{ }_a}{2}
 + \frac{5 \muY^2}{8}
 \biggr]
 \nn 
 & + & 
 \sum_{i = u,c,t}
 \chi^{ }_\rmii{F}(m_i) 
 \, \biggl[ 
   3 \mu_q^2
  + \frac{5 \muY^{ }\mu_q}{2}
  + \frac{17\muY^2}{24}
 \biggr]
  +  
 \sum_{i = d,s,b}
 \chi^{ }_\rmii{F}(m_i) 
 \, \biggl[ 
   3 \mu_q^2
  - \frac{\muY^{ } \mu_q}{2}
  + \frac{5\muY^2}{24}
 \biggr]
 \nn 
 & + & 
  \Bigl[ \chi^{ }_\rmii{B}(m_\phi) + \chi^{ }_\rmii{B}(\mZ)
  + 2   \chi^{ }_\rmii{B}(\mW)
  \Bigr]
  \, \frac{\muY^2}{16}
 \;. \la{pressure_highT}
\ea
Here $\mu^{ }_q = \mu^{ }_\rmii{\!$B$}/3$, and 
masses have been retained as reminders of the origins 
of the contributions. 
For obtaining $n^{ }_a - n^{ }_\rmii{$B$}/3$ and 
$n^{ }_\rmii{$B$} + \sum_a n^{ }_a $, 
we follow the procedure described in \se{4.3} of ref.~\cite{cptheory}, 
writing $\mu^{ }_a = \tilde{\mu}^{ }_a + \tmuBL^{ }$ and
$
 \mu^{ }_\rmii{$B$} = 
 \tmuBL^{ } - \sum_a \tilde{\mu}^{ }_a/3
$ 
(sphaleron equilibrium corresponds to $ \tmuBL^{ } = 0 $, see below).  
Extremizing with respect to $\muY^{ }$, and going to 
the massless limit, when 
$ \chi^{ }_\rmii{F} $ and $ \chi^{ }_\rmii{B} $ can
be approximated according to \eq\nref{chi}, we obtain
\be
 \muY^{ } \; = \; 
 \frac{8}{33} \Bigl( \sum_a \tilde{\mu}^{ }_a +
 \frac{ 3 \tmuBL^{ }}{2} \Bigr) + \rmO(g)
 \;.
\ee
Derivatives with respect to $\tilde{\mu}^{ }_a, \tmuBL^{ }$
yield $n^{ }_a - n^{ }_\rmii{$B$}/3$, 
$n^{ }_\rmii{$B$} + \sum_a n^{ }_a $, respectively, 
and inverting these relations results in
\be
 \left( 
  \begin{array}{c}
    \tilde{\mu}^{ }_1 \\
    \tilde{\mu}^{ }_2 \\ 
    \tilde{\mu}^{ }_3 \\
    \tmuBL^{ }   
  \end{array}
 \right)
 \; \stackrel{v\ll T}{=} \; \frac{1}{144 T^2} 
 \left( 
   \begin{array}{rrrr} 
     319 & 31 & 31 & -23 \\  
     31 & 319 & 31 & -23 \\ 
     31 & 31 & 319 & -23 \\
     -23 & -23 & -23 & 79
   \end{array}
 \right)
 \left( 
  \begin{array}{c}
    n^{ }_1 - \frac{n^{ }_\rmii{\!$B$}}{3}  \\
    n^{ }_2 - \frac{n^{ }_\rmii{\!$B$}}{3} \\ 
    n^{ }_3 - \frac{n^{ }_\rmii{\!$B$}}{3} \\
    n^{ }_\rmii{\!$B$} + \sum_a n^{ }_a  
  \end{array} 
 \right) + \rmO(g)
 \;. \la{susc_highT}
\ee
Subsequently, 
$\mu^{ }_a = \tilde{\mu}^{ }_a + \tmuBL^{ }$ and
$
 \mu^{ }_\rmii{$B$} = 
 \tmuBL^{ } - \sum_a \tilde{\mu}^{ }_a/3
$.

At low temperatures, when $v \gg T$, inserting
$
 \muA^{ } 
$
and  
$
 \muY^{ } 
$
from \eq\nref{broken2} and omitting terms 
proportional to $\muZ^{ }$ leads to
\ba
 && \hspace*{-1.0cm}
 p(T,\mu) - p(T,0) \Big|^{ }_\rmi{$v \gg T$}
 \nn 
 & \approx & 
 \sum_a
 \chi^{ }_\rmii{F}(m_{\nu_a}) 
 \, \biggl[ 
  \frac{\mu_a^2}{2}
 \biggr]
  +  
 \sum_a
 \chi^{ }_\rmii{F}(m_{e_a}) 
 \, \Bigl[ 
   ( \mu^{ }_a -  \muQ^{ } )^2
 \Bigr]
 \nn 
 & + & 
 \sum_{i = u,c,t}
 \chi^{ }_\rmii{F}(m_i) 
 \, \biggl[ 
   \frac{ ( \mu^{ }_\rmii{\!$B$} + 2 \muQ^{ } )^2 }{3}
 \biggr]
  +  
 \sum_{i = d,s,b}
 \chi^{ }_\rmii{F}(m_i) 
 \, \biggl[ 
   \frac{ ( \mu^{ }_\rmii{\!$B$} - \muQ^{ } )^2 }{3}
 \biggr]
 \; + \; 
  \chi^{ }_\rmii{B}(\mW)
 \, \biggl[
   \frac{3\muQ^2}{2}
 \biggr]
 \;. \hspace*{3mm} \la{pressure_lowT}
\ea
Considering for simplicity temperatures $T \gsim 50$~GeV 
so that susceptibilities can still be set 
to their massless values (cf.\ \eq\nref{chi}), 
extremization with respect to $\muQ^{ }$ yields
\ba
 \muQ^{ } \; = \; \frac{4}{33} \Bigl( \sum_a \tilde{\mu}^{ }_a +
 \frac{ 3 \tmuBL^{ }}{2} \Bigr) 
 + \rmO(g) 
 \;. \hspace*{8mm} \la{muQ_low}
\ea
The chemical potentials appearing here can be obtained from 
\be
 \left( 
  \begin{array}{c}
    \tilde{\mu}^{ }_1 \\
    \tilde{\mu}^{ }_2 \\ 
    \tilde{\mu}^{ }_3 \\
    \tmuBL^{ }   
  \end{array}
 \right)
 \; \stackrel{v\gg T \gsim \rmi{50~GeV}}{=} \; \frac{1}{204 T^2} 
 \left( 
   \begin{array}{rrrr} 
     407 & -1 & -1 & -39 \\  
     -1 & 407 & -1 & -39 \\ 
     -1 & -1 & 407 & -39 \\
     -39 & -39 & -39 & 111
   \end{array}
 \right)
 \left( 
  \begin{array}{c}
    n^{ }_1 - \frac{n^{ }_\rmii{\!$B$}}{3}  \\
    n^{ }_2 - \frac{n^{ }_\rmii{\!$B$}}{3} \\ 
    n^{ }_3 - \frac{n^{ }_\rmii{\!$B$}}{3} \\
    n^{ }_\rmii{\!$B$} + \sum_a n^{ }_a  
  \end{array} 
 \right) + \rmO(g)
 \;. \la{susc_lowT}
\ee
In the numerical solution,  
we include dependences
on top, bottom, $W^{\pm}_{ }$, $Z^0_{ }$ and Higgs masses,
whereby the equations become a bit more complicated.  

In the intermediate regime $v \sim T$, we employ
the full \eq\nref{pressure} rather than \nref{pressure_highT} 
or \nref{pressure_lowT}, and both $\muA^{ }$ and $\muY^{ }$
need to be extremized simultaneously~\cite{khlebnikov}, 
which leads to a smooth interpolation
between \eqs\nref{susc_highT} and \nref{susc_lowT}. The price to pay 
is that when neither \eq\nref{symmetric} nor
\eq\nref{broken} is satisfied, perturbation theory
becomes complicated due to the coupling of gauge and 
scalar modes (cf.\ \se\ref{se:ensemble}). 
To understand when we find ourselves in this situation, 
we note that the extremal value of $\muZ^{ }$ is given by 
\ba
 \muZ^{ } & = & 
 \frac{
 \sum_a [2 n^{ }_{\nu_a} - (1 - 4s^2) n^{ }_{e_a}]
 + (1 - \frac{8s^2}{3})  n^{ }_{u,c,t} 
 - (1 - \frac{4 s^2}{3}) n^{ }_{d,s,b} 
 + (\frac{10}{3} - 4 s^2 ) n^{ }_\rmii{$W$}  
 }{v^2 + \chi^{ }_\rmi{eff}}
 \;, \hspace*{4mm} \la{muZ}
 \\[2mm] 
 \chi^{ }_\rmi{eff} & \equiv & 
 \chi^{ }_\rmii{F}(0)\, \biggl( 18  - 36 s^2 + \frac{152 s^4}{3} \biggr)
 + \chi^{ }_\rmii{F}(m^{ }_t) \, \biggl( 3 - 8 s^2 + \frac{32 s^4}{3} \biggr)
 \la{chi_eff} \\
 & + &
   \chi^{ }_\rmii{F}(m^{ }_b) \, \biggl( 3 - 4 s^2 + \frac{8 s^4}{3} \biggr)
 + \chi^{ }_\rmii{B}(\mW^{ }) \bigl( 9 - 20 s^2 + 12 s^4 \bigr)
 + \frac{ \chi^{ }_\rmii{B}(\mH^{ }) + \chi^{ }_\rmii{B}(\mZ^{ })}{2}
 \;, \nonumber 
\ea
where $n^{ }_{u,c,t} \equiv \sum_{i=u,c,t} n^{ }_i$; 
$s^2 \equiv \sin^2(\tilde{\theta})$ with $\tilde{\theta}$ from 
\eq\nref{thetat}; and 
\ba
 &&
 n^{ }_{u} \; \equiv \; 
 2 \chi^{ }_\rmii{F}( m^{ }_{u})\, (\muB^{ } + 2 \muQ^{ })
 \;, \quad
 n^{ }_{d} \; \equiv \; 
 2 \chi^{ }_\rmii{F}( m^{ }_{d})\, (\muB^{ } - \muQ^{ })
 \;, 
 \\ 
 &&
 n^{ }_{\nu_a} \; \equiv \; 
 \chi^{ }_\rmii{F}( m^{ }_{\nu_a})\, \mu^{ }_a
 \;, \quad 
 n^{ }_{e_a} \; \equiv \; 
 2 \chi^{ }_\rmii{F}( m^{ }_{e_a})\, (\mu^{ }_a - \muQ^{ }) 
 \;, \quad 
 n^{ }_\rmii{$W$} \; \equiv \;
 3 \chi^{ }_\rmii{B}(\mW^{ }) \, \muQ^{ }
 \;. \la{n_nua}
\ea
Eqs.~\nref{muZ}--\nref{n_nua} show that the assumption in \eq\nref{broken}
is valid for $v^2 \gg \chi^{ }_\rmi{eff} \sim T^2$. 
For $v \ll T$, $s^2\to 0$ and $m/T\to 0$, 
and \eq\nref{muZ} then implies that $\muZ^{ }\to \muQ^{ }$.

Let us now turn to the implications of sphaleron equilibrium on this 
discussion. 
The sphaleron rate falls out of  
equilibrium in the intermediate domain 
$v\sim T$~\cite{krs}, and in this regime neither 
\eq\nref{susc_highT} nor \nref{susc_lowT} is accurate. 
As long as the sphaleron rate is fast, 
$
    \tmuBL^{ }   
$
re-adjusts itself to zero on a time scale much shorter than we can resolve, 
so \eqs\nref{susc_highT} and \nref{susc_lowT} show that the ``would-be''
equilibrium state has 
\be
 \YBpL^\rmi{eq}\bigr|^{ }_\rmi{$v \ll T$}
 \;\equiv\; \frac{23}{79} 
 \sum_a \biggl( Y^{ }_a - \frac{\YB^{ }}{3} \biggr)
 \;, \quad
 \YBpL^\rmi{eq}\bigr|^{ }_\rmi{$v \gg T$}
 \;\equiv\; \frac{13}{37} 
 \sum_a \biggl( Y^{ }_a - \frac{\YB^{ }}{3} \biggr)
 \la{YBpLeq}
 \;. 
\ee
The corresponding rate equation,  
{\it viz.}
$
 \YBpL'  \; = \; 
 \sum_a F_a^{ } - \gamma\, (\YBpL^{ } - \YBpL^\rmi{eq}) 
$,
where we employ the notation of 
\eq(4.2) of ref.~\cite{cpnumerics}, 
contains the coefficient
($\nG^{ }\equiv3$, $\Gamma^{ }_\rmi{diff}$ is from ref.~\cite{sphaleron})
\be
 \gamma \bigr|^{ }_\rmi{$v \ll T$} 
 \; = \;
 \frac{79 \nG^2 \Gamma^{ }_\rmi{diff}}{216 c_s^2 H T^3} 
 \;, \quad
 \gamma \bigr|^{ }_\rmi{$v \gg T$} 
 \; = \;
 \frac{37 \nG^2 \Gamma^{ }_\rmi{diff}}{102 c_s^2 H T^3} 
 \;. \la{evol_YBpL}  
\ee
The sphaleron rate is in equilibrium 
when $\gamma\gg 1$. According to \eq\nref{YBpLeq},  
a sudden switch from one limiting treatment 
to the other would insert a discontinuity in $\YBpL^{ }$
if  $\gamma\gg 1$.
In order to avoid this, we have derived the analogues
of \eqs\nref{susc_highT}, \nref{susc_lowT}, 
\nref{YBpLeq}, \nref{evol_YBpL} from an extremization of 
the full \eq\nref{pressure} with respect to both 
$\muA^{ }$ and $\muY^{ }$. It is straightforward to verify
that the resulting expressions interpolate continuously 
between the limiting values. Our numerical
results make use of this continuous interpolation. 

%
\section{Thermally modified weak mixing angles}
\la{se:mixing}

A thermal medium modifies the weak (Weinberg) mixing angle between 
neutral gauge field components. Furthermore, the mixing angle 
becomes momentum-dependent~\cite{broken}. In \se\ref{ss:gauge}
we addressed interaction rates for two different
helicity states, denoted by 
$
 \Gamma^\rmii{HTL}_{\! u} 
$ 
and
$
 \Gamma^\rmii{HTL}_{\! u} + 2 k \Gamma^\rmii{HTL}_{\!\mathcal{K}} 
$, 
cf.\ \eqs\nref{Gamma_HTL_u} and \nref{Gamma_HTL_upK}, 
respectively. 
These turned out to be sensitive to different momentum ranges of
the gauge bosons exchanged in soft $2\leftrightarrow 2$ 
scatterings: ``static'' momenta 
$p^{ }_0, p^{ }_\parallel \ll g T$ or
``hard'' momenta 
$g T \ll p^{ }_0, p^{ }_\parallel \ll \pi T$, 
where $\vec{p}^{ }_\parallel \parallel \vec{k}$
and $\vec{k}$ is the neutrino momentum. Here we 
specify the mixing angles relevant for these cases, 
obtained from the HTL-resummed
gauge propagators given in appendix~B of ref.~\cite{broken}. 

The HTL self-energies are parametrized by Debye masses, 
\be
 m^{2}_\rmii{E1} 
 \; \equiv \; 
 \Bigl( \fr{\nS}6 + \frac{5\nG}{9} \Bigr) g_1^2 T^2 
 \;, \quad
 m^{2}_\rmii{E2} \; \equiv \; 
 \Bigl( \fr23 + \fr{\nS}6 + \frac{\nG}{3} \Bigr) g_2^2 T^2
 \;, \la{Debye}
\ee
where $\nS \equiv 1$ is the number of 
Higgs doublets and $\nG \equiv 3$ 
is the number of fermion generations.
Like in \eq\nref{gauge_prop}, two different HTL self-energies 
play a role. Here we need their limiting values: 
\ba
 && \lim_{p^{ }_0\to 0} \Pi^{ }_\rmii{E$i$}(p^{ }_0,{p}^{ }_\perp,p^{ }_0) 
 \; = \; 
 m_\rmii{E$i$}^2 
 \;, \quad
 \lim_{p^{ }_0\to \infty} 
 \Pi^{ }_\rmii{E$i$}(p^{ }_0,{p}^{ }_\perp,p^{ }_0) 
 \; = \; 
 0
 \;, \la{limits_PiE} \\
 && \lim_{p^{ }_0\to 0} \Pi^{ }_\rmii{T$i$}(p^{ }_0,{p}^{ }_\perp,p^{ }_0) 
 \; = \; 
 0 
 \;, \hspace*{7mm}
 \lim_{p^{ }_0\to \infty} 
 \Pi^{ }_\rmii{T$i$}(p^{ }_0,{p}^{ }_\perp,p^{ }_0) 
 \; = \; 
 \frac{m_\rmii{E$i$}^2}{2}
 \;. \la{limits_PiT} 
\ea
As discussed around \eq\nref{Phi_HTL_3}, we have here set
$p^{ }_\parallel = p^{ }_0$.
The medium modifies the mixing angles in the limits 
where these self-energies differ from zero. 

For $p^{ }_0 \to 0$, it is the ``electric'' 
components whose mixing is modified, 
cf.\ \eq\nref{limits_PiE}. 
Given the standard vacuum 
angle
$
 \sin(2\theta) \equiv 2 g^{ }_1 g^{ }_2/(g_1^2 + g_2^2)
$, 
the relevant angle is 
\be
 \sin(2\tilde\theta) \; \equiv \;
 \frac{\sin(2\theta) \mZ^2 }{\sqrt{\sin^2(2\theta) \mZ^4 +
 [\cos(2\theta) \mZ^2 + m_\rmii{E2}^2 - m_\rmii{E1}^2]^2}}
 \;, \la{thetat}
\ee
and we also need the corresponding mass eigenvalues, 
\ba
  \mWt^2 & \equiv & 
  \mW^2 + m_\rmii{E2}^2 
  \;, \quad
  \mZt^2 \; \equiv \;  \tilde{m}_{+}^2 
  \;, \quad
  \mQt^2 \; \equiv \;  \tilde{m}_{-}^2  
  \;, \\ 
 \tilde{m}_{\pm}^2 & \equiv & 
 \frac{1}{2} 
 \Bigl\{
   \mZ^2 + m_\rmii{E1}^2 + m_\rmii{E2}^2 \pm 
   \sqrt{\sin^2(2\theta) \mZ^4 +
   [\cos(2\theta) \mZ^2 + m_\rmii{E2}^2 - m_\rmii{E1}^2]^2}
 \Bigr\}
 \;. 
\ea
For $p^{ }_0 \to \infty$, 
the ``transverse'' polarizations are affected, 
cf.\ \eq\nref{limits_PiT}. 
We denote
\be
 \sin(2\bar\theta) \; \equiv \;
 \frac{\sin(2\theta) \mZ^2 }{\sqrt{\sin^2(2\theta) \mZ^4 +
 [\cos(2\theta) \mZ^2 + (m_\rmii{E2}^2 - m_\rmii{E1}^2)/2]^2}}
 \;, \la{thetab}
\ee
and the corresponding mass eigenvalues read 
\ba
  \mWb^2 & \equiv & 
  \mW^2 + \frac{m_\rmii{E2}^2}{2} 
  \;, \quad
  \mZb^2 \; \equiv \;  \bar{m}_{+}^2 
  \;, \quad
  \mQb^2 \; \equiv \;  \bar{m}_{-}^2  
  \;, \\ 
 \bar{m}_{\pm}^2 & \equiv & 
 \frac{1}{2} 
 \biggl\{
   \mZ^2 + \frac{m_\rmii{E1}^2 + m_\rmii{E2}^2}{2} \pm 
   \sqrt{\sin^2(2\theta) \mZ^4 +
   \Bigl[\cos(2\theta) \mZ^2 + \frac{ m_\rmii{E2}^2 - m_\rmii{E1}^2}{2} 
   \Bigr]^2}
 \biggr\}
 \;. 
\ea

%
\section{Coefficients for the Fermi limit of the active neutrino width}

We list here the values of the sums appearing in 
\eq\nref{Gammaminus_Fermi} when the integrals
of \eqs\nref{Xi_t} and \nref{Xi_s} are expanded to linear order 
in chemical potentials; 
coefficients are inserted from
table~\ref{table:channels}; 
and \eq\nref{broken} is made use of 
(we write the result in terms of 
$\bmuB = \Nc^{ }\,\bar{\mu}^{ }_q$ and 
make use of the symmetry of $\Xi^s_{(\tau)}$
in $\mu^{ }_{\alpha} \leftrightarrow \mu^{ }_{\beta}$): 
\ba
  \Gamma^\rmi{Fermi}_{(\tau)}
  & = & 
  \sum_{\rmi{channels}}
  \Bigl\{  
   ( c^{ }_\rmii{1L}
  - c^{ }_\rmii{2}
  + c^{ }_\rmii{1R}) 
  \Bigl[  
  2 \Xi^t_{(\tau)}
 +   \Xi^s_{(\tau)}
  \Bigr]^{ }_{\mu^{ }_i = 0} 
 \nn 
 &  &
 \; + \, 
  \bigl[ 
    ( c^{ }_\rmii{1L}
  - c^{ }_\rmii{2} )\,(\mu^{ }_1 + \mu^{ }_3)
  + c^{ }_\rmii{1R}\, (\mu^{ }_1 + \mu^{ }_2) 
  \bigr] 
 \Bigl[ 
  \bigl(  \partial^{ }_{\mu^{ }_{\alpha}}
        + \partial^{ }_{\mu^{ }_{\gamma}} \bigr)\,
  \Xi^t_{(\tau)} 
 + 
  \partial^{ }_{\mu^{ }_{\alpha}}
  \Xi^s_{(\tau)} 
 \Bigr]^{ }_{\mu^{ }_i = 0}
 \nn 
 &  & 
 \; + \, 
  \bigl[ 
    ( c^{ }_\rmii{1L}
  - c^{ }_\rmii{2} )\,(\mu^{ }_2)
  + c^{ }_\rmii{1R}\, (\mu^{ }_3) 
  \bigr] 
  \Bigl[ 
  2 \partial^{ }_{\mu^{ }_{\beta}}
   \Xi^t_{(\tau)} 
  + 
  \partial^{ }_{\mu^{ }_{\gamma}}
  \Xi^s_{(\tau)}
  \Bigr]^{ }_{\mu^{ }_i = 0}
  \; \Bigr\} 
 \; + \; \rmO(\mu^2)  
 \\
 & = & 
 \biggl[ 
  \frac{15}{2} - 2 s^2 + 12 s^4 
 + 2\Nc \, \biggl(
  \frac{5}{4} - \frac{7 s^2}{3} + \frac{22 s^4}{9} 
  + 
   |V^{ }_{\!ud}|^2 + |V^{ }_{\!us}|^2
 + |V^{ }_{\!cd}|^2 + |V^{ }_{\!cs}|^2
 \biggr)
 \biggr] 
\nn 
 &  &
 \; \times \,  
 \Bigl[ \Xi^s_{(\tau)} + 2 \Xi^t_{(\tau)}  
 \Bigr]^{ }_{\mu^{ }_i = 0} 
 \nn 
 & + & 
 \bar{\mu}^{ }_a 
 \biggl\{ 
 \biggl( \frac{3}{2} - 4 s^2 \biggr)
 \Bigl[
    2 \partial^{ }_{\mu^{ }_{\beta}}\Xi^t_{(\tau)}  
     + 
    \partial^{ }_{\mu^{ }_{\gamma}}\Xi^s_{(\tau)} 
 \Bigr]^{ }_{\mu^{ }_i = 0} 
 \nn 
  &  &  
 \; + \, 
 2 
 \biggl[ 
  3 + s^2 + 6 s^4 
  + \Nc^{ }
  \biggl(
    \frac{5}{4} - \frac{7s^2}{3} + \frac{22s^4}{9}
 + |V^{ }_{\!ud}|^2 + |V^{ }_{\!us}|^2
 + |V^{ }_{\!cd}|^2 + |V^{ }_{\!cs}|^2
  \biggr)
 \biggr]
\nn
 &  &
 \; \times \,  
 \Bigl[
   \bigl( \partial^{ }_{\mu^{ }_{\alpha}}
             + \partial^{ }_{\mu^{ }_{\gamma}} \bigr)\, \Xi^t_{(\tau)}  
     +
  \partial^{ }_{\mu^{ }_{\alpha}}\Xi^s_{(\tau)} 
 \Bigr]^{ }_{\mu^{ }_i = 0} 
 \biggr\} 
 \nn 
 & + & \biggl\{\, - \bmuQ^{ } 
 \biggl[ 
   \frac{11}{2} -2 s^2 + 
  \frac{2 \Nc}{3} \, \biggl(
   -\frac{1}{4} + \frac{5 s^2}{3} 
   +  |V^{ }_{\!ud}|^2 + |V^{ }_{\!us}|^2
  + |V^{ }_{\!cd}|^2 + |V^{ }_{\!cs}|^2
      \biggr)
 \biggr]
\nn
 & & \; + \,  
 \bigl( {\textstyle \sum_b}\, \bar{\mu}^{ }_b\bigr)
 \bigl( 3 - 2s^2\bigr)
 \; + \; 
 2 \bmuB^{ }
 \biggl( 
 \frac{5}{4} - \frac{7 s^2}{3} 
  +  |V^{ }_{\!ud}|^2 + |V^{ }_{\!us}|^2
 + |V^{ }_{\!cd}|^2 + |V^{ }_{\!cs}|^2
 \biggr)
 \biggr\} 
\nn
 & & 
 \; \times \, 
 \Bigl[
 \bigl( \partial^{ }_{\mu^{ }_{\alpha}}
              - 2 \partial^{ }_{\mu^{ }_{\beta}} 
             + \partial^{ }_{\mu^{ }_{\gamma}}
           \bigr)\, \Xi^t_{(\tau)}  
     +
 \bigl( \partial^{ }_{\mu^{ }_{\alpha}} - 
        \partial^{ }_{\mu^{ }_{\gamma}}
 \bigr)\, \Xi^s_{(\tau)} 
 \Bigr]^{ }_{\mu^{ }_i = 0} 
 + \rmO(\mu^2)
 \;. \la{Fermi_simplified}
\ea
\normalsize

%
\section{Phase space integrals for 
direct $2\leftrightarrow 2$ scatterings}

We list here the phase space integrals
appearing in \eq\nref{direct_2to2_hard}. 
The associated coefficients $c^{ }_i$ and chemical potentials 
are listed in table~\ref{table:channels_direct}. The five cases read
\small 
\ba
 \Xi_{(+)}^{\,t1} 
  & \equiv & 
 \frac{ \nF^{-1}\bigl(k^{ }_{ } - \Sigma^{ }_i \mu^{ }_i\bigr) }{2k}
 \int \!{\rm d}\Omega^{ }_{2\leftrightarrow 2} \, 
 \nB^{ }(p^{ }_1 - \mu^{ }_1)\,
 \nB^{ }(p^{ }_2 - \mu^{ }_2)\,
 \bigl[ 1 - \nF^{ }(p^{ }_3 + \mu^{ }_3) \bigr] 
 \, \biggl( \, \frac{u}{t} \biggr) 
 \nn 
 & = & 
 \frac{1}{(4\pi)^3 k^2 }
 \int_{0}^{ k } \! {\rm d} \qp
 \int_{-\infty}^{0} \! {\rm d} \qm  
  \bigl[1 - \nF^{ }(p^{ }_0 - \mu^{ }_1 - \mu^{ }_3)
 + \nB^{ }(k - p^{ }_0 - \mu^{ }_2) \bigr]
 \nn 
 & \times & 
 \biggl\{ 
 \frac{(k - \qp)T}{p} 
 \Bigl[\, 
 \lnf(\mu^{ }_3 -\qm ) - \lnb(\qp - \mu^{ }_1) 
 \,\Bigr]  
 + 
 \frac{(p^{ }_0 - 2k)T^2}{p^2}
 \Bigl[\, 
 \lif( \mu^{ }_3 -\qm ) -  \lib(\qp - \mu^{ }_1) 
 \,\Bigr]
 \biggr\} 
 \;, \nn \la{channels_direct_t1} \\
 \Xi_{(+)}^{\,u1} 
  & \equiv & 
 \frac{ \nF^{-1}\bigl(k^{ }_{ } - \Sigma^{ }_i \mu^{ }_i\bigr) }{2k}
 \int \!{\rm d}\Omega^{ }_{2\leftrightarrow 2} \, 
 \nB^{ }(p^{ }_1 - \mu^{ }_1)\,
 \nF^{ }(p^{ }_2 - \mu^{ }_2)\,
 \bigl[ 1 + \nB^{ }(p^{ }_3 + \mu^{ }_3) \bigr]
 \, \biggl( - \frac{s}{u} \biggr)
 \nn 
 & = & 
 \frac{1}{(4\pi)^3 k^2 }
 \int_{0}^{ k } \! {\rm d} \qp
 \int_{-\infty}^{0} \! {\rm d} \qm  
  \bigl[1 - \nF^{ }(p^{ }_0 - \mu^{ }_2 - \mu^{ }_3)
 + \nB^{ }(k - p^{ }_0 - \mu^{ }_1) \bigr]
 \nn 
 & \times & 
 \biggl\{ 
 \frac{(k - \qm)T}{p}  
 \Bigl[\, 
 \lnf(\qp - \mu^{ }_2) - \lnb( \mu^{ }_3 -\qm ) 
 \,\Bigr]  
 + 
 \frac{(p^{ }_0 - 2k)T^2}{p^2}
 \Bigl[\, 
 \lif(\qp - \mu^{ }_2) -  \lib( \mu^{ }_3 -\qm ) 
 \,\Bigr]
 \biggr\} 
 \;, \nn \la{channels_direct_u1} \\
 \Xi_{(+)}^{\,s1} 
  & \equiv & 
 \frac{ \nF^{-1}\bigl(k^{ }_{ } - \Sigma^{ }_i \mu^{ }_i\bigr) }{2k}
 \int \!{\rm d}\Omega^{ }_{2\leftrightarrow 2} \, 
 \nB^{ }(p^{ }_1 - \mu^{ }_1)\,
 \nF^{ }(p^{ }_2 - \mu^{ }_2)\,
 \bigl[ 1 + \nB^{ }(p^{ }_3 + \mu^{ }_3) \bigr]
 \, \biggl( - \frac{u}{s} \biggr)
 \nn 
 & = & 
 \frac{1}{(4\pi)^3 k^2 }
 \int_{k}^{ \infty } \! {\rm d} \qp
 \int_{0}^{k} \! {\rm d} \qm  
  \bigl[\nF^{ }(p^{ }_0 - \mu^{ }_1 - \mu^{ }_2)
 + \nB^{ }(p^{ }_0 - k + \mu^{ }_3) \bigr]
 \nn 
 & \times & 
 \biggl\{ \frac{p}{2} + 
 \frac{(k - \qm)T}{p}  
 \Bigl[\, 
 \lnf(\qp - \mu^{ }_2) - \lnb(\qm - \mu^{ }_1) 
 \,\Bigr]  
 + 
 \frac{(k - \qp)T}{p}  
 \Bigl[\, 
 \lnf(\qm - \mu^{ }_2) - \lnb(\qp - \mu^{ }_1) 
 \,\Bigr]  
 \nn 
 & + &
 \frac{(p^{ }_0 - 2k)T^2}{p^2}
 \Bigl[\, 
  \lif(\qp - \mu^{ }_2)
 - \lif(\qm - \mu^{ }_2)
 + \lib(\qm - \mu^{ }_1) 
 -  \lib(\qp - \mu^{ }_1) 
 \,\Bigr]
 \biggr\} 
 \;, \hspace*{5mm} \la{channels_direct_s1} \\[3mm]
 \Xi_{(+)}^{\,t0} 
  & \equiv & 
 \frac{ \nF^{-1}\bigl(k^{ }_{ } - \Sigma^{ }_i \mu^{ }_i\bigr) }{2k}
 \int \!{\rm d}\Omega^{ }_{2\leftrightarrow 2} \, 
 \nF^{ }(p^{ }_1 - \mu^{ }_1)\,
 \nF^{ }(p^{ }_2 - \mu^{ }_2)\,
 \bigl[ 1 - \nF^{ }(p^{ }_3 + \mu^{ }_3) \bigr]
 \nn 
 & = & 
 \frac{1}{(4\pi)^3 k^2 }
 \int_{0}^{k} \! {\rm d} \qp
 \int_{-\infty}^{0} \! {\rm d} \qm  
  \bigl[1 + \nB^{ }(p^{ }_0 - \mu^{ }_1 - \mu^{ }_3)
 - \nF^{ }(k - p^{ }_0 - \mu^{ }_2) \bigr]
 \nn 
 & \times & 
 \biggl\{  
 T  
 \Bigl[\, 
 \lnf( \mu^{ }_3 - \qm ) 
 - \lnf(\qp - \mu^{ }_1) 
 \,\Bigr]  
 \biggr\} 
 \;, \hspace*{5mm} \la{channels_direct_t0} \\[3mm]
 \Xi_{(+)}^{\,s0} 
  & \equiv & 
 \frac{ \nF^{-1}\bigl(k^{ }_{ } - \Sigma^{ }_i \mu^{ }_i\bigr) }{2k}
 \int \!{\rm d}\Omega^{ }_{2\leftrightarrow 2} \, 
 \nF^{ }(p^{ }_1 - \mu^{ }_1)\,
 \nF^{ }(p^{ }_2 - \mu^{ }_2)\,
 \bigl[ 1 - \nF^{ }(p^{ }_3 + \mu^{ }_3) \bigr]
 \nn 
 & = & 
 \frac{1}{(4\pi)^3 k^2 }
 \int_{k}^{ \infty } \! {\rm d} \qp
 \int_{0}^{k} \! {\rm d} \qm  
  \bigl[\nB^{ }(p^{ }_0 - \mu^{ }_1 - \mu^{ }_2)
 + \nF^{ }(p^{ }_0 - k + \mu^{ }_3) \bigr]
 \nn 
 & \times & 
 \biggl\{ {p} + 
 T  
 \Bigl[\, 
 \lnf(\qp - \mu^{ }_1) 
 - \lnf(\qm - \mu^{ }_1) 
 + \lnf(\qp - \mu^{ }_2)
 - \lnf(\qm - \mu^{ }_2) 
 \,\Bigr]  
 \biggr\} 
 \;, \hspace*{5mm} \la{channels_direct_s0} 
\ea \normalsize
where $u,t,s$ are the Mandelstam variables, 
the polylogarithmic functions appearing on the right-hand sides 
have been defined 
in \eqs\nref{lnb}--\nref{lnf}, 
$p^{ }_0 = \qp + \qm$, 
$p = \qp - \qm$, 
and
\be
 \int \!{\rm d}\Omega^{ }_{2\leftrightarrow 2}
 \; \equiv \; 
 \int_{\vec{p}^{ }_1\vec{p}^{ }_2\,\vec{p}^{ }_3}
 \hspace*{-1mm} 
 \frac{
 \deltabar\bigl( 
  \mathcal{P}^{ }_{\!1}
 + \mathcal{P}^{ }_{\!2}
 - \mathcal{P}^{ }_{\!3}
 - \mathcal{K}^{ }_{ }
 \bigr)
 }{8 p^{ }_1 p^{ }_2 p^{ }_3}
 \;. 
\ee
The function $\deltabar$ is defined such that 
$\int_\mathcal{P}\deltabar(\mathcal{P}) = 1$.
Even if not obvious from the right-hand sides
of the expressions, the definitions and numerical values
of 
$ \Xi_{(+)}^{\,t0} $
and 
$ \Xi_{(+)}^{\,s0} $ 
coincide. 

\small{
%

}

\end{document}